\documentclass[namedreferences,linksfromyear]{solarphysics}
\usepackage[optionalrh,natbib]{spr-sola-addons} 
\usepackage{graphicx}                    
\usepackage{color}                       
\usepackage{url}                         
\usepackage{soul}
\usepackage{hyperref}
\usepackage{afterpage}
\usepackage{longtable}
\usepackage{lscape}

\usepackage{xcolor}

\begin{document}

\begin{article}

\begin{opening}

\title{Relationship between Solar Energetic Particles and Properties of Flares
  and CMEs: Statistical Analysis of Solar Cycle 23 Events}

%

\author{M.~\surname{Dierckxsens}$^{1}$\sep 
        K.~\surname{Tziotziou}$^{2}$\sep		     
        S.~\surname{Dalla}$^{3}$\sep 
        I.~\surname{Patsou}$^{2}$\sep 
        M.~S.~\surname{Marsh}$^{3}$\sep 
        N.~B.~\surname{Crosby}$^{1}$\sep
        O.~\surname{Malandraki}$^{2}$\sep
        G.~\surname{Tsiropoula}$^{2}$
       }

%

%
\institute{$^{1}$ Belgian Institute for Space Aeronomy (BIRA-IASB), Ringlaan 3, 1180 Brussels, Belgium; 
  email: \url{Mark.Dierckxsens@aeronomie.be}\\
           $^{2}$ IAASARS, National Observatory of Athens, GR-15236 Penteli, Greece \\
           $^{3}$ Jeremiah Horrocks Institute, University of Central Lancashire, Preston, PR1 2HE, UK \\          }

\begin{abstract}

A statistical analysis of the relationship between solar energetic particles
(SEPs) and properties of solar flares and coronal mass ejections (CMEs) is
presented. 
SEP events during Solar Cycle 23 are selected which are
associated with solar flares originating on the visible hemisphere of the Sun
and at least of magnitude M1. 
Taking into account all flares and CMEs that occurred
during this period, the probability for the occurrence of a SEP event near
Earth is determined. 
A strong rise of this probability is observed for increasing flare
intensities, more western locations, larger CME speeds and halo CMEs.
The correlations between the proton peak flux and these solar
parameters are derived for a low ($>$10~MeV) and high ($>$60~MeV) energy
range excluding any flux enhancement due to the passage of fast
interplanetary shocks.
The obtained correlation coefficients are: $0.55 \pm 0.07$ ($0.63 \pm 0.06$)
with flare intensity and $0.56 \pm 0.08$ ($0.40 \pm 0.09$) with the CME speed
for $E>$10~MeV ($E>$60~MeV).  
For both energy ranges, the correlations with flare longitude and CME width
are very small or non-existent. 
Furthermore, the occurrence probabilities, correlation coefficients and
mean peak fluxes are derived in multi-dimensional bins combining the
aforementioned solar parameters.  
The correlation coefficients are also determined in different proton energy
channels ranging from 5 to 200~MeV.
The results show that the correlation between the proton peak flux and the CME
speed decreases with energy, while the correlation with the flare intensity
shows the opposite behavior.
Furthermore, the correlation with the CME speed is stronger than the correlation
with the flare intensity below 15~MeV and becomes weaker above 20~MeV.
When the enhancements in the flux profiles due to interplanetary shocks are not 
excluded, only a small but not very significant change is observed in
the correlation coefficients between the proton peak flux below 7~MeV
and the CME speed.

\end{abstract}

%

\end{opening}

%

\section{Introduction}

Solar energetic particles (SEPs) are accelerated at the Sun or in the
interplanetary medium during flares and coronal mass ejections (CMEs) reaching 
energies up to several GeV.
SEP events consist of electrons, protons and ions (up to Fe) that propagate
along the interplanetary magnetic field causing sudden and often long lasting
particle radiation enhancements. 
SEP events show significant variability in their relative abundance and
intensity profile, with the maximum of the proton flux near Earth reached in
timescales ranging from less than an hour to longer than a day.  
The characteristics of SEPs are obtained by {\it in situ} singular or multiple 
spacecraft observations. 
Some SEP profiles exhibit a short period of flux increase varying from a few
minutes to half a day when a fast
interplanetary shock passes the observation point, called an energetic storm
particle (ESP) event, indicating {\it in situ} acceleration of particles by the
shock \citep{Can1995}. 

Statistical studies of the properties of SEP events have been
carried out for decades (starting with \citet{vHo1975}), with the purpose
of broadening our understanding of particle acceleration and propagation.  
One of the outcomes of such studies has been the suggestion, in the 1990s, of
the so-called two-class paradigm for SEPs \citep{Cane1986,Ream1988,Ream1999,Ream2013}.  
The first class of events is associated with impulsive acceleration and
generally characterized by a short rise time of the flux
time profiles, observation point magnetically well connected with the location
of the associated flare, no clear association with a CME or interplanetary
shock, high electron/proton and $^3$He/$^4$He ratios and enhanced heavier
ion content.  
The second class is referred to as gradual events and predominantly described
by a slow rising particle flux, an association with a long lasting flare and
CME originating anywhere on the Sun or interplanetary shock, and a relative
composition similar to that of coronal material. 
Later, \citet{Can2010} studied 280 solar proton events and 
their associated flares and CMEs, and found a continuum of
event properties as opposed to the two classes of the standard SEP paradigm.

In recent years, statistical studies have also become important for space
weather forecasting by identifying which types of solar events are most
efficient at producing SEPs.
\citet{Lau2009} carried out an analysis of data from the 
{\it Geostationary Operational Environmental Satellite} (GOES) to develop a
technique for short-term warning of SEP events using flare properties and type
III radio emissions. 
Several other authors have studied the dependence of SEP characteristics on
various flare parameters \citep{Kur2004,Bel2005,Par2010,Cli2012}, CME
characteristics \citep{Kah2001,Gop2008,Par2012} or both
\citep{Wan2006,Can2010,Hwa2010,Mit2013,Ric2014}.  
Some of these studies also included information on type II 
\citep{Wan2006,Gop2008,Ric2014} and type III \citep{Can2010,Ric2014} radio
bursts. 

This article presents a statistical analysis of SEP events and their
associated flares and CMEs during Solar Cycle 23.  
The purpose of this study is twofold: 1) to characterize the probabilities
of SEP occurrence based on a number of characteristics of flares and CMEs 
and 2) for the solar events that did result in observed SEPs, to identify
quantitatively the relationship between the associated flare and CME
properties and the peak value of the proton flux during the SEP event. 
This analysis was performed as part of the EU FP7 project COMESEP
(COronal Mass Ejections and Solar Energetic Particles; \citealp{Cro2012};
http://www.comesep.eu). 
The main objectives of this project were the development of
tools to forecast geomagnetic and SEP radiation storms, and
the integration of these tools into an operational, fully automated space
weather alert system.  
The results of this statistical analysis are incorporated into the SEP
forecast tools of COMESEP.

Flare intensity and location as well as CME speed and angular width are known
to influence the occurrence and characteristics of SEP events, as demonstrated
in the aforementioned studies.  
As far as the location of the flare is concerned, the dependence on flare
longitude has been known since the earliest studies
({\it e.g.}\ \citet{vHo1975}; \citet{Can1988}), while any influence of latitude 
for near-Earth SEP events is unclear \citep{Dal2010}.  
Radio data is not considered in our analysis as there are currently no near
real-time radio burst detection tools available that can be used for forecast
purposes.  
While most studies have analyzed SEP events for protons within a
specific energy range, typically around 10 or 20~MeV, 
here we also consider a higher energy range dominated by protons of about
60~MeV ({\it i.e.}\ the range $E>$60~MeV), in addition to a low energy range
($E>$10~MeV). 
These two energy ranges are chosen to ensure adequate coverage of different
types of technological and biological effects induced by SEP radiation.
Hence, we investigate the dependence of the SEP peak flux in these two energy
ranges and the occurrence probability on flare intensity and longitude, and
CME speed and width. 
In this work, we go a step further by examining how the quantities
under study vary when combining flare and CME parameters, by exploring the
energy dependence of the correlation coefficients between the peak flux and
these solar parameters, and by assessing how not excluding the flux
originating from an ESP influences these values.

The event lists and SEP data used for this analysis are described in
Section~\ref{sec.listsdata} and our methodology is explained in detail in
Section~\ref{sec.method}.
Section~\ref{sec.results} presents our results which are discussed and
compared with previous works in Section~\ref{sec.discussion}.

\section{Event Lists and Associated Data}\label{sec.listsdata}

Our statistical studies cover the years 1997--2006, {\it i.e.}\ mostly Solar
Cycle 23, to take advantage of the {\it Large Angle and Spectrometric
  Coronagraph} (LASCO) on-board the {\it Solar and Heliospheric Observatory}
satellite (SOHO). 
The SEP dataset and the comprehensive lists of events (SEPs, solar flares
and associated CMEs) used for this analysis are described in the following
sections. 

\subsection{SEP Data} \label{subsec.sepdata}

For a statistical study of the characteristics of SEP events, it is
essential to use a well verified dataset. 
Therefore, we use the Solar Energetic Particle Environment Modelling
(SEPEM) reference proton dataset (\citet{Cro2014};
http://dev.sepem.oma.be) from the European Space Agency (ESA).    
Within this project, data from the {\it Space Environment Monitor} (SEM)
on GOES and the {\it Goddard Medium Energy} (GME) instrument on the
{\it Interplanetary Monitoring Platform} (IMP-8) for the period 1973 to 2013 were
cleaned, rebinned, cross-calibrated and merged to provide a uniform dataset of
ten differential energy channels exponentially distributed in the range from 5 to
200~MeV.
The SEP data used in our analysis constitute, thus, a more consistent dataset
than those used in previous studies that relied on data obtained by single or
multiple satellites and instruments.

\subsection{SEP Event Lists}\label{subsec.eventlists}

Our study of SEP occurrence probabilities and peak fluxes relies on
two SEP event lists as will be described in Section~\ref{sec.results}. 
In both lists, only events associated with
solar flares of at least magnitude M1  and with an identified source longitude
within the visible earthward solar hemisphere $[-90^{\circ},90^{\circ}]$ 
were retained. 

\paragraph*{CRR2010 List}
The first SEP list is a subset of the list of \citet{Can2010} (hereinafter
referred to as CRR2010) which they compiled from the analysis of $>25$~MeV proton
data from the IMP8/GME and SOHO/ERNE ({\it Energetic and Relativistic Nuclei and
Electron}) instruments.
The information on the solar flare and CME associated with each event was
also obtained from CRR2010 and verified with the original sources (see
Section~\ref{subsec.solarlists}).  
The values for the angular width were re-evaluated by CRR2010 for
asymmetric halo CMEs and in some cases differ substantially from those
provided in the SOHO/LASCO CME catalogue.  
After applying the aforementioned criteria to the associated flare, 
the subset contains 160 events which will be referred to in the
following as the CRR2010 list.

\begin{figure} 
\centerline{\includegraphics[width=0.95\textwidth,clip=]{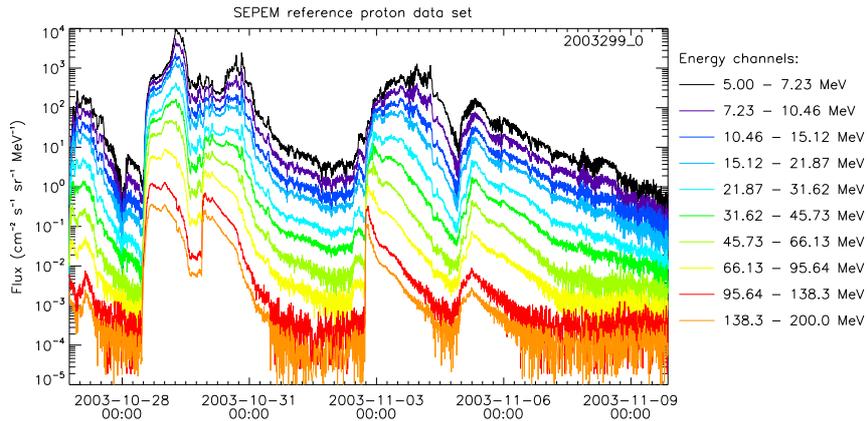}}
\caption{The flux-time profile  in the SEPEM reference proton data
  channels of the event starting on October 26, 2003. Separate enhancements
  during the event duration can be observed corresponding to seven events
  identified by CRR2010. }\label{fig.fluxprofile_multi} 
\end{figure}

\paragraph*{SSE List}
The second SEP event list is a subset of the reference proton event list
derived within the SEPEM project based on the proton energy range
$7.23-10.45$~MeV \citep{Jig2012}. 
The majority of events in this list have a clear counterpart in the CRR2010
list.
In some cases SEPEM events are in fact the sum of several
physical SEP events due to the requirement of a minimum dwell time of 24 hours
between consecutive events. 
These SEPEM events were split up into separate events based on the start times
of events in CRR2010. 
An example of a SEPEM event containing more than one CRR2010 event can be seen
in Figure~\ref{fig.fluxprofile_multi}.
The information of the associated flare and CME was obtained from the CRR2010
list.  
For the few events without a clear counterpart in this list, we
associated solar activity using the sources mentioned in
Section~\ref{subsec.solarlists}. 
This resulted in a total of 90 events satisfying the flare criteria and will
be referred to in the following as the SEPEM Sub Event (SSE) list. 
The full list of SSE events, including the parameters of the
  associated flares and CMEs, can be found in the Appendix in
  Table~\ref{tab.sselist}. 

\paragraph*{} The difference between the number of SEP events in the two
lists, 160 and 90 SEPs respectively, stems from both the different energy
ranges used for the compilation of the two lists, resulting from different
satellites/instruments, and the different threshold criteria used for
classifying a flux enhancement as a SEP event. 

\subsection{Solar Eruptive Event Lists}\label{subsec.solarlists}

In order to study the SEP occurrence probability, several solar eruptive event
lists (flares and CMEs) were collected. 
These lists and their original sources were also used to verify and complement
the associations with the SEP events as described in
Section~\ref{subsec.eventlists}. 

\paragraph*{Flare List}

A subset of the NOAA GOES X-ray solar flare catalogue (available at
http://www.ngdc.noaa.gov/stp/spaceweather.html)
was obtained by applying the same time range,
flare intensity and location criteria as for the SEP lists. 
This resulted in a sample of 1298 flares.

\paragraph*{Flare-CME List}

Events from the SOHO/LASCO CME Catalogue (available at
http://cdaw.gsfc.nasa.gov/CME\_list/) were associated by \citet{Dum2014}  
with solar flares from the NOAA X-ray solar flare catalogue using both
temporal and spatial criteria.  
All flares within one hour of each CME liftoff were selected.
For non-halo CMEs and flares with source location information, the difference 
between the CME and flare position angles was required to be less than half the
CME angular width.
In cases where multiple flares satisfy the above criteria, only the strongest flare was
chosen as the CME-associated flare.
Applying the same criteria as for the
flare list above, a total of 438 flare-CME pairs were obtained for our analysis.

\section{Methodology}\label{sec.method}

This section describes the methodology used for the
derivation of SEP occurrence probabilities in solar eruptive events, the
identification of the ESP events, the fitting to the
energy spectra of the proton peak flux during the SEP event, and the
derivation of integral fluxes. 

\subsection{Probabilities of SEP Occurrence in Solar Eruptive Events}\label{subsec.prob}

Given a subset $j$ of flare or CME events characterized by certain parameters
({\it e.g.} flare intensity or CME speed in a specified range), the probability of
SEP occurrence $P_j$ for subset $j$ is defined as: 
\begin{equation}
P_j = \frac{N_j^{SEP}}{ N_j},
\end{equation}
where $N_j$ is the total number of solar eruptive events in the subset $j$ and
$N_j^{SEP}$ the number of those events that resulted in an observed SEP event.

The uncertainty associated with the probability $P_j$ is derived
according to the binomial proportion confidence interval as
$\pm \sqrt{P_j(1-P_j)/N_j}$.
This corresponds to a $68\%$ confidence level when the sample
is described by a normal distribution, which is however a weak assumption for
the bins containing a small number of events.

\subsection{Identification of ESP Events}\label{subsec.esp}

An ESP is an enhancement in particle fluxes caused by a
local interplanetary shock wave. 
In the interplanetary medium, shocks accelerate lower energy particles more
efficiently, making ESP enhancements usually much more apparent in the lower
energy channels (see {\it e.g.} \citet{Cha2005}). 
For each event in our SSE list, the time profiles of the proton flux
in all energy channels were examined for ESP-like increases during the decay
phase of the initial event.  
Lists of interplanetary shocks observed at the {\it Advanced Composition
  Explorer} (ACE) and {\it Wind} spacecraft 
(http://www.cfa.harvard.edu/shocks) were used to verify
whether a shock was observed at 1 AU around the time of these ESP-like enhancements. 
An ESP could be identified during 45 out of the 90 events in our SSE list.
The ESP onset is determined as the time when the proton flux starts to
rise again after the initial decay.

\subsection{Proton Peak Fluxes and Energy Spectrum Fit} \label{subsec.energy_fit}

To obtain the peak flux in the ten differential SEPEM energy channels, the
maximum observed flux is determined before any identified ESP onset time 
during the event (see Section \ref{subsec.esp}).
The energy in each channel is determined as the square root of the product of
the lower and upper limits of the corresponding channel. 

\begin{figure} 
\mbox{\includegraphics[width=0.49\textwidth,clip=]{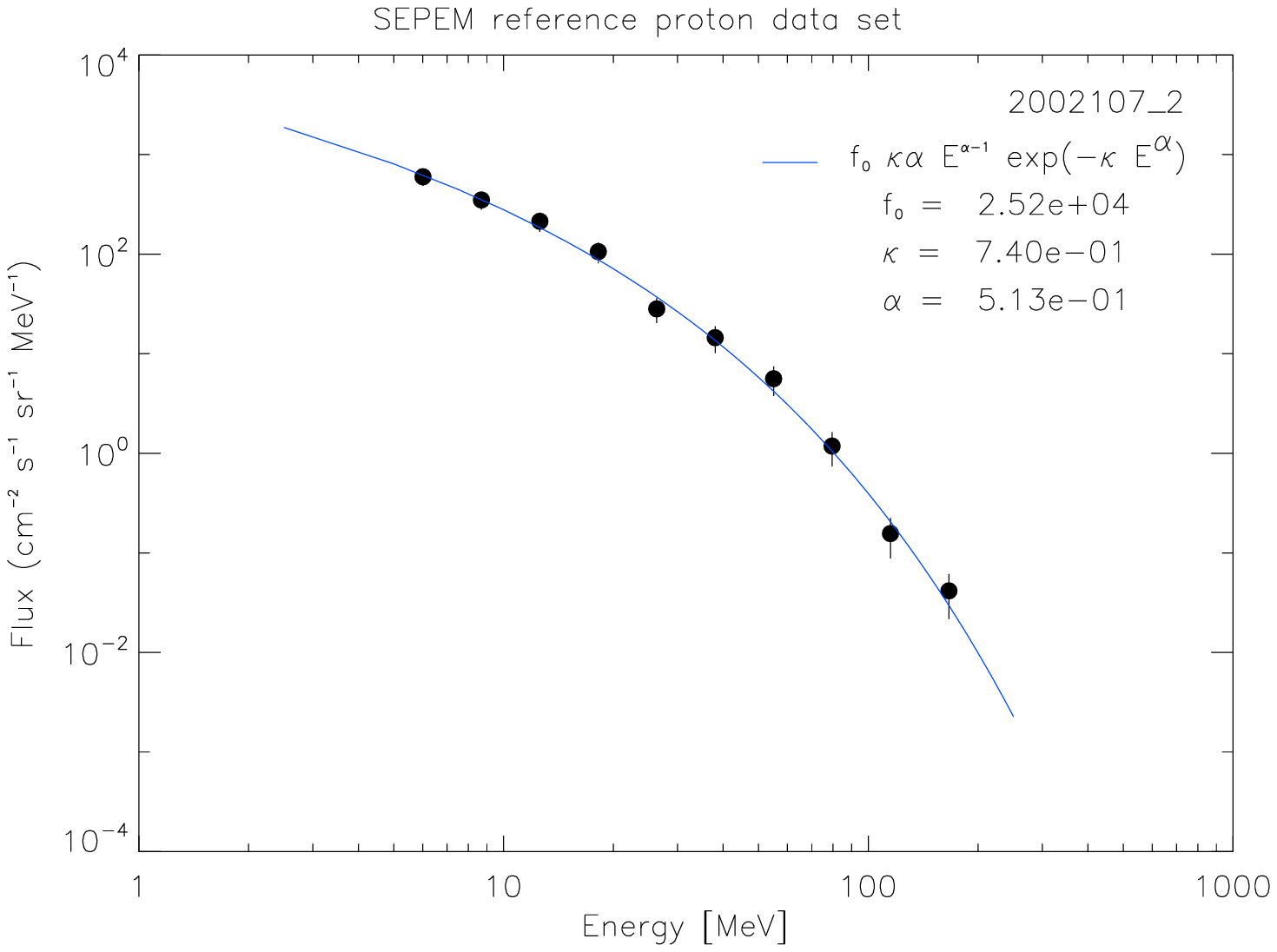}
  \includegraphics[width=0.49\textwidth,clip=]{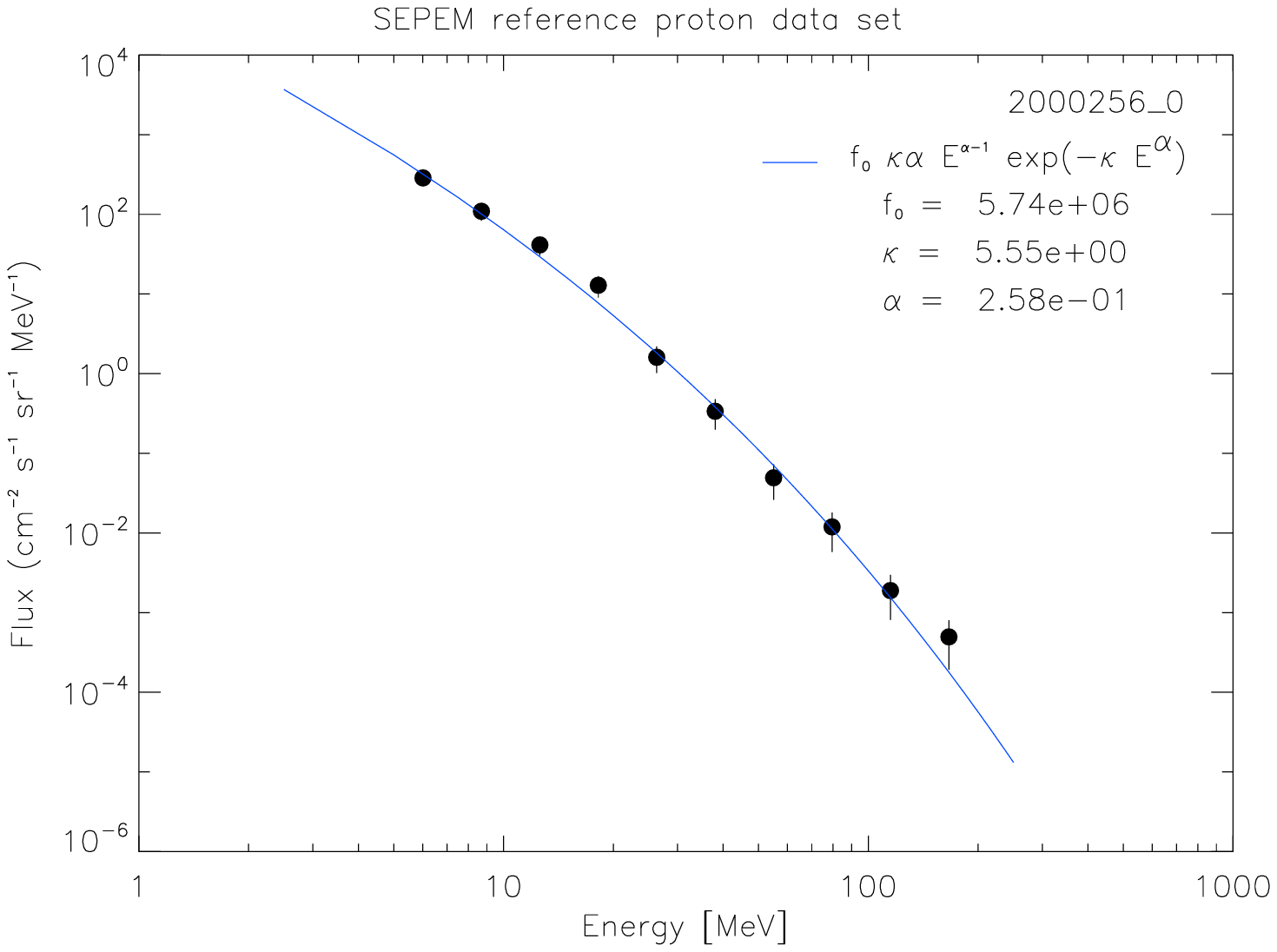}}
\caption{The energy spectrum of the proton peak flux for
the event on 21 April 2002 (left) and on 12 September 2000 (right) showing
two different but typical shapes.  
The black dots show the differential flux measured in each energy
channel, while the blue line represents the fit with a Weibull function. The
resulting fit parameters are also shown.}\label{fig.spectralfit}
\end{figure}

The obtained spectrum for each event is fitted with a Weibull function, which
was shown to describe SEP spectra accurately up to GeV energies
\citep{Xap2000} and has the form: \begin{equation}\label{eq.weibull}
\frac{d \phi}{d E} = f_0 \, \kappa \, \alpha \, E^{\alpha-1} e^{-\kappa  E^{\alpha}},
\end{equation}
with $\phi$ representing the particle flux, $E$ the energy of the particle
measured in MeV/nucleon, $f_0$ a normalization factor, and $\kappa$ and
$\alpha$ two parameters that determine the shape of the distribution. An
example of this fit for two different spectral shapes can be seen in
Figure~\ref{fig.spectralfit}.  

An uncertainty on the peak flux values was required in order for the fits
  to properly converge. Unfortunately, these values are not provided in 
  the SEPEM data. We have chosen to let the relative uncertainty on the flux
  value increase with the square root of the energy to reflect the increase in
  the uncertainty with the decrease in flux (particle count).
  Since the resulting uncertainties on the fit parameters are not used in the
  subsequent analysis, an arbitrary uncertainty of $10\%$ was assigned to the
  SEPEM reference energy channel ($7.23-10.45$~MeV).
  Various different dependencies on the flux and energy
  were explored, but it was found that this only had a very small effect on
  the derived integral peak fluxes that are described in
  Section~\ref{subsec.integral_flux}.

\subsection{Determination of Integral Peak Fluxes} \label{subsec.integral_flux}

Within a forecasting framework, it is useful to provide forecasts of peak
fluxes for a set of standard integral energy ranges. 
Here we have chosen to derive the dependence of the impact on the solar
parameters for the energy ranges  $E>$10~MeV and $E>$60~MeV.  
To construct an integral peak flux from the SEPEM fluxes, the results
from the fit to the peak flux spectrum as described in 
Section~\ref{subsec.energy_fit} are used. 
By integrating Equation~\ref{eq.weibull} and filling
in the fit parameters, the integral peak fluxes in a desired energy range can
be easily derived.
The resulting values of the integral peak flux in the aforementioned
energy ranges for each event in the SSE list can be found in
Table~\ref{tab.sselist} in the Appendix.

\section{Results} \label{sec.results}

This section presents our results on the probabilities of SEP occurrence
in solar eruptive events and the SEP event characteristics based on the
methodologies described in Section~\ref{sec.method}. 
We remind that the results described in
Section~\ref{subsec.results.prob} 
were derived using the larger CRR2010 list, since in this analysis we are
mainly interested in SEP occurrence probabilities as a function of
flare and CME characteristics, irrespective of the particular SEP profile
characteristics. 
The results described in Section~\ref{subsec.results.impact}  were derived
using the SSE list since it contains the detailed SEP profile
characteristics based on the cleaned dataset.

\subsection{Probabilities of SEP Occurrence}\label{subsec.results.prob}

We have derived the probabilities of SEP occurrence (see
section~\ref{subsec.prob}) by determining the fraction of flares in the Flare
or Flare-CME list resulting in an entry in the CRR2010 list as a function of
several solar event parameters.
Since the COMESEP SEP forecast tools are only triggered following the
observation of a soft X-ray solar flare, the combinations chosen in the
following sections are based on the potential availability of data after the
measurement of the flare magnitude.
For comparison, the same quantities have been derived as a function of the flare
parameters for the SSE list used to derive the proton peak fluxes.

\subsubsection{Probabilities as a Function of Flare Parameters}\label{sec.probflares}

\begin{figure}[t]
\mbox{\includegraphics[width=0.49\textwidth]{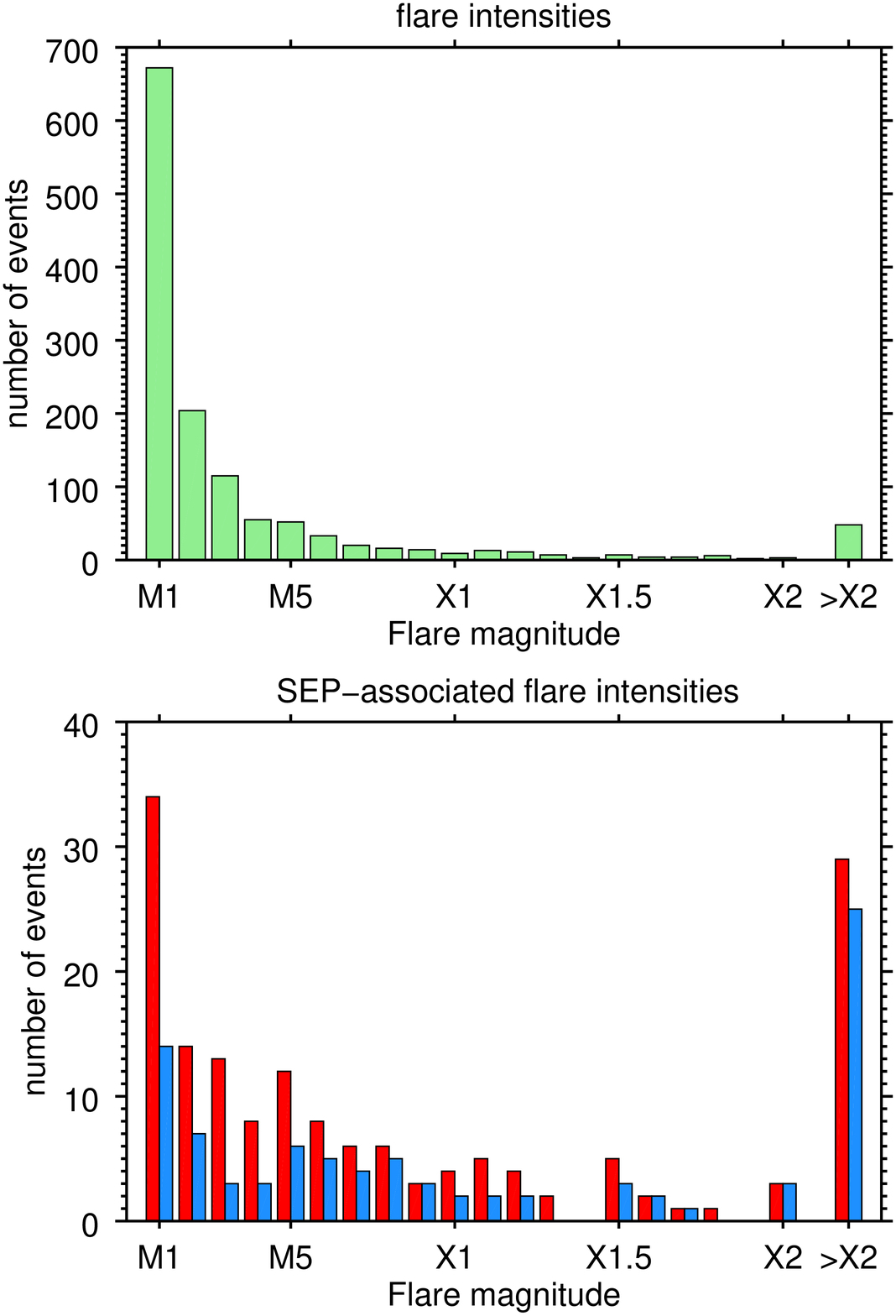}
\includegraphics[width=0.49\textwidth]{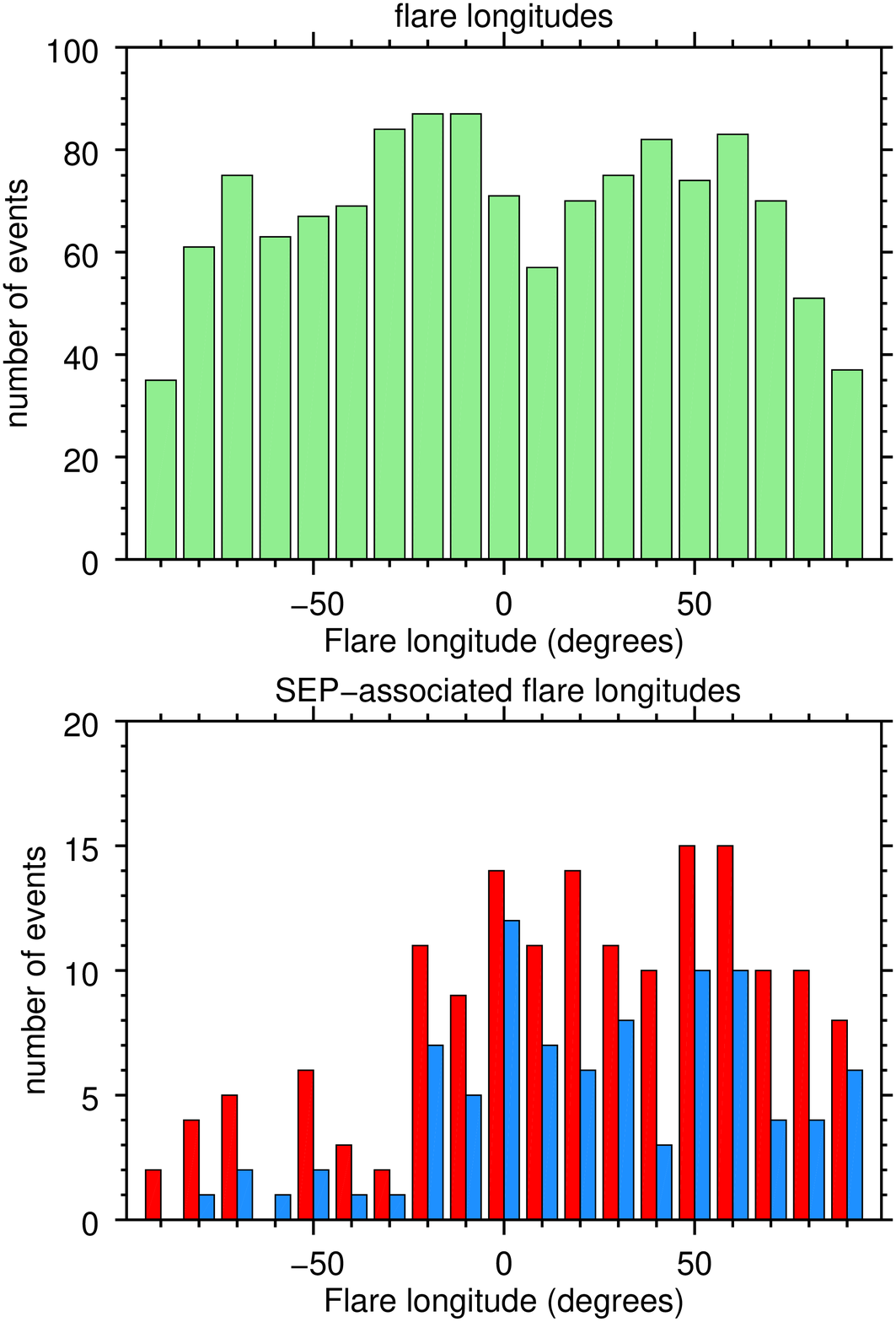}}
\caption{Top row: distributions of flare magnitude (left) and flare longitude
  (right) of the sample of 1298 flares used for the analysis of Section
  \ref{sec.probflares}. Bottom row: distributions of the respective parameters
  of the flares associated to the events in the CRR2010 list (red bars) and in the
  SSE list (blue bars).}\label{fig.flare_charact}
\end{figure}

Figure~\ref{fig.flare_charact} shows the distributions of flare magnitude and
longitude of the samples containing the 1298 flares and 160
SEP-associated flares from the CRR2010 list used for the
derivation of SEP occurrence probabilities.
The distributions for the 90 SEP-associated flares of the SSE list are also
shown.
In the CRR2010 list, 104 SEP events are related to M-class flares and 56 events
to X-class flares, while the respective numbers in the SSE list
are 50 and 40.
Within a solar cycle, the frequency of flares drops considerably with increasing
magnitude while there is no preference for the longitude location
on the solar disk, as our distributions clearly demonstrate.
As for flares resulting in an observed SEP event, the decline in the
number of events as a function of flare magnitude is less steep, while
there is a clear westward preference in location.

Probabilities are derived as a function of flare magnitude only, and
as a function of both flare magnitude and longitude.
Given the relatively small number of SEP events (160) compared to the total
flare sample (1298), the analysis
is performed within a limited number of flare magnitude and flare
longitude bins: five flare magnitude bins, namely
[M1-M3.9], [M4-M6.9], [M7-M9.9], [X1-X4.9] and [$\ge$X5], and five
longitude bins, namely [-90$^{\circ}$,-71$^{\circ}$],
[-70$^{\circ}$, -31$^{\circ}$], [-30$^{\circ}$,29$^{\circ}$],
[30$^{\circ}$,69$^{\circ}$] and [70$^{\circ}$,90$^{\circ}$].
The bin widths were defined by taking into account the
distributions presented in Figure~\ref{fig.flare_charact}. 
They are equidistant for longitude, as the distribution is rather flat,
while for flare magnitude the choice was based both on the presence of
sufficient number of events {\it per} bin and the importance of specific
flare magnitudes for SEP occurrence. It should be noted that a
different choice of binning resulted in qualitatively similar results as
presented here.
Figure~\ref{fig.probflareint} (and Table~\ref{tab.probflareint} in the Appendix)
exhibit a clear rise in SEP occurrence probabilities for increasing flare
magnitude.
The same probabilities derived for the SSE list are also shown.
Although probabilities are smaller compared to the ones obtained with the
CRR2010 list as a result of the lower number of selected SEP events,
their flare magnitude dependence shows a rather good qualitative agreement.

\begin{figure}[tp]
  \centering
  \includegraphics[width=0.49\textwidth]{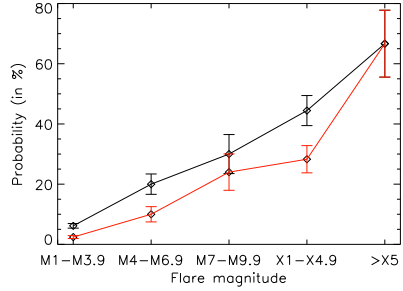}
  \caption{SEP occurrence probabilities (in \%) and their respective errors as a
    function of flare magnitude derived from the CRR2010 list (black line) and
    the SSE list (red line) for five different flare magnitude
    bins: [M1,M3.9], [M4,M6.9], [M7,M9.9], [X1,X4.9] and
    [$\ge$X5].}\label{fig.probflareint}
\end{figure}

Figure~\ref{fig.probflareintlon} (and
Table~\ref{tab.probflareintlon} in the Appendix) show that the derived SEP
occurrence probabilities increase for both stronger and more westward flares.
The probabilities derived with the SSE list 
exhibit a very similar behavior to the values derived with the CRR2010 
list although they are generally somewhat lower.
Some of the most significant decreases in the SSE SEP probabilities
compared to the respective CRR2010 SEP probabilities are present in the
[X1,X4.9] flare magnitude bin for both
east [-90$^{\circ}$,-71$^{\circ}$] and west [70$^{\circ}$,90$^{\circ}$] longitude bins. 
A detailed investigation of the events not present in the SSE list that
fall in these bins has shown that they are mainly
electron dominated events, which are not included due to the different
selection criteria.

\begin{figure}[tp]
  \centering
  \includegraphics[width=0.49\textwidth]{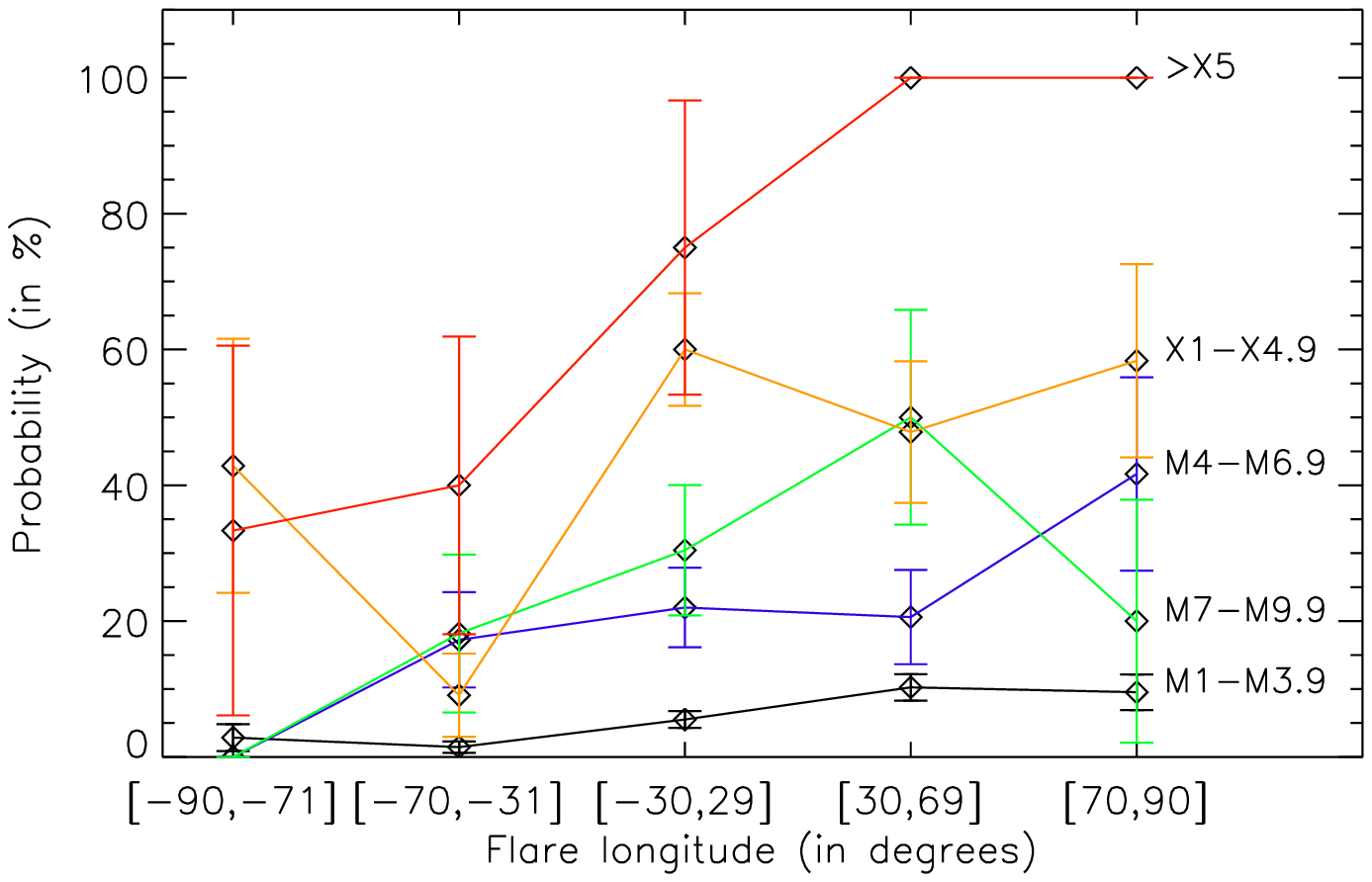}
  \includegraphics[width=0.49\textwidth]{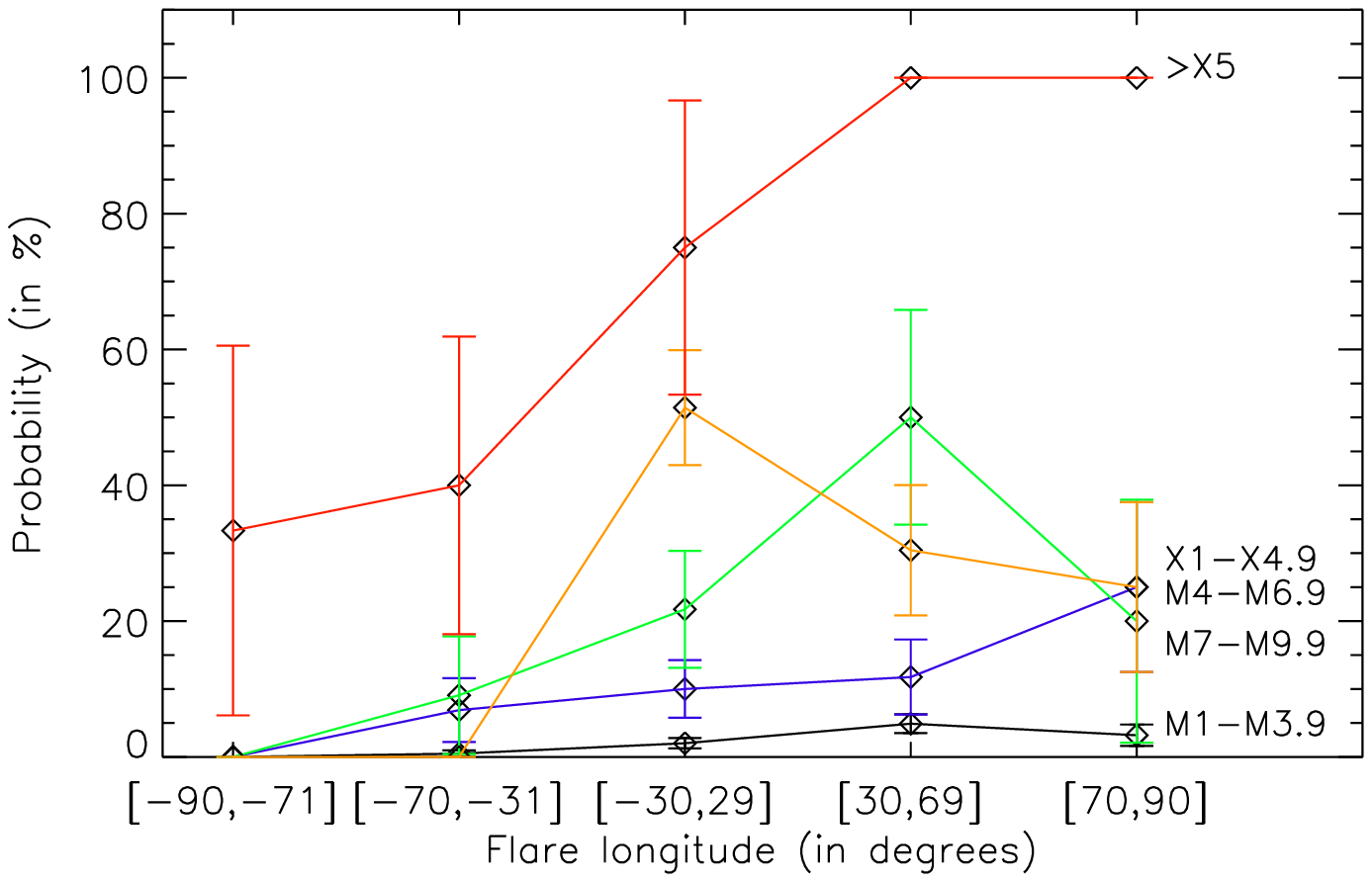}
  \caption{SEP occurrence probabilities (in \%) and their respective errors as a
    function of flare magnitude and location derived from the CRR2010 list (left)
    and the SSE list (right) for five longitude bins: [-90$^{\circ}$,-71$^{\circ}$],
    [-70$^{\circ}$,-31$^{\circ}$], [-30$^{\circ}$,29$^{\circ}$],
    [30$^{\circ}$,69$^{\circ}$] and [70$^{\circ}$,90$^{\circ}$], and five flare
    magnitude bins: [M1,M3.9] (black), [M4,M6.9] (blue), [M7,M9.9] (green),
    [X1,X4.9] (orange) and [$\ge$X5] (red).}
  \label{fig.probflareintlon}
\end{figure}

\subsubsection{Probabilities as a Function of Flare and CME Characteristics}

\begin{figure}[tbp]
\mbox{\includegraphics[width=0.49\textwidth]{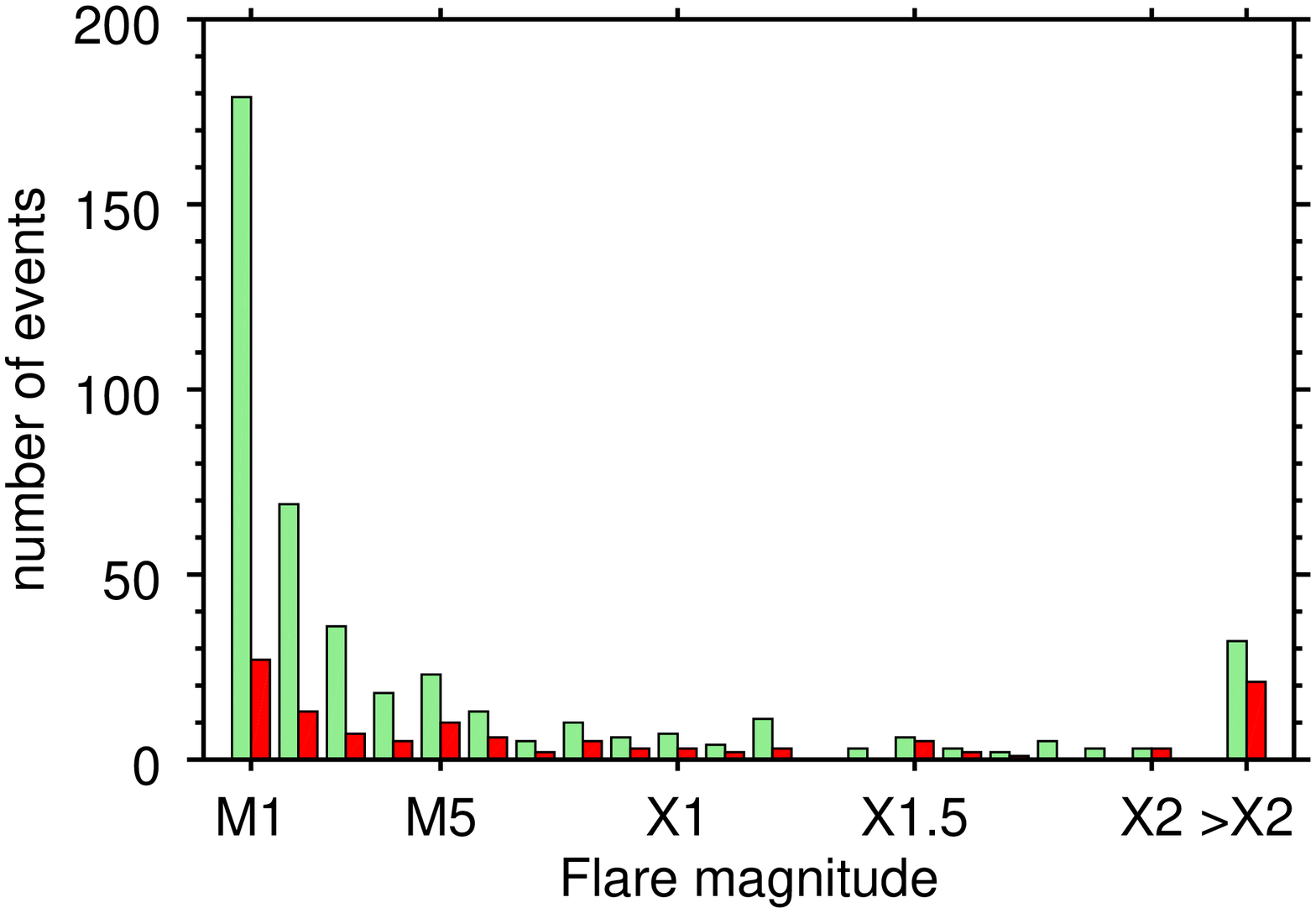}
\includegraphics[width=0.49\textwidth]{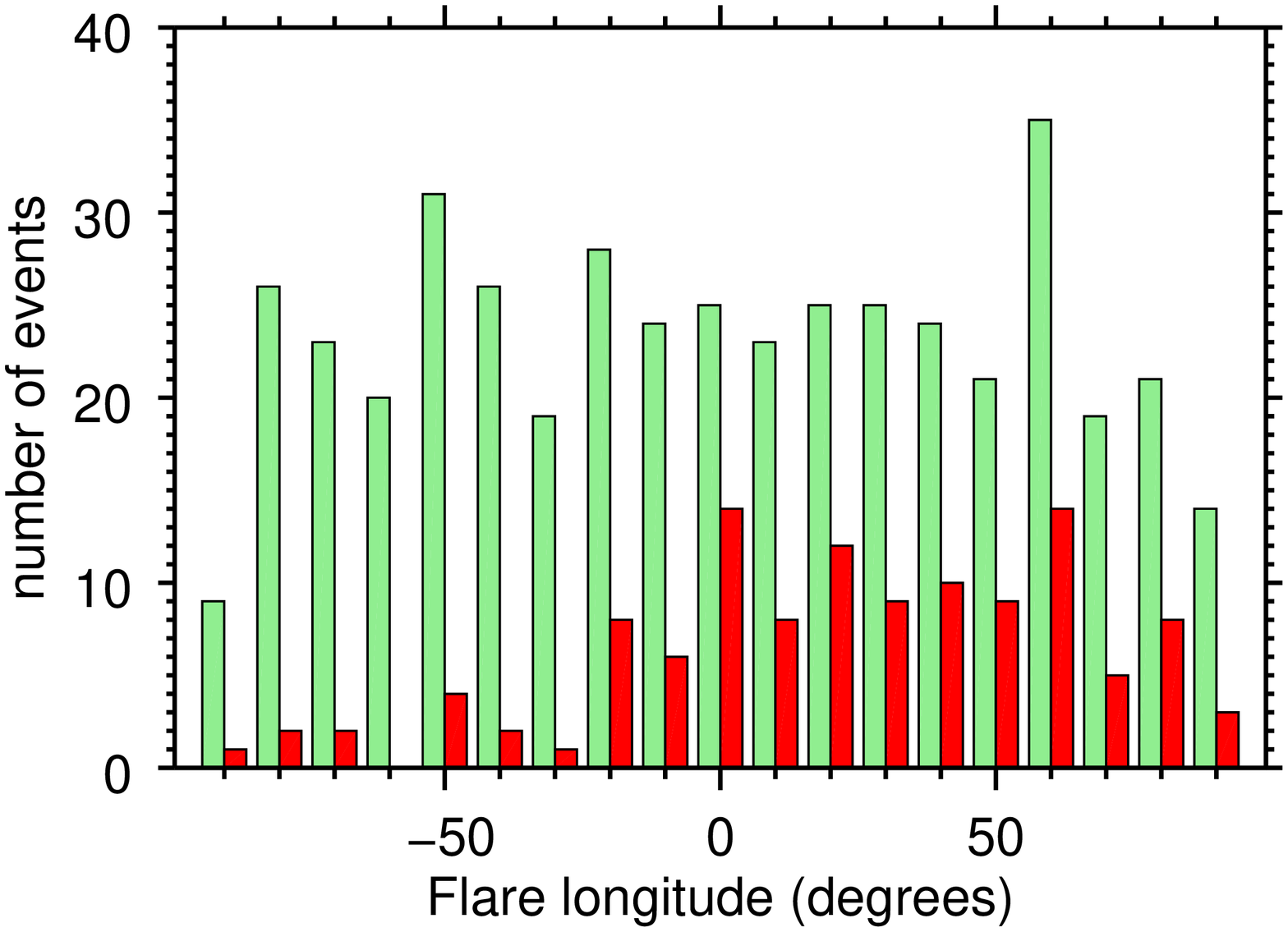}}\\
\mbox{\includegraphics[width=0.49\textwidth]{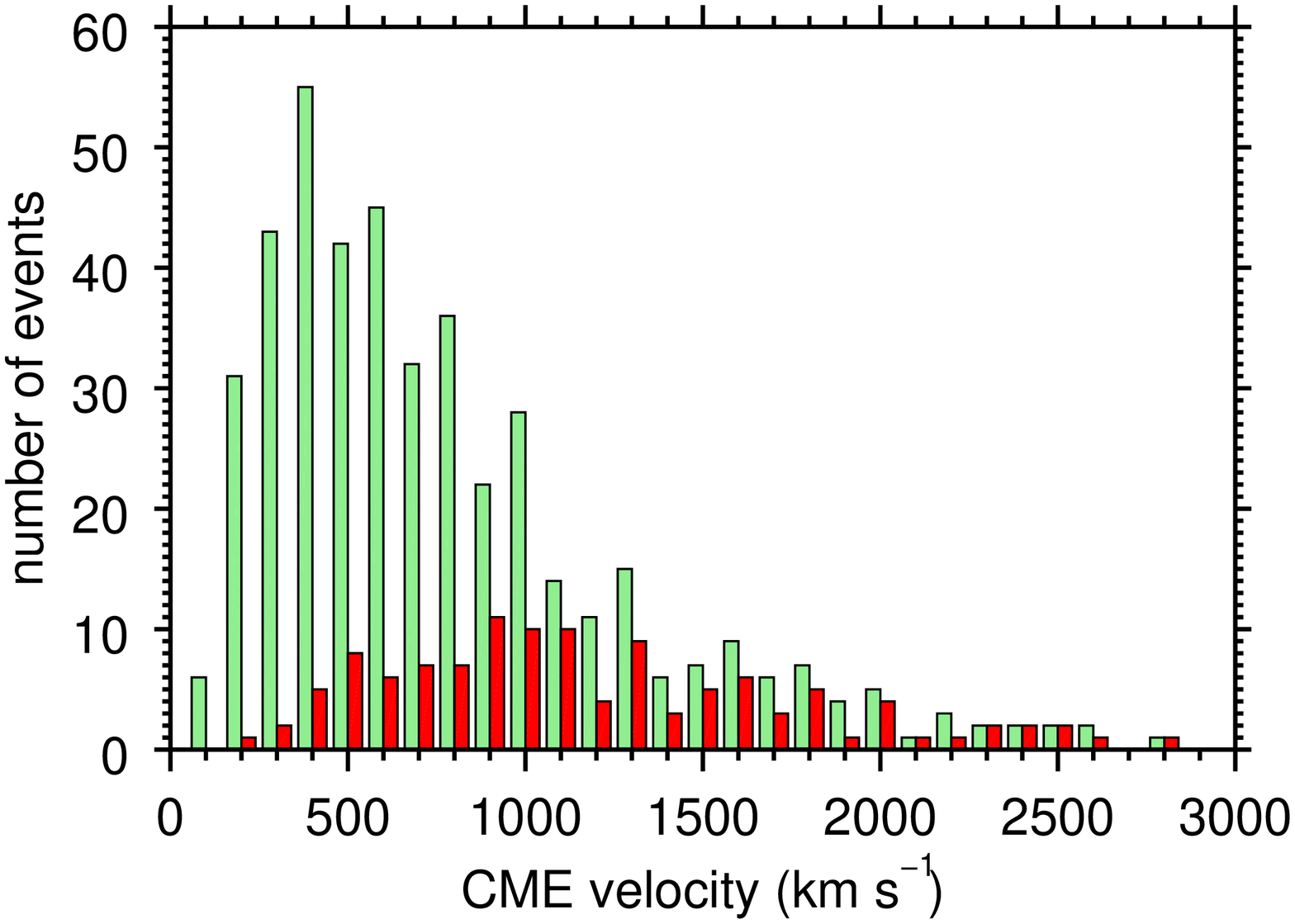}
\includegraphics[width=0.49\textwidth]{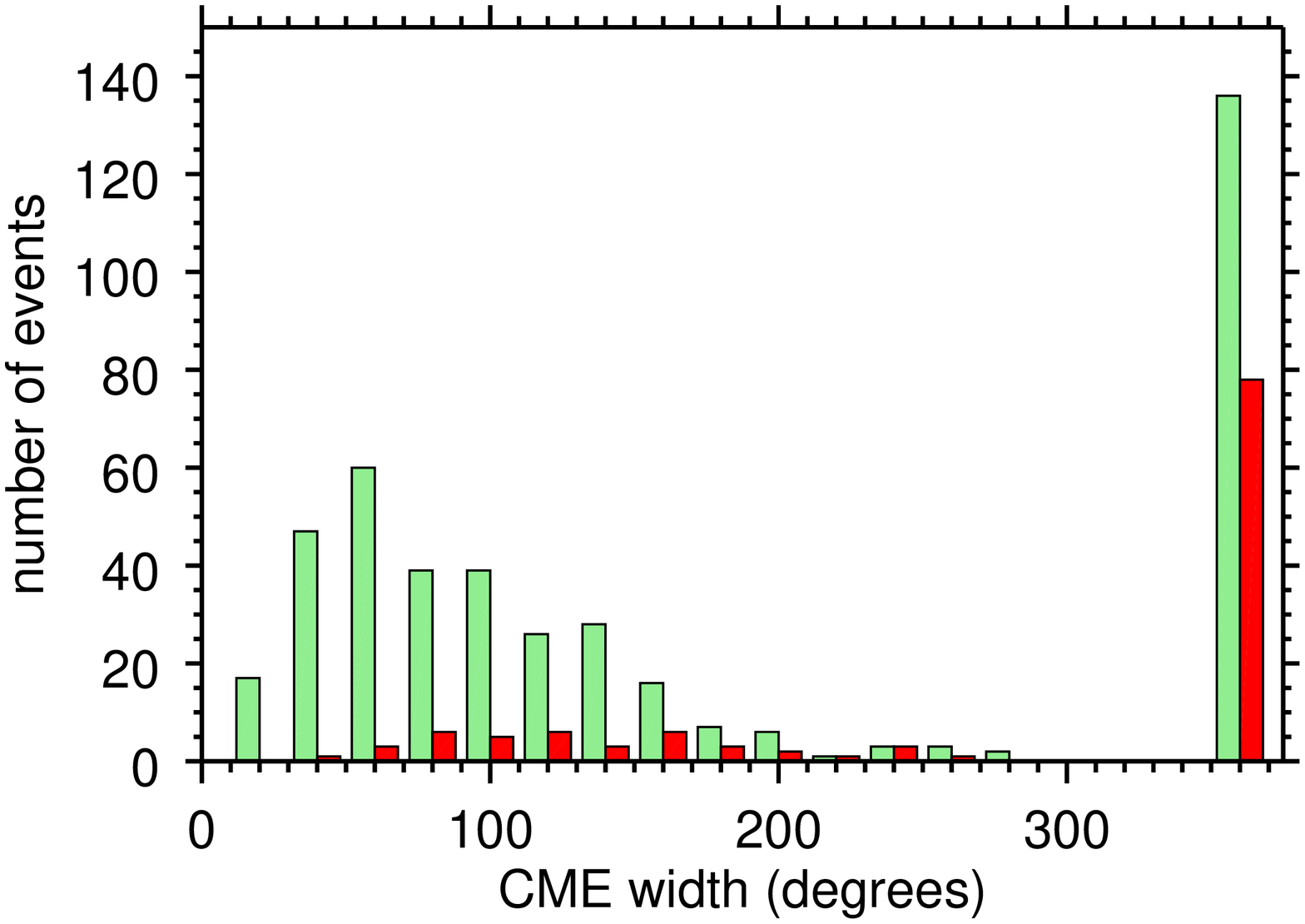}}
\caption{Distributions of flare magnitude, flare longitude, CME velocity
and CME width for the 438 CME-associated flare sample (green bars) and the 118
SEP-associated events (red bars). In the CME width distribution the 360$^\circ$ bin
represents halo CMEs.}\label{fig.flarecme_charact}
\end{figure}

Figure~\ref{fig.flarecme_charact} shows the distributions of flare magnitude,
flare longitude, CME velocity and CME width for the 438 flare-CME
sample and the 118 SEP-associated events
used to derive the SEP probabilities as a function of flare and CME
characteristics.
Out of these 118 SEP events, 78 are
associated with M-class flares while 40 are associated with X-class flares.
The flare magnitude and longitude distributions show similarities with
the respective distributions of the sample of 1298 flares discussed in the
previous section (see  Figure~\ref{fig.flare_charact}, top row), suggesting that
this CME-associated flare sample is quite representative of the initial
flare sample.
The majority of CME velocities are in the range from 0 to 1500~km~s$^{-1}$,
with a tail in the distribution extending to almost 3000~km~s$^{-1}$.
In the CME width distribution, there seem to be two distinct and similarly
sized populations: one of CMEs with a width in the range $[0^{\circ},
300^{\circ}]$ and one of halo CMEs (360$^{\circ}$).
When only the subset resulting in SEP events is considered (118 events) we
notice again a clear preference for westward longitudes
and a much higher mean CME velocity of the order of 1000~km~s$^{-1}$.

\begin{figure}[tbp]
\centering
\includegraphics[width=0.49\textwidth]{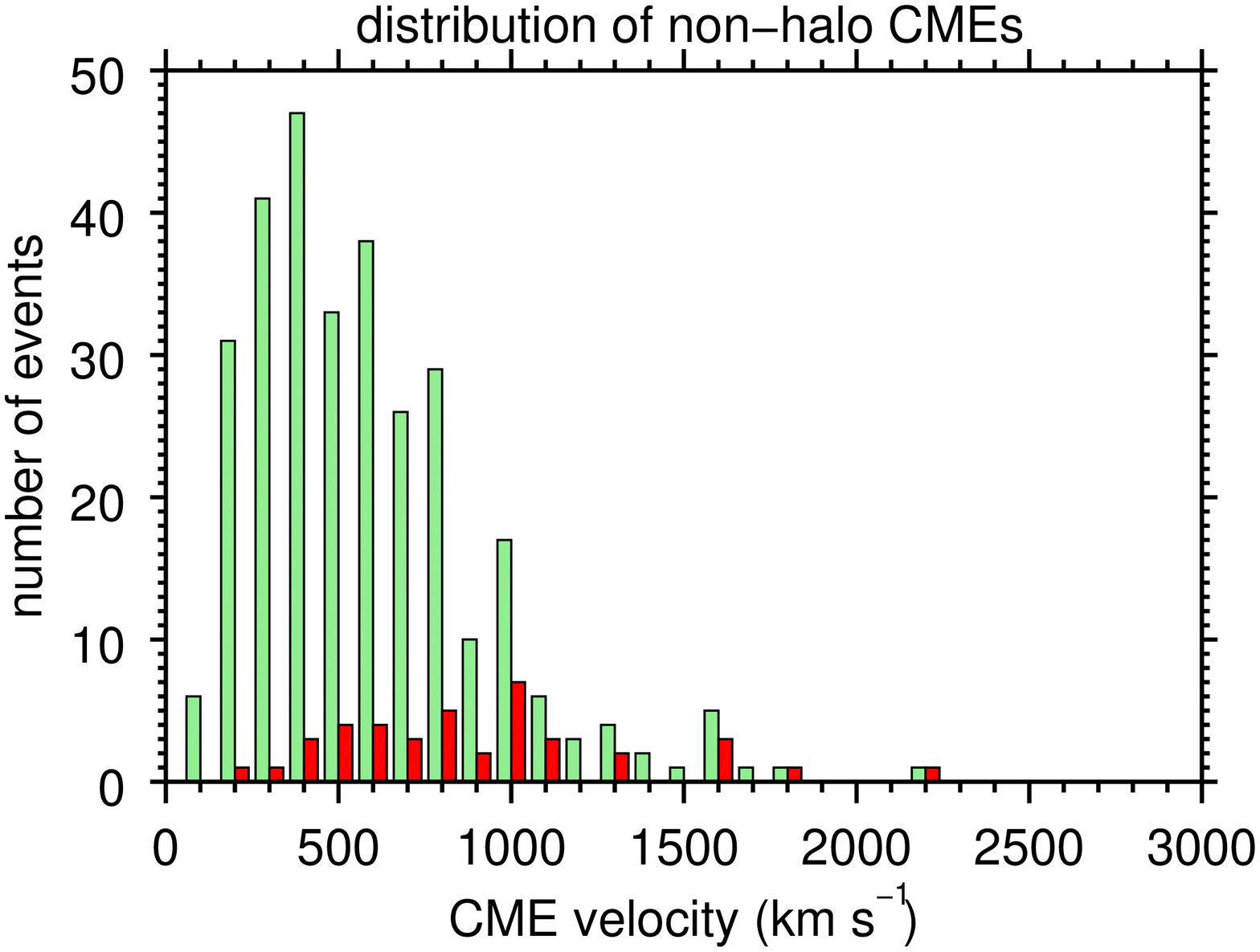}
\includegraphics[width=0.49\textwidth]{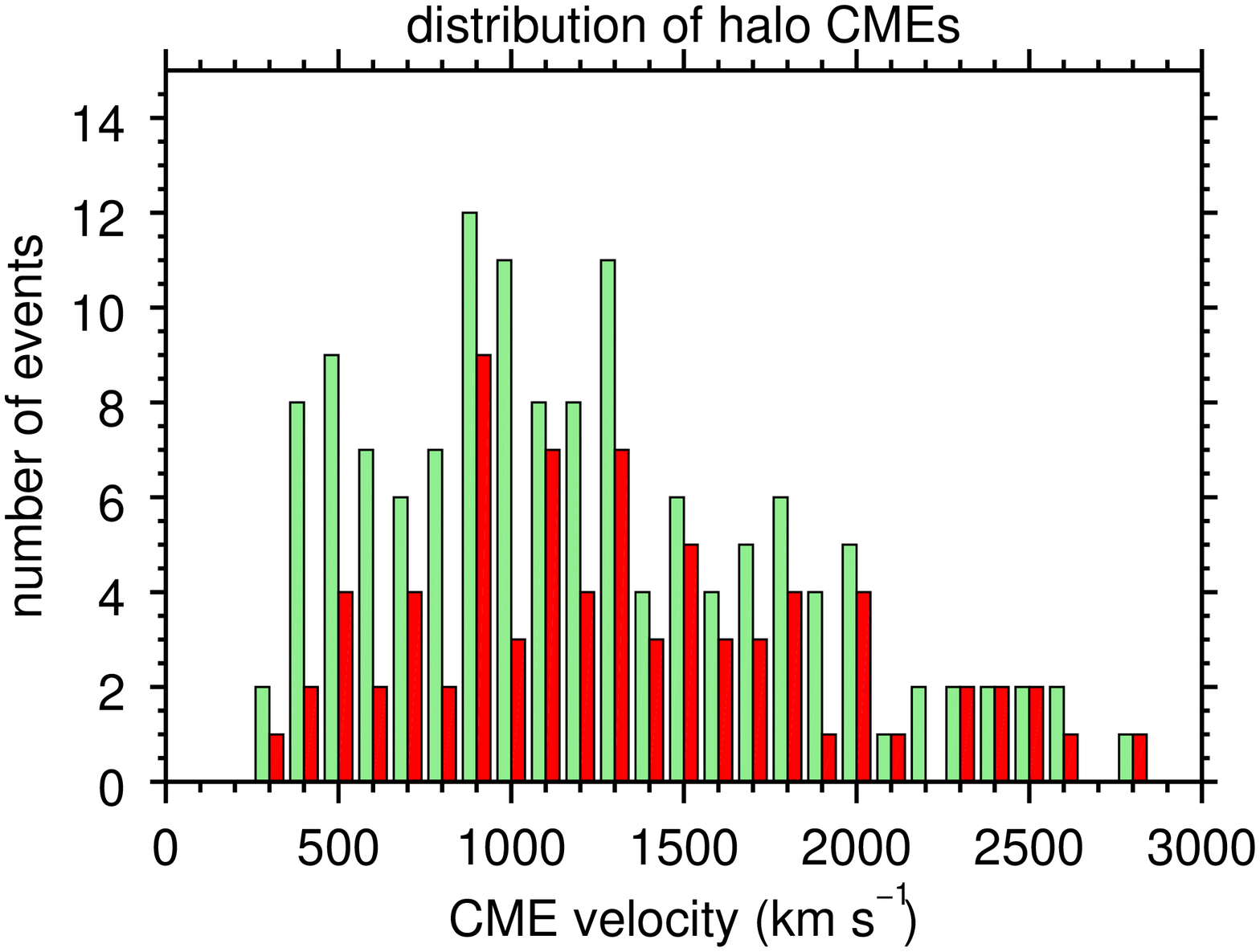}
\caption{Distribution of non-halo (left panel) and the halo (right panel)
    CME velocity for the CME-associated flare sample 
  (green bars) and for the corresponding SEP-associated sample (red
  bars).}\label{fig.flarecme_characthalo}
\end{figure}

Given the small number of both SEPs (118) and CME-associated flares (438), the
derivation of SEP occurrence probabilities has been performed for a small
number of flare magnitude, flare longitude, and CME velocity bins to limit
statistical fluctuations: three flare magnitude bins, namely
[M1-M3.9], [M4-M9.9] and
[$\ge$X1]; two longitude bins, namely [-90$^{\circ}$,-1$^{\circ}$] and
[0$^{\circ}$, 90$^{\circ}$]; and three CME velocity bins, namely
[0,499], [500,999] and [$\ge$1000]~km~s$^{-1}$.
Again the choice of binning is based on the distributions
of Figure~\ref{fig.flarecme_charact} and the requirement to have sufficient
number of events {\it per} bin.
As for the dependence on CME width,
two distinct cases are considered: non-halo ($<360^{\circ}$) and halo
(360$^{\circ}$) CMEs.
As Figure~\ref{fig.flarecme_characthalo} clearly demonstrates, the ratio
between the distribution of the SEP-associated sample and the 
full sample is rather constant for halo CMEs, except for the very
fast CMEs ($\ge$2000~km~s$^{-1}$). 
For the very fast halo CMEs the sample numbers are, however, very
small. 
Since there is no particularly strong dependence on the CME velocity for
halo CMEs, the SEP occurrence probabilities will only be derived as a
function of flare characteristics.
This is in contrast to the non-halo CMEs which clearly show a strong
dependence on the CME velocity. 
The difference in the dependency on the CME velocity between the two CME
categories is further explored in Section~\ref{sec.probcmes}.

CME velocity and width are both derived simultaneously following the detection
of a CME in SOHO/LASCO.
This is in contrast to the flare parameters since the location is not always
known when a flare is observed as they are obtained from different
instruments.
We therefore consider the following cases of SEP occurrence probabilities: 
a) as a function of flare magnitude alone for halo CMEs, 
b) as a function of flare magnitude and CME velocity for non-halo CMEs,
c) as a function of both flare magnitude and longitude for halo CMEs,
and d) as a function of flare magnitude and longitude and CME velocity
for non-halo CMEs. 

Figure~\ref{fig.probflarecmeinthalo} (and Table~\ref{tab.probflarecmeinthalo}
in the Appendix) show the derived SEP occurrence
probabilities as a function of flare magnitude for halo CMEs while
Figure~\ref{fig.probflarecmeintnhalo} (and
Table~\ref{tab.probflarecmeintnhalo} in the Appendix)
show these quantities as a function of flare magnitude
and CME velocity for non-halo CMEs.
For halo CMEs, the SEP occurrence probability increases with flare
magnitude but seems to stay almost constant for flares with magnitudes
larger than M5.
For non-halo CMEs, with the exception of SEPs associated with
flares in the [M4-M9.9] magnitude bin and CMEs faster than 1000~km~s$^{-1}$,
there is a clear monotonic dependence on both flare magnitude and CME velocity.

\begin{figure}[t]
  \centering
  \includegraphics[width=0.49\textwidth]{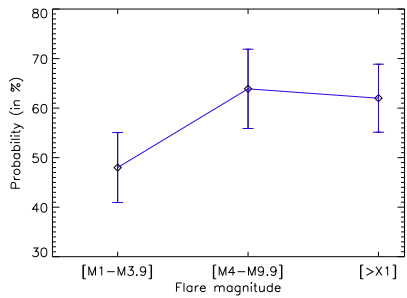}
  \caption{SEP occurrence probabilities (in \%) and their respective errors as a
    function of flare magnitude for halo CMEs.}\label{fig.probflarecmeinthalo}
\end{figure}

\begin{figure}[t]
  \centering
  \includegraphics[width=0.49\textwidth]{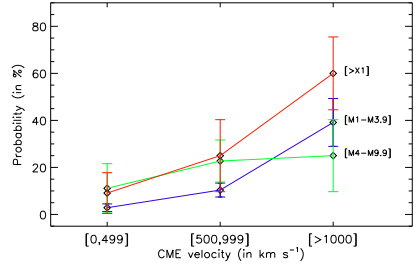}
  \caption{SEP occurrence probabilities (in \%) and their respective errors as a
    function of flare magnitude and CME velocity for non-halo CMEs.}
  \label{fig.probflarecmeintnhalo}
\end{figure}

Figure~\ref{fig.probflarecmeintlonhalo} (and
Table~\ref{tab.probflarecmeintlonhalo} in the Appendix) show the derived SEP
occurrence 
probabilities as a function of flare magnitude and longitude for halo CMEs.
Probabilities do generally show a monotonic
dependence with flare magnitude and are much higher for events originating at
the western solar hemisphere as noted before.
Figure~\ref{fig.probflarecmeintlonnhalo} (and
Table~\ref{tab.probflarecmeintlonnhalo} in the Appendix) show the derived SEP
occurrence 
probabilities as a function of flare magnitude, flare longitude and CME
velocity for non-halo CMEs.
Again, probabilities are generally higher for larger flares and
westward flare longitudes, while there also exists a clear monotonic
dependence on CME velocity,
with fast CMEs resulting in higher SEP probabilities.
There is still an exception for the fastest CMEs in the [M4,M9.9] magnitude
bin as previously noted, which is now clearly associated with westward
flares.

\begin{figure}[t]
  \centering
  \includegraphics[width=0.49\textwidth]{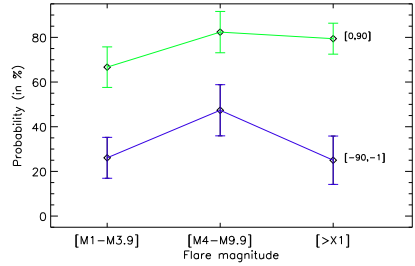}
  \caption{SEP occurrence probabilities (in \%) and their respective errors as a
    function of flare magnitude and longitude for halo
    CMEs.}\label{fig.probflarecmeintlonhalo}
\end{figure}

\begin{figure}[t]
  \centering
  \includegraphics[width=1.\textwidth]{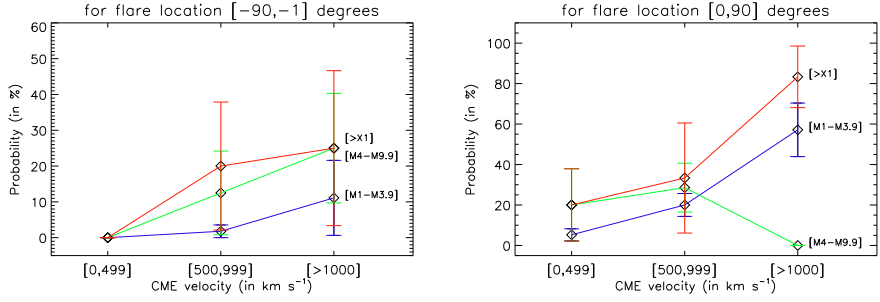}
  \caption{SEP occurrence probabilities (in \%) and their respective errors as a
    function of flare magnitude, flare longitude and CME velocity for non-halo
    CMEs.}\label{fig.probflarecmeintlonnhalo}
\end{figure}

\subsubsection{Probabilities as a Function of CME Characteristics} \label{sec.probcmes}

All CMEs of the studied sample are associated with flares and
hence SEP occurrence probabilities were derived by
taking into account both flare and CME characteristics. 
As a complementary analysis to the aforementioned combinations, the
dependency on CME characteristics only was also studied.  
Figure~\ref{fig.probcme} (and Table~\ref{tab.probcme} in the Appendix) show
the derived SEP 
occurrence probabilities as a function of CME velocity only for the 438
flare-CME sample, regardless of flare magnitude and location, and for the
two considered categories defined by CME width: non-halo and halo CMEs. 
For all cases, the SEP occurrence probability increases with CME
velocity, where the behavior is much more pronounced for non-halo CMEs
compared to the halo CMEs.
For the latter category, the probability is considerable even for
slow CMEs while for non-halo CMEs, the probability becomes considerable
only for the fastest CMEs. 
Figure~\ref{fig.probcme} also shows that regardless of CME
velocity, the probability for a SEP event to occur following a 
halo CME is significantly larger than after a non-halo CMEs.

\begin{figure}[t]
  \centering
  \includegraphics[width=0.49\textwidth]{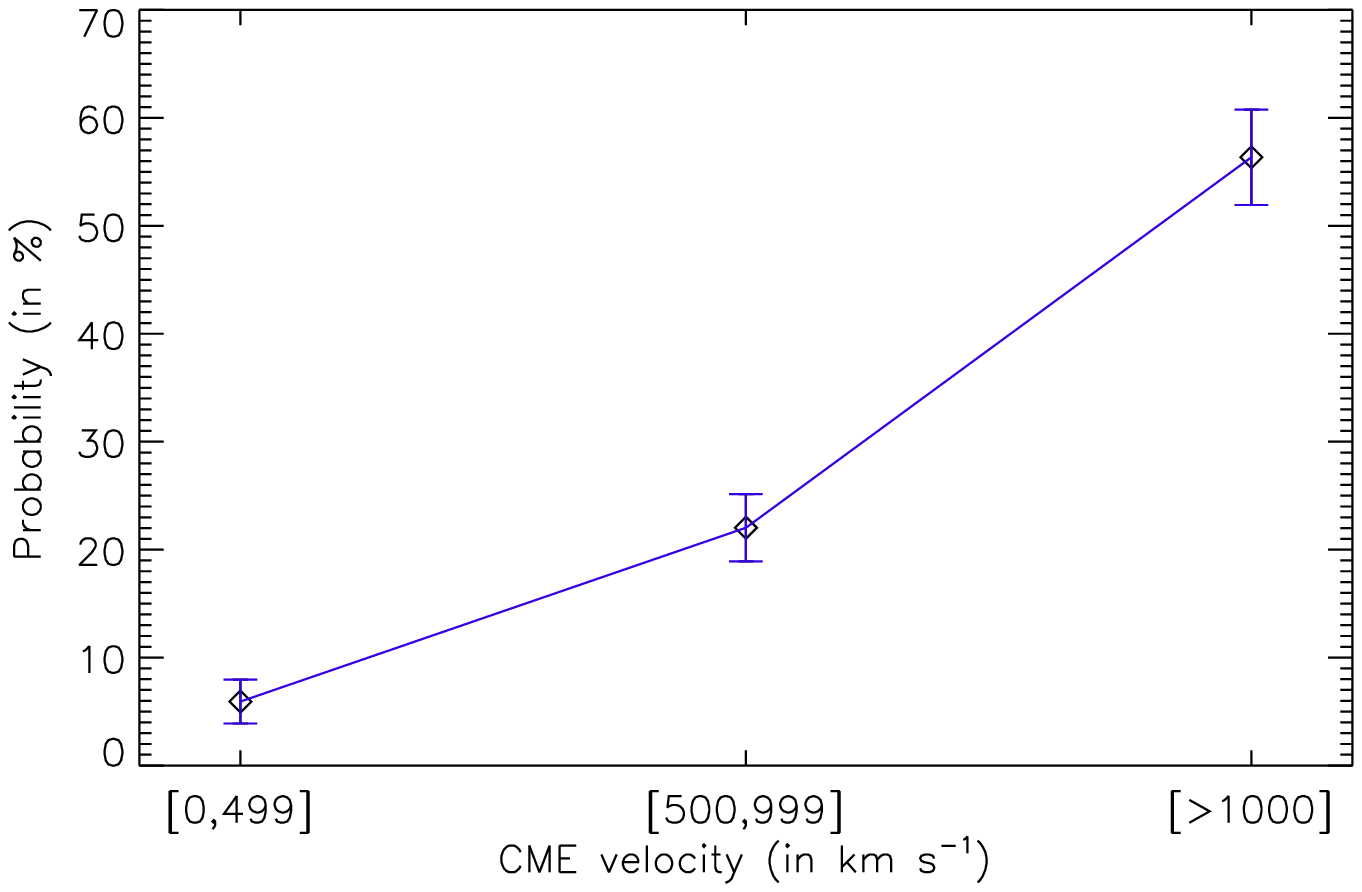}
  \includegraphics[width=0.49\textwidth]{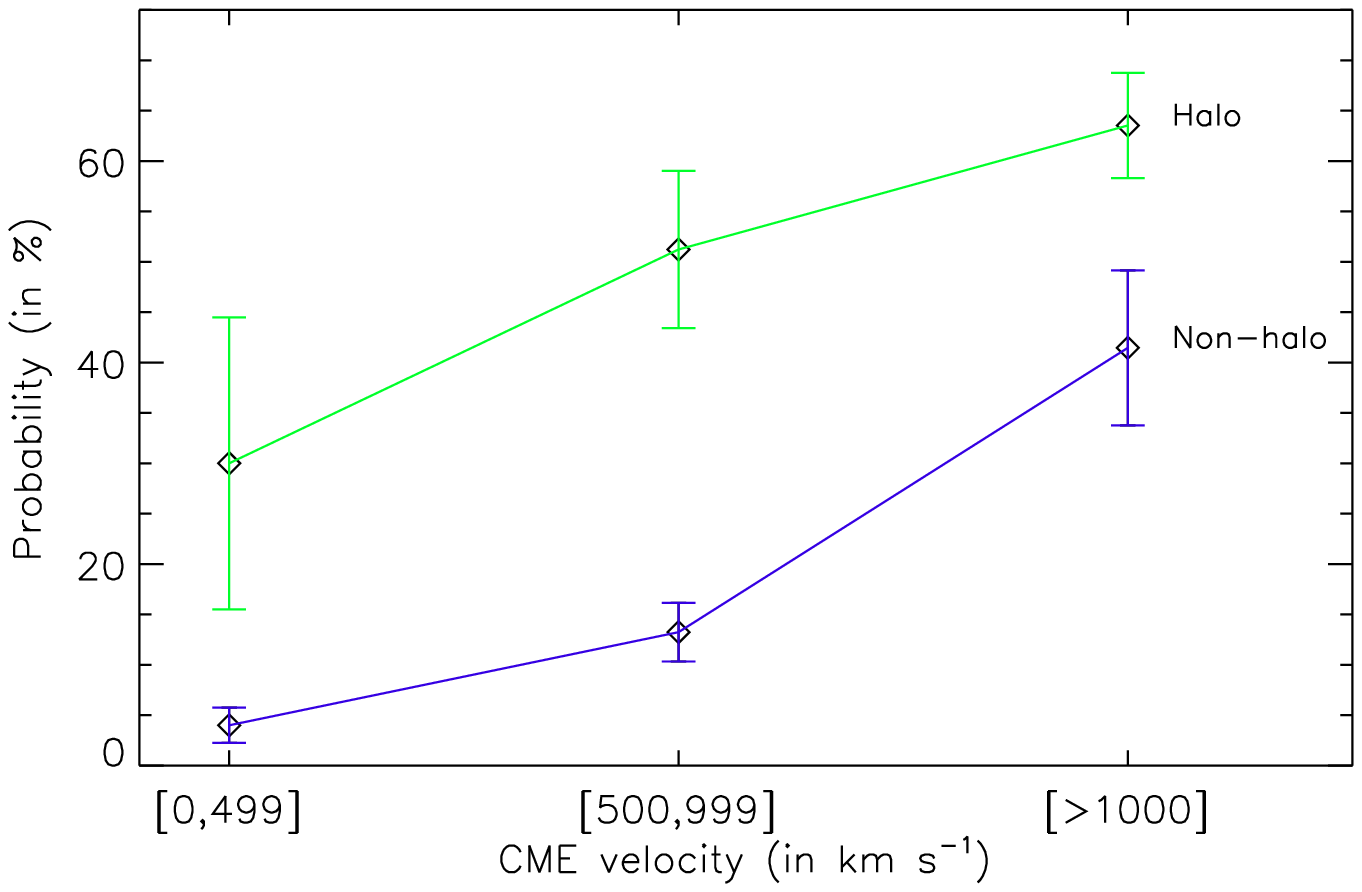}
  \caption{SEP occurrence probabilities (in \%) and their respective errors as a
    function of CME velocity for all CMEs (left panel) and for non-halo and
    halo CMEs separately (right panel).}
  \label{fig.probcme}
\end{figure}

\subsection{Proton Peak Flux}\label{subsec.results.impact}

\begin{figure}[tp]
\mbox{\includegraphics[angle=-90,width=0.49\textwidth]{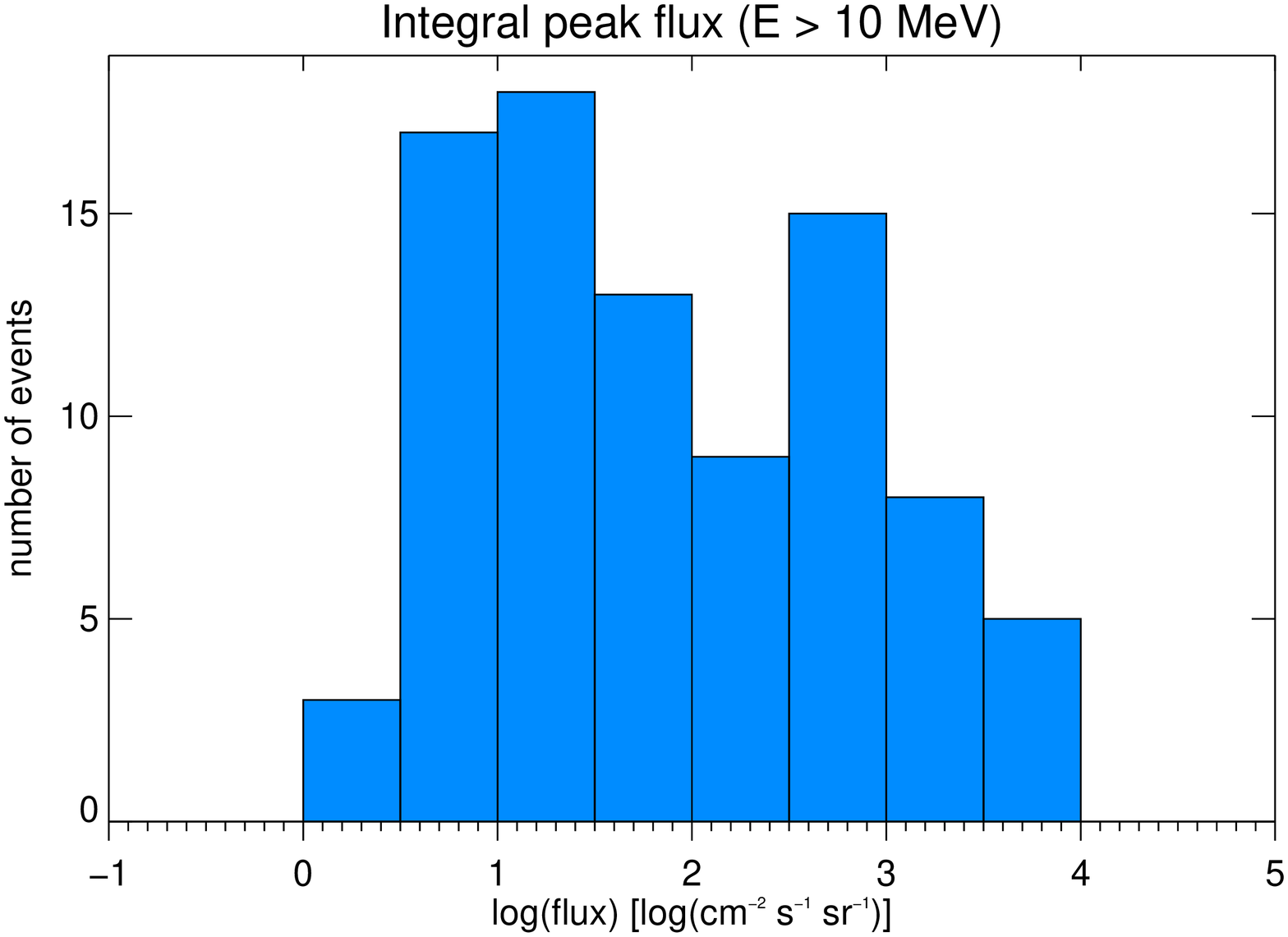}
\includegraphics[angle=-90,width=0.49\textwidth]{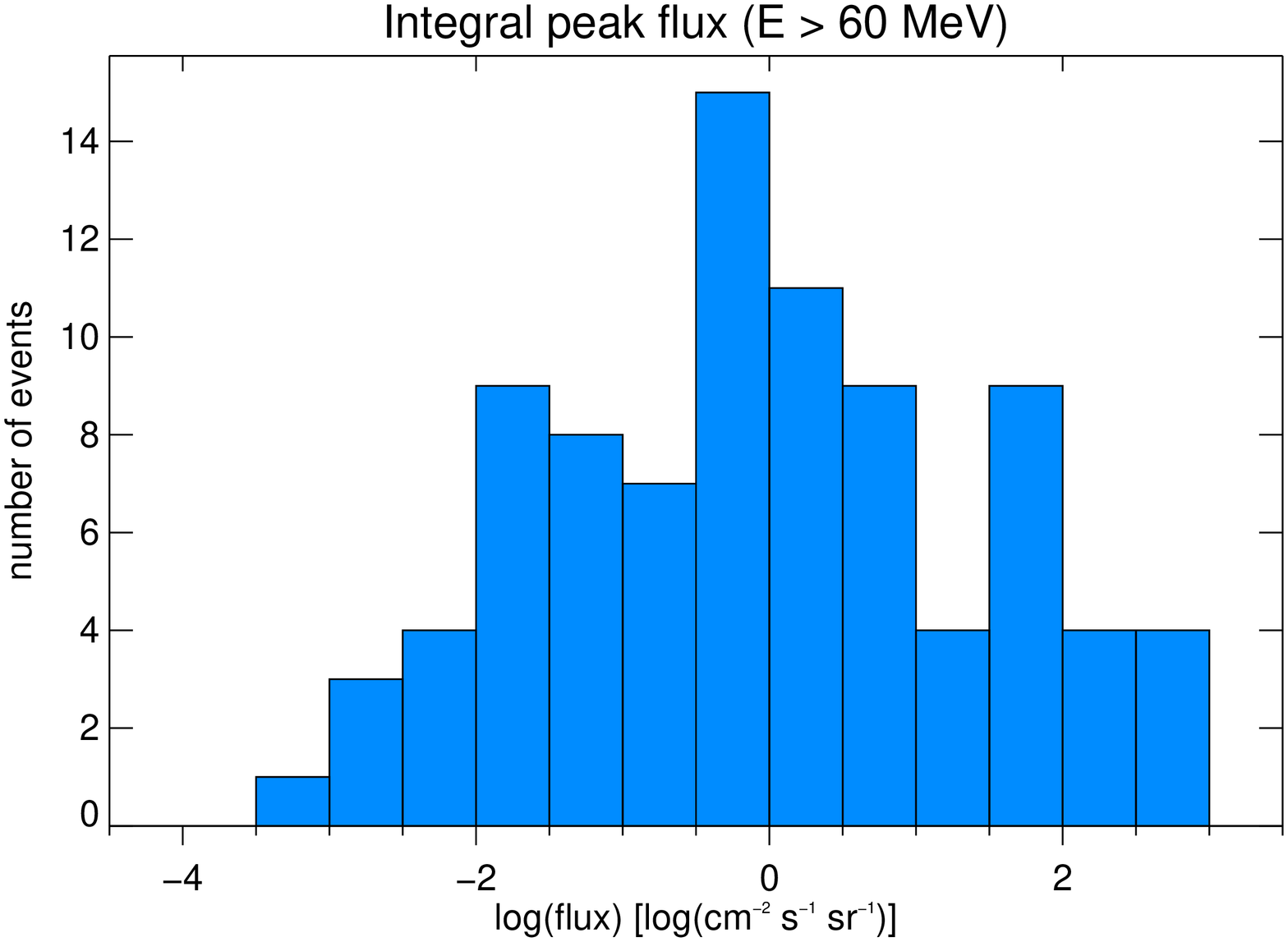}}
\caption{Distributions of SEP integral peak fluxes (in logarithmic scale) for
  protons with $E>$10~MeV (left) and  $E>$60 MeV (right) for the events in the
  SSE list.} 
\label{fig.flux_distributions} 
\end{figure}

In this section, the relationships between the proton peak fluxes of the SEP
events and their associated flares and/or CMEs are derived.  
The integral peak fluxes in the energy ranges $E>$10~MeV and
$E>$60~MeV as described in Section~\ref{subsec.integral_flux} for the events
in the SSE list are used.
Out of the 90 events in this list, two do not surpass the original
flux threshold used to construct the SEPEM reference proton event list and are
therefore not considered in this part of the analysis.
Their distributions can be seen in Figure~\ref{fig.flux_distributions} and
show that the larger events occur less frequently. 
The sharp rise in the distribution for $E>$10~MeV is due to the threshold
applied to define a SEP event.
The distribution for $E>$60~MeV shows a much slower rise which is a
consequence of the variety in spectral shapes of SEP events. 

\begin{figure}[tp]
\mbox{\includegraphics[angle=-90,width=0.49\textwidth]{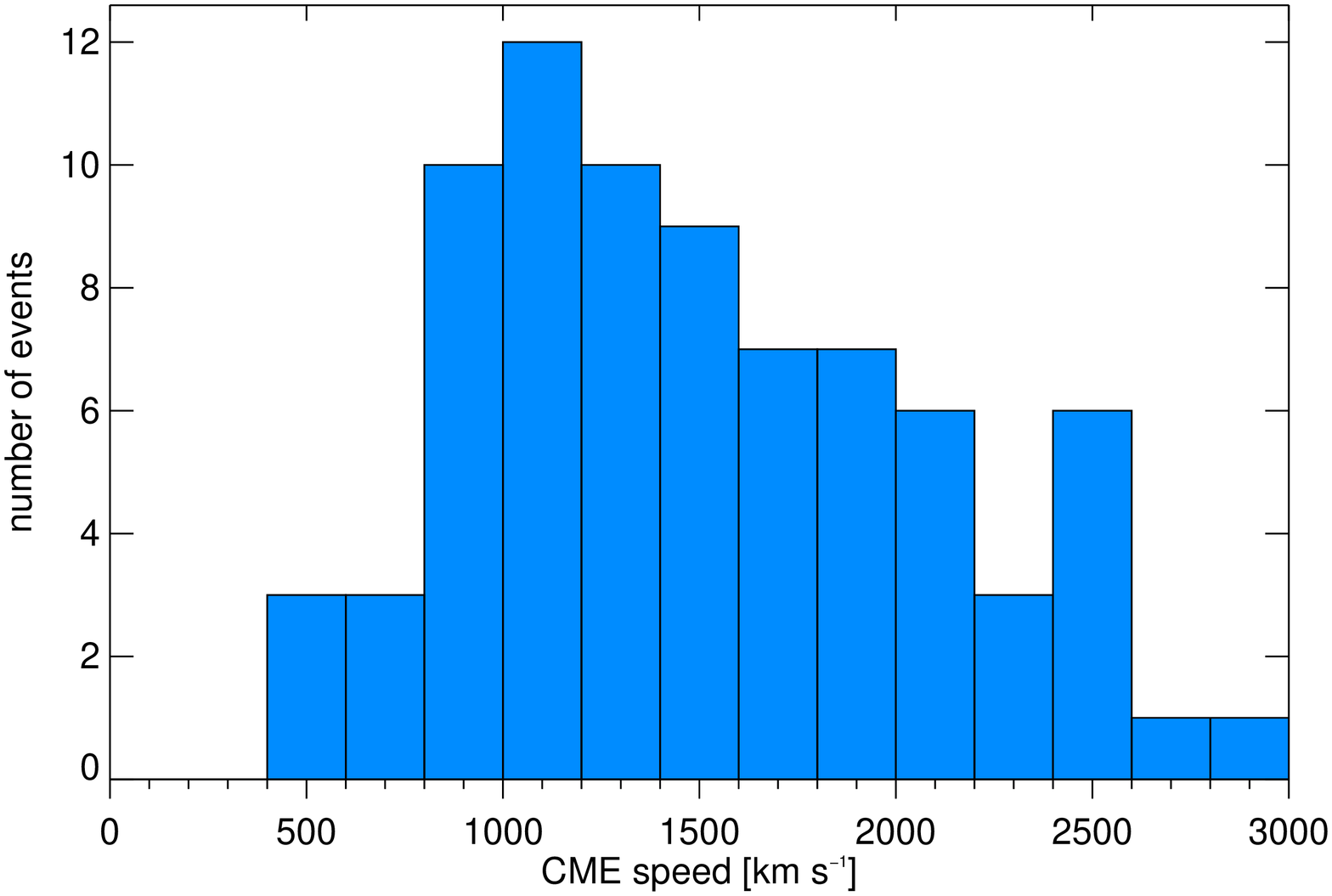}
\includegraphics[angle=-90,width=0.49\textwidth]{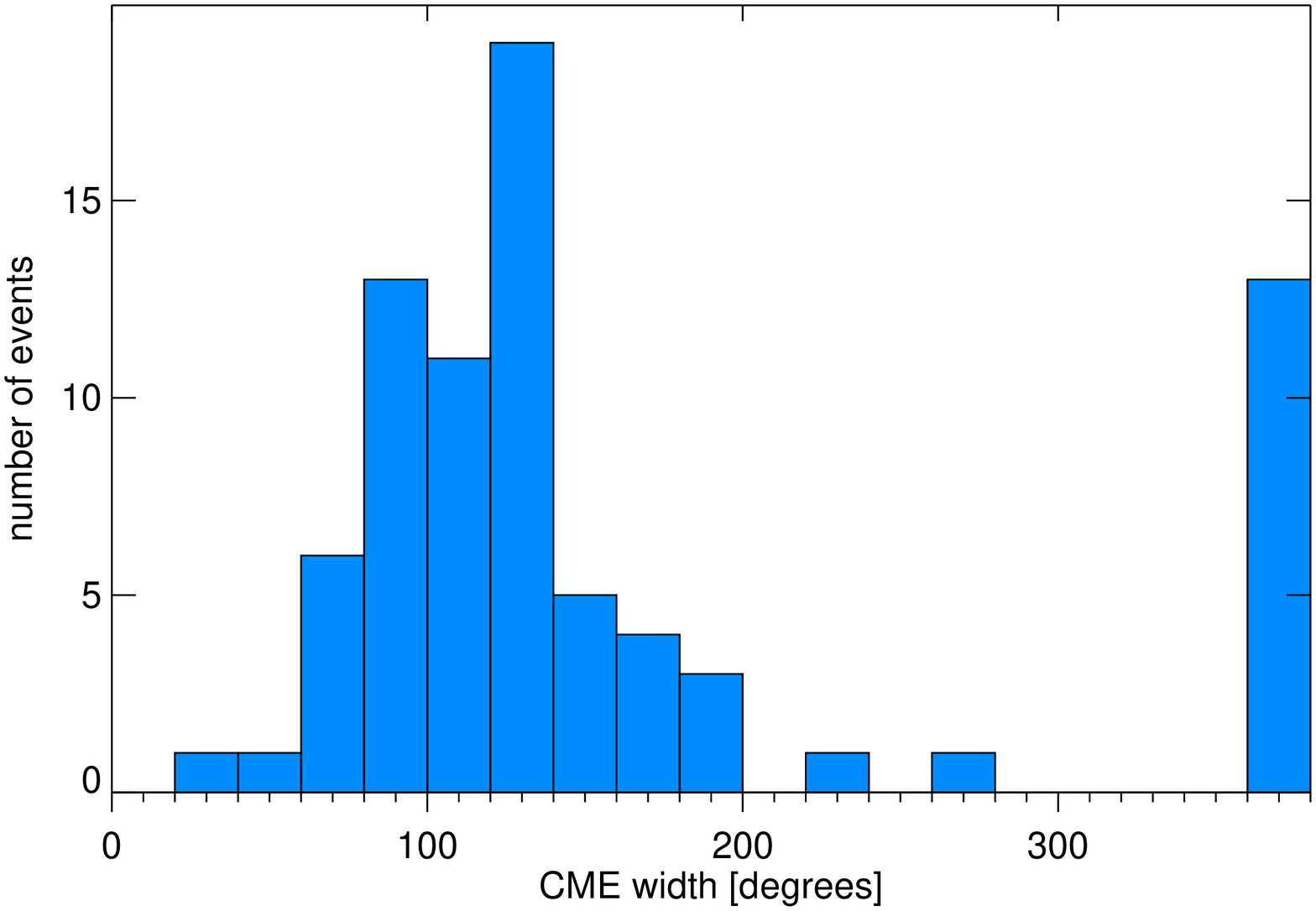}}
\caption{Distributions of the speed (left) and angular width (right) of the
  CMEs associated with the events in the SSE
  list.}\label{fig.flarecmeparams_distributions}  
\end{figure}

The solar eruptive event parameters considered in this analysis are the same
as in Section~\ref{subsec.results.prob}: flare intensity and longitude, and CME
speed and angular width.  
Out of the 88 events in the SSE list which are related
to $\ge$M1 flares on the visible disk, 78 were associated with a CME.
The distributions of the flare parameters associated with SEPs are shown in
the bottom row of Figure~\ref{fig.flare_charact}, while the CME parameters can
be seen in  Figure~\ref{fig.flarecmeparams_distributions}.
To study the dependence of the peak flux on these solar parameters, the
Pearson product-moment correlation coefficient between these two quantities
is calculated. 
The uncertainties on these coefficients are determined using the bootstrap
method \citep{wall2003}. 
In total N events are randomly drawn from a sample of N events, where the same event
can be chosen multiple times. 
This process is repeated many times, and for each new sample the correlation
coefficient is calculated. 
The average of the resulting distribution is very close to the correlation
coefficient of the original sample and the standard deviation provides an
estimate of the uncertainty.

\subsubsection{Peak Flux as a Function of Individual Flare and CME Parameters}

\begin{figure}[t]
\mbox{\includegraphics[angle=-90,width=0.49\textwidth]{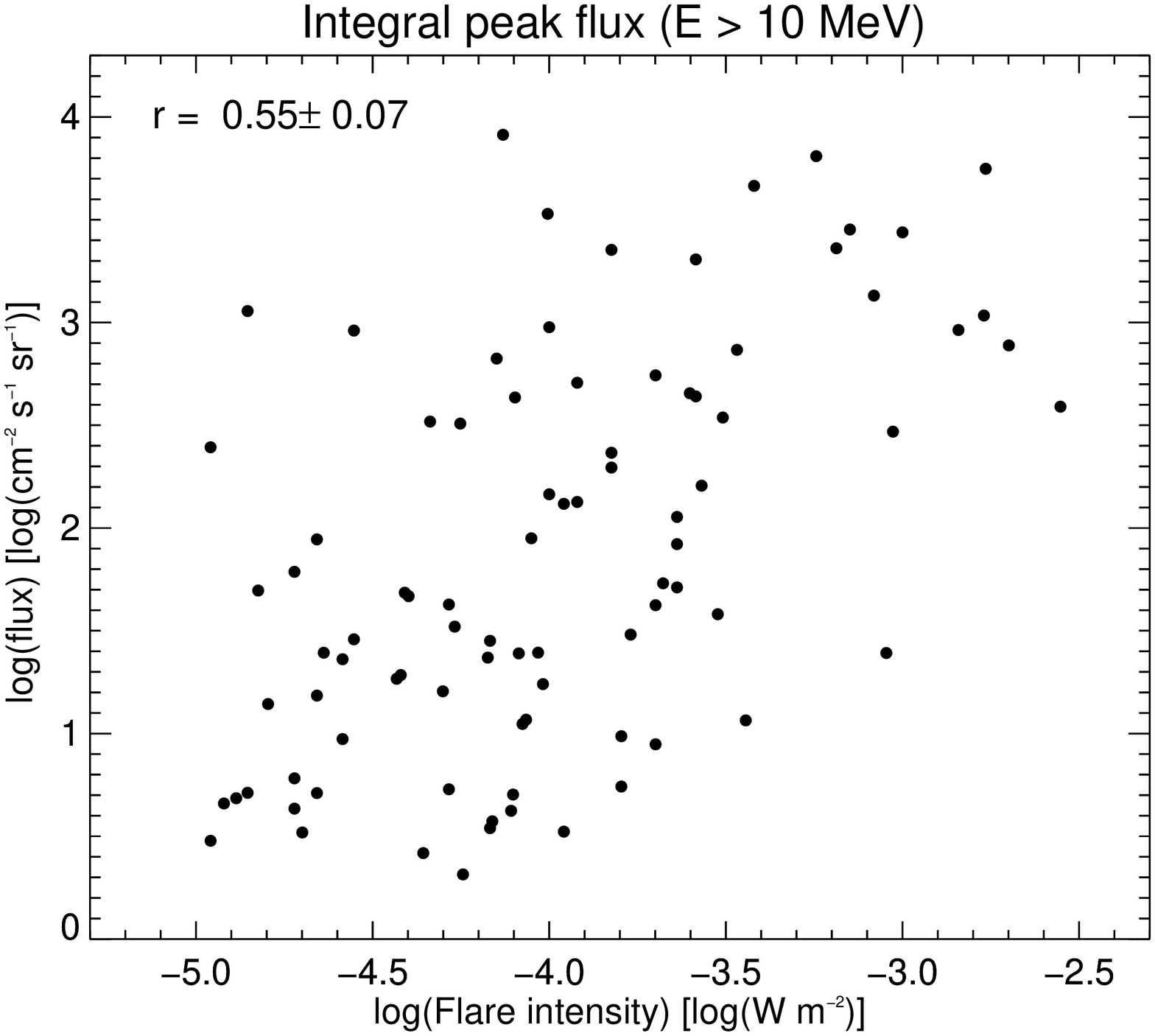}
\includegraphics[angle=-90,width=0.49\textwidth]{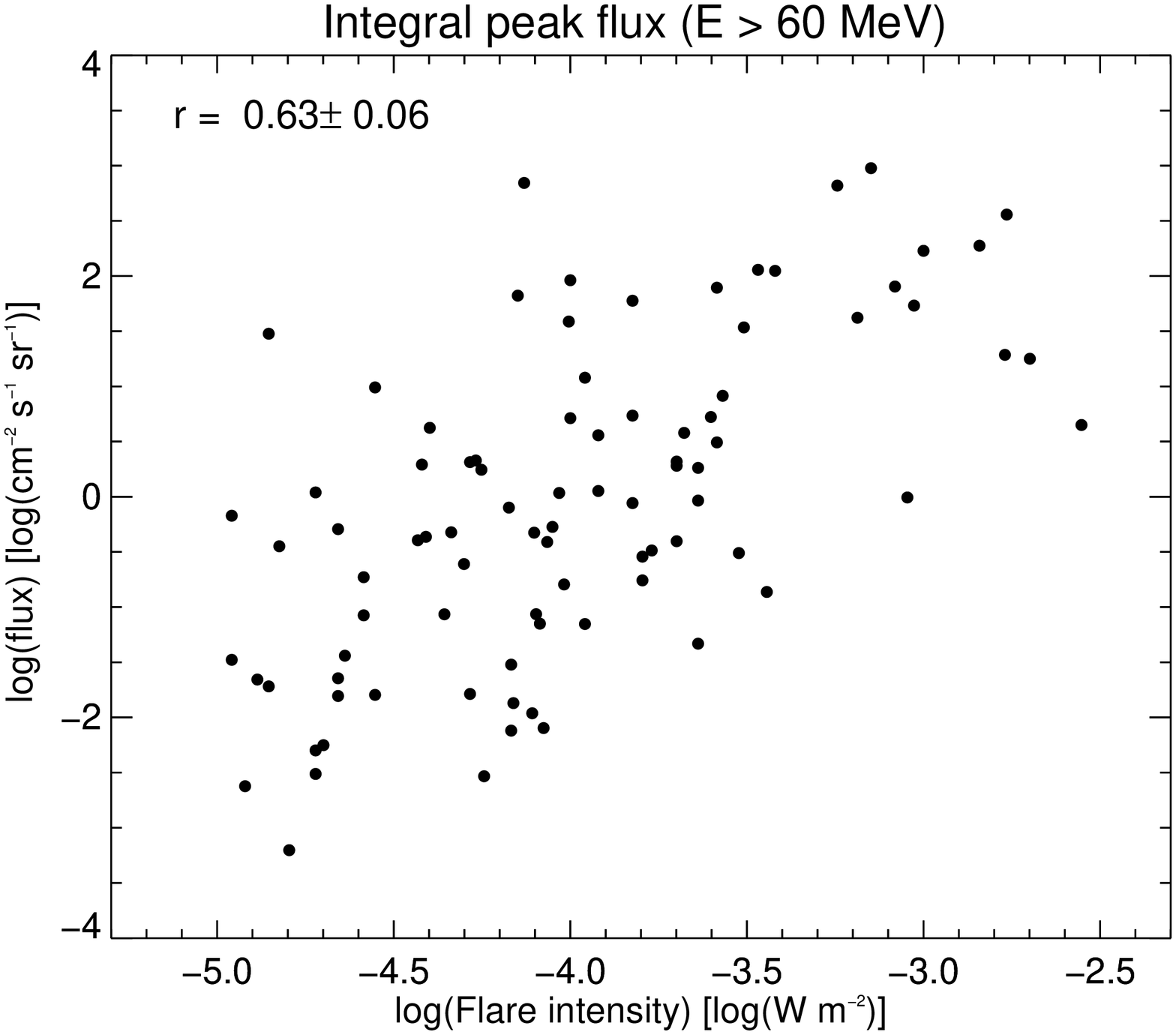}}
\caption{Scatter plots of the logarithm of the proton peak flux as a
  function of the logarithm of the flare intensity for $E>$10~MeV (left) and $E>$60~MeV
  (right). The correlation coefficient r and its uncertainty is also shown.}\label{fig.flux_vs_flareintens} 
\end{figure}

\begin{figure}[t]
\mbox{\includegraphics[angle=-90,width=0.49\textwidth]{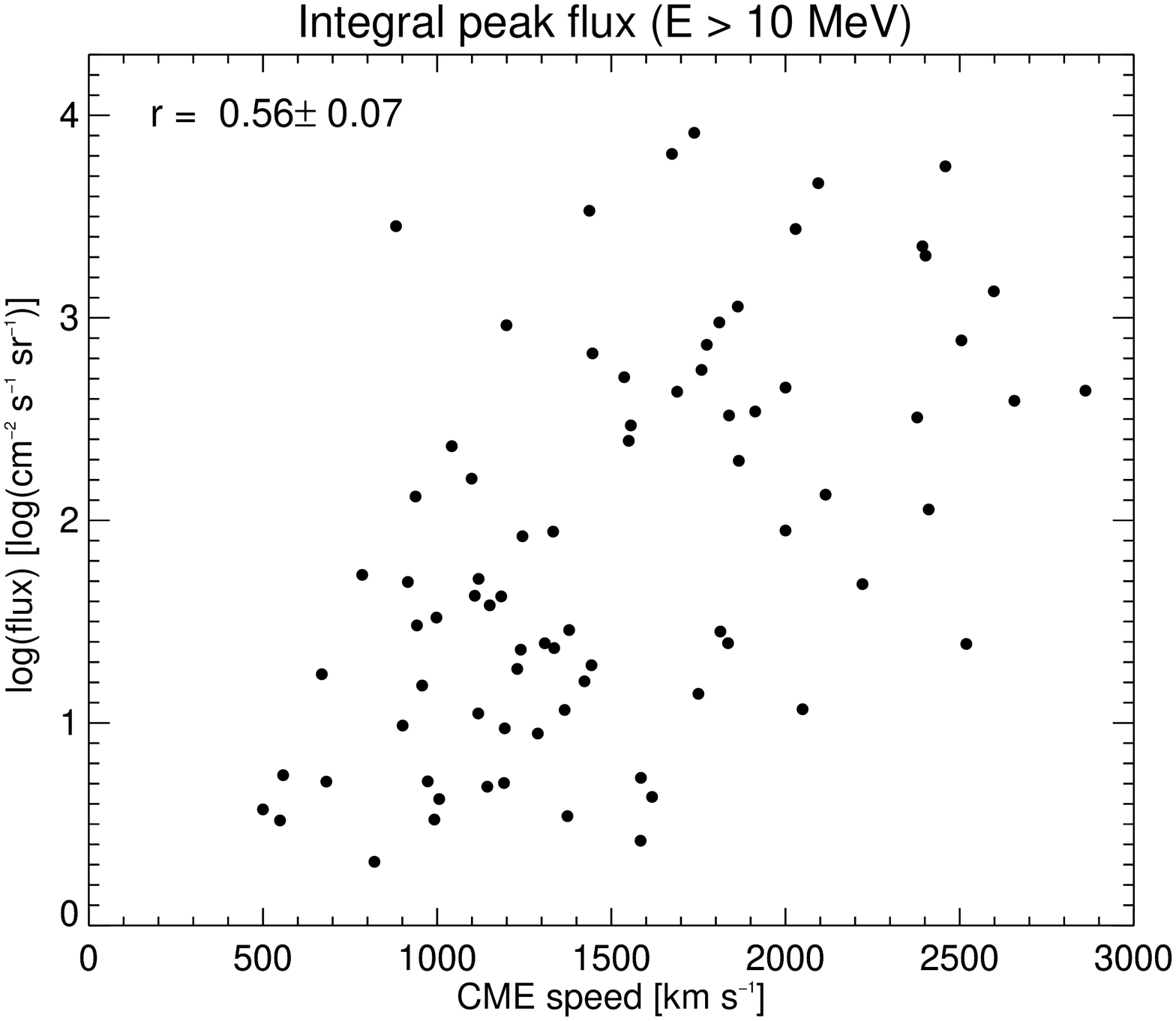}
\includegraphics[angle=-90,width=0.49\textwidth]{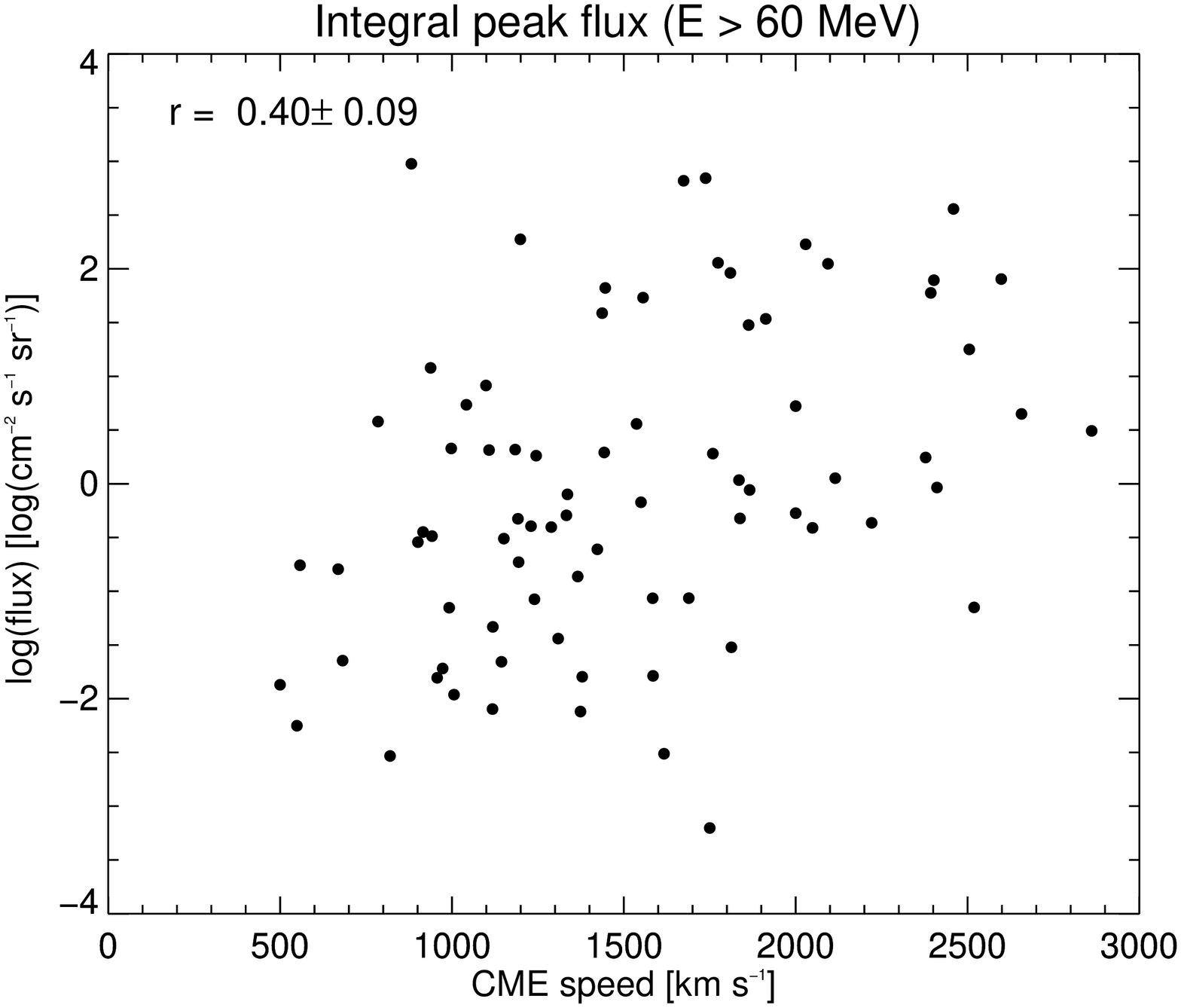}}
\caption{Scatter plots of the logarithm of the proton peak flux as a
  function of the CME speed for $E>$10~MeV (left) and  $E>$60~MeV (right).
  The correlation coefficient r and its uncertainty is also shown.}
\label{fig.flux_vs_cmespeed} 
\end{figure}

\begin{figure}[t]
\mbox{\includegraphics[angle=-90,width=0.49\textwidth]{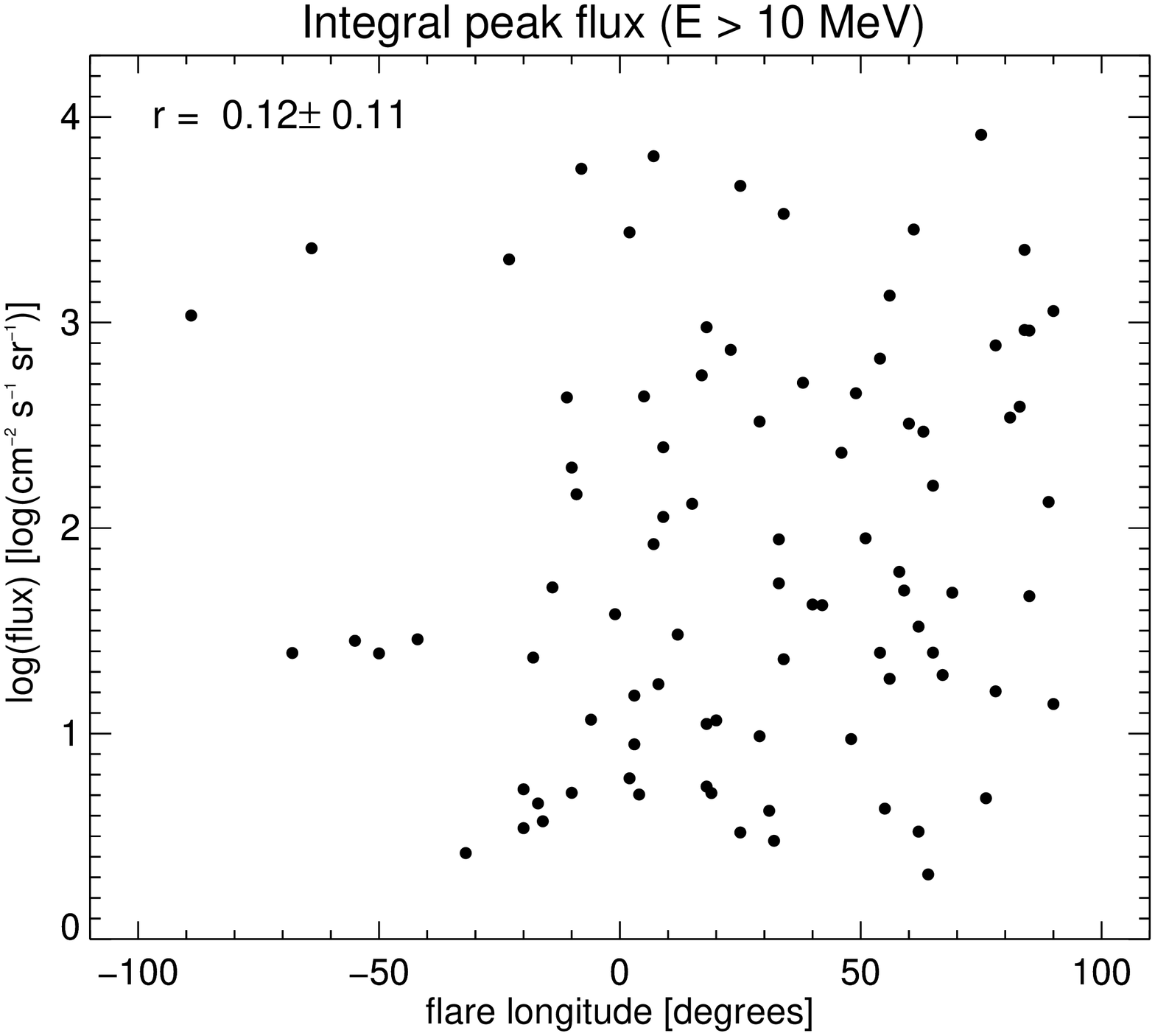}
\includegraphics[angle=-90,width=0.49\textwidth]{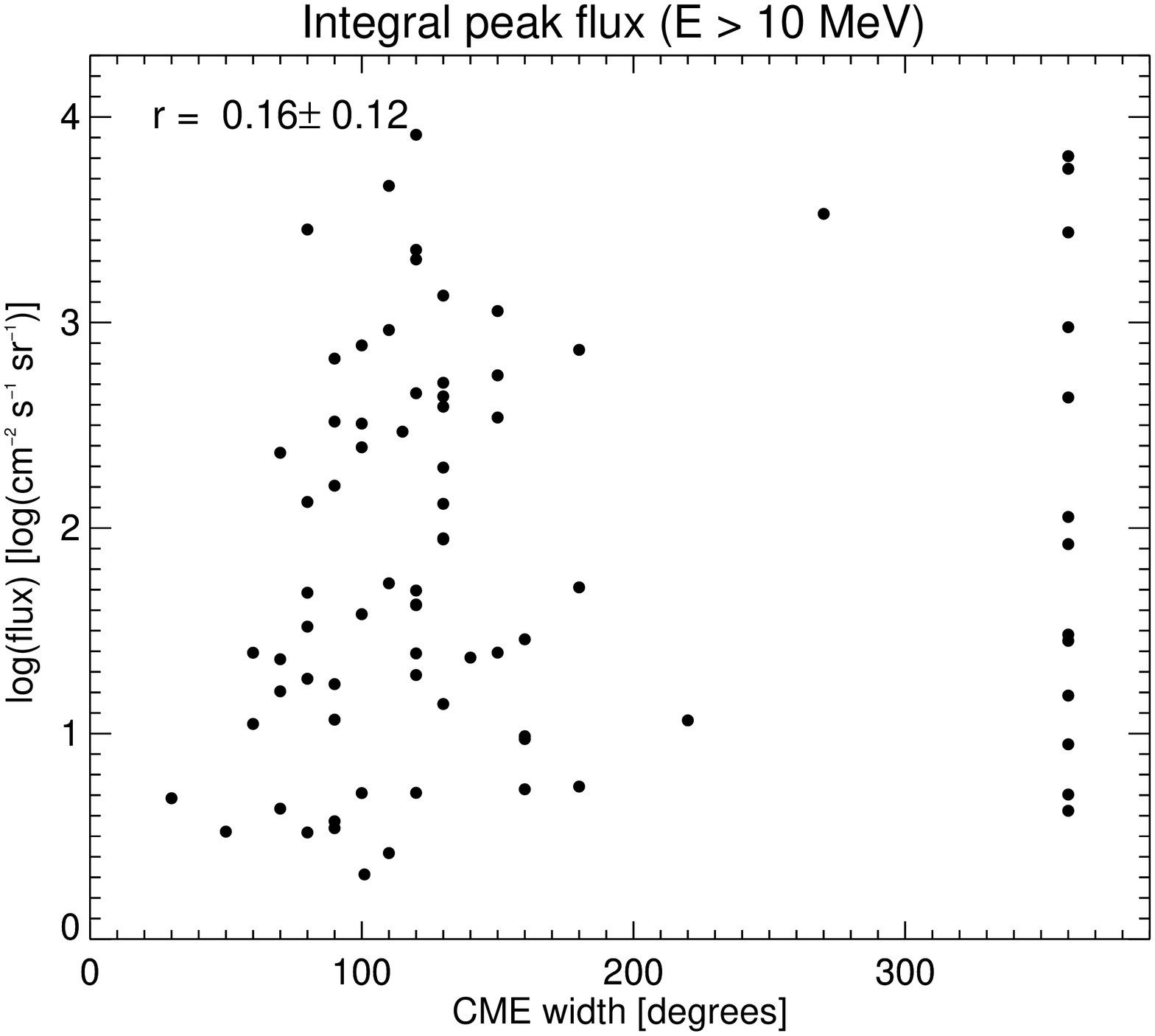}}
\caption{Scatter plots of the logarithm of the proton peak flux for $E>$10~MeV as a
  function of the flare location (left) and the CME angular width (right). The
  data points with a width of 360$^{\circ}$ correspond to halo CMEs.
  The correlation coefficient r and its uncertainty is also shown.}
\label{fig.flux_vs_flarelong_cmewidth} 
\end{figure}

Figure~\ref{fig.flux_vs_flareintens} shows that a moderate
correlation exists between the proton peak flux and flare intensity, and is 
slightly larger for the higher energy range.
A moderate correlation can also be observed in
Figure~\ref{fig.flux_vs_cmespeed} between the proton peak flux and CME speed, 
which is larger for the lower energy range. 
The integral peak flux for $E>$10~MeV versus the flare longitude and CME 
angular width is shown in
Figure~\ref{fig.flux_vs_flarelong_cmewidth}, and in both cases, no significant
correlation is observed. 
All derived correlation coefficients for both peak fluxes 
of $E>$10~MeV and $E>$60~MeV protons are given in Table~\ref{table.corrcoeff}.

\begin{table}[t]
\caption{The correlation coefficients between the logarithm of the proton peak
  flux (for $E>$10~MeV and $E>$60~MeV) and the following
  characteristics of the 
  associated solar eruptive events: logarithm of flare intensity
  $log(I_{\mathrm{f}})$, flare longitude $L_{\mathrm{f}}$, CME 
  speed $v_{\mathrm{CME}}$ and CME angular width $w_{\mathrm{CME}}$.
  The correlations with flare (CME) parameters are based on 88 (78) events.}
\label{table.corrcoeff}
\begin{tabular}{ccccc}
  \hline                   
 $ $ & $log(I_{\mathrm{f}})$ & $L_{\mathrm{f}}$ & $v_{\mathrm{CME}}$ & $w_{\mathrm{CME}}$ \\
  \hline
 $E>$10~MeV  & $0.55\pm0.07$ & $0.12\pm0.11$ & $0.56\pm0.08$ & $0.16\pm0.12$ \\ 
 $E>$60~MeV  & $0.63\pm0.06$ & $0.19\pm0.11$ & $0.40\pm0.09$ & $0.13\pm0.12$ \\
  \hline
\end{tabular}
\end{table}

\begin{figure}[tp]
\centering
\mbox{\includegraphics[angle=-90,width=0.80\textwidth]{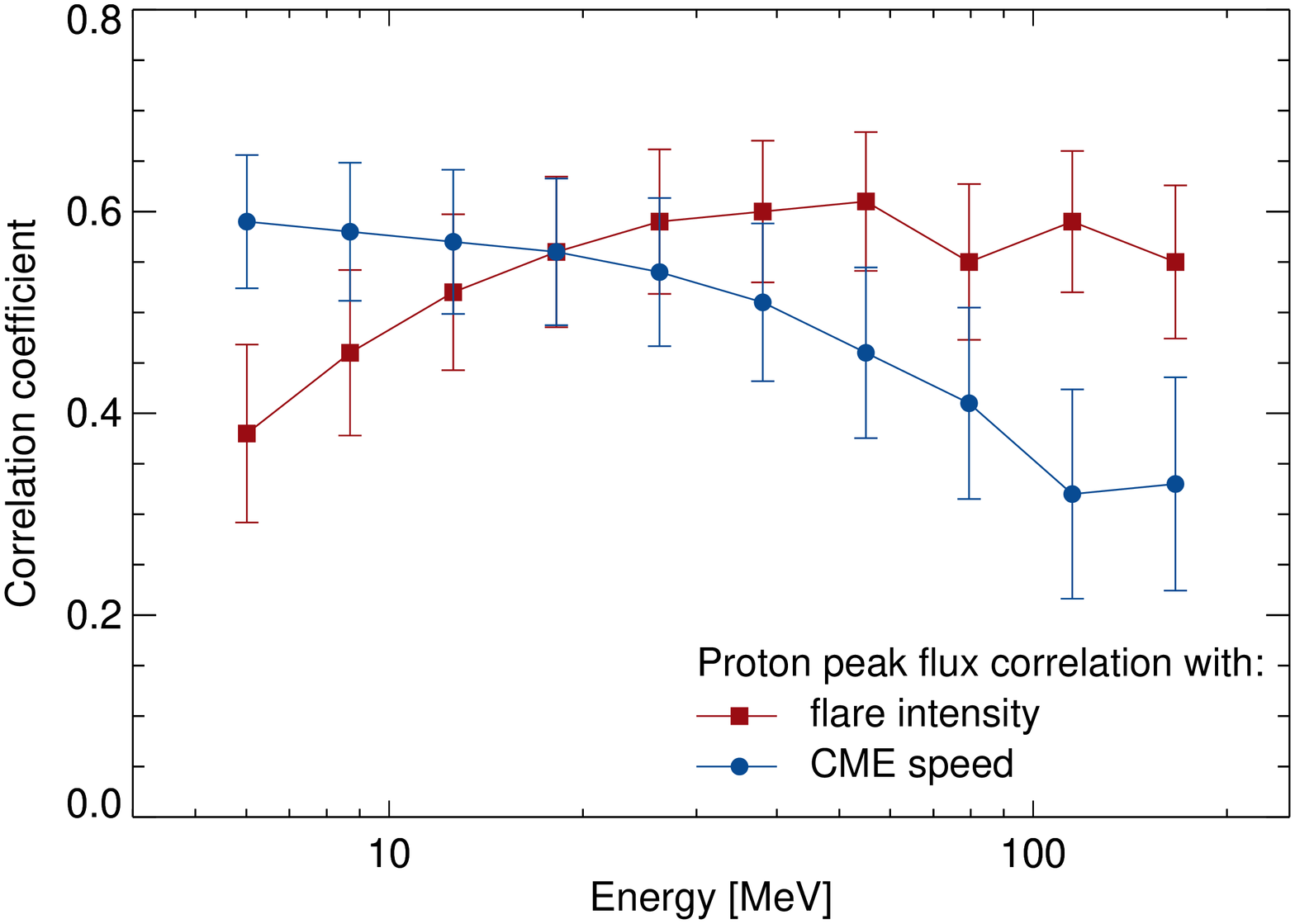}}
\caption{The correlation coefficients between the proton peak flux
  and the flare intensity (red squares) and CME speed (blue circles)
  as a function of the proton energy.}\label{fig.corr_spectrum} 
\end{figure}

Since there is a difference between the values of the different energy ranges,  
a potential energy dependence of these quantities was further investigated. 
The correlation coefficients between the peak flux and flare intensity and CME
speed were derived for each individual SEPEM energy channel and  are listed in
Table~\ref{table.corrcoeff_edep} and shown in Figure~\ref{fig.corr_spectrum}.
It can be seen that the correlation between the peak flux and flare
intensity increases with energy up to 30~MeV and then stays roughly constant. 
The correlation with the CME speed shows the opposite behavior: the value
decreases with energy at least until above 100~MeV.
It is also interesting to observe that the correlation with
CME speed is larger than the correlation with flare intensity for the
lowest energies, while they are smaller for the higher energies.

\begin{table}[tp]
\caption{The correlation coefficients between the logarithm of the proton peak flux
  as measured in each SEPEM
  energy channel and the logarithm of flare intensity $log(I_{\mathrm{f}})$
  and CME speed $v_{\mathrm{CME}}$, excluding and including the ESP contribution in
  the flux profile.} 
\label{table.corrcoeff_edep} 
\begin{tabular}{ccccc}                                
  \hline                   
  SEPEM proton & \multicolumn{2}{c}{ESP excluded} & \multicolumn{2}{c}{ESP included} \\  
  energy channel & $log(I_{\mathrm{f}})$ & $v_{\mathrm{CME}}$ & $log(I_{\mathrm{f}})$ & $v_{\mathrm{CME}}$ \\
  \hline
  5.00 -- 7.23 MeV   & $0.38\pm0.09$ & $0.59\pm0.06$& $0.41\pm0.09$ & $0.52\pm0.07$  \\
  7.23 -- 10.46 MeV  & $0.46\pm0.08$ & $0.58\pm0.07$& $0.46\pm0.08$ & $0.56\pm0.07$  \\
  10.46 -- 15.12 MeV & $0.52\pm0.08$ & $0.57\pm0.07$& $0.50\pm0.08$ & $0.57\pm0.07$  \\
  15.12 -- 21.87 MeV & $0.56\pm0.07$ & $0.56\pm0.07$& $0.54\pm0.08$ & $0.56\pm0.07$  \\
  21.87 -- 31.62 MeV & $0.59\pm0.07$ & $0.54\pm0.07$& $0.57\pm0.07$ & $0.55\pm0.07$  \\
  31.62 -- 45.73 MeV & $0.60\pm0.07$ & $0.51\pm0.08$& $0.59\pm0.07$ & $0.52\pm0.07$  \\
  45.73 -- 66.13 MeV & $0.61\pm0.07$ & $0.46\pm0.09$& $0.60\pm0.07$ & $0.46\pm0.08$  \\
  66.13 -- 95.64 MeV & $0.55\pm0.08$ & $0.41\pm0.09$& $0.55\pm0.08$ & $0.41\pm0.09$  \\
  95.64 -- 138.3 MeV & $0.59\pm0.07$ & $0.32\pm0.10$& $0.60\pm0.07$ & $0.33\pm0.10$  \\
  138.3 -- 200.0 MeV & $0.55\pm0.08$ & $0.33\pm0.11$& $0.53\pm0.07$ & $0.32\pm0.10$  \\
  \hline
\end{tabular}
\end{table}

Table~\ref{table.corrcoeff_edep} also lists the correlation coefficients when
the enhancement in flux due to an ESP is not excluded from the peak flux 
determination. 
Taking into account that values within the same energy channels are highly
correlated, a small but not very significant difference can only be observed for the
correlation between the peak flux in the lowest energy channel
($5.00-7.23$~MeV) and the CME speed. 
The negligible effect on the correlations is not surprising as only half of
the events in the SSE list contain an identified ESP, and often the
secondary flux enhancement due to the shock passage does not exceed the
initial peak flux. 
This is in particular the case for higher energy channels.

\begin{figure}[t]
\mbox{\includegraphics[angle=-90,width=0.49\textwidth]{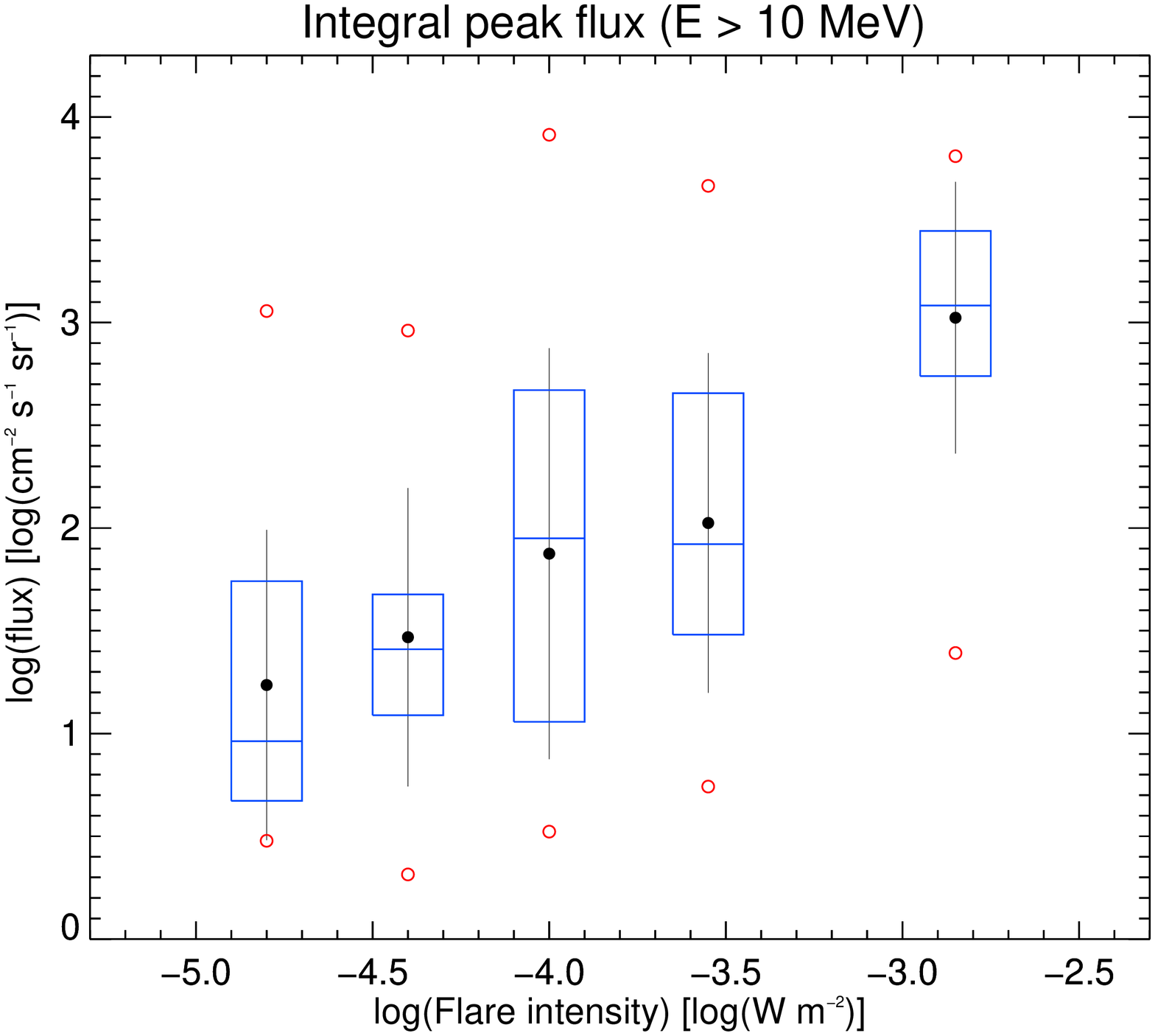}
\includegraphics[angle=-90,width=0.49\textwidth]{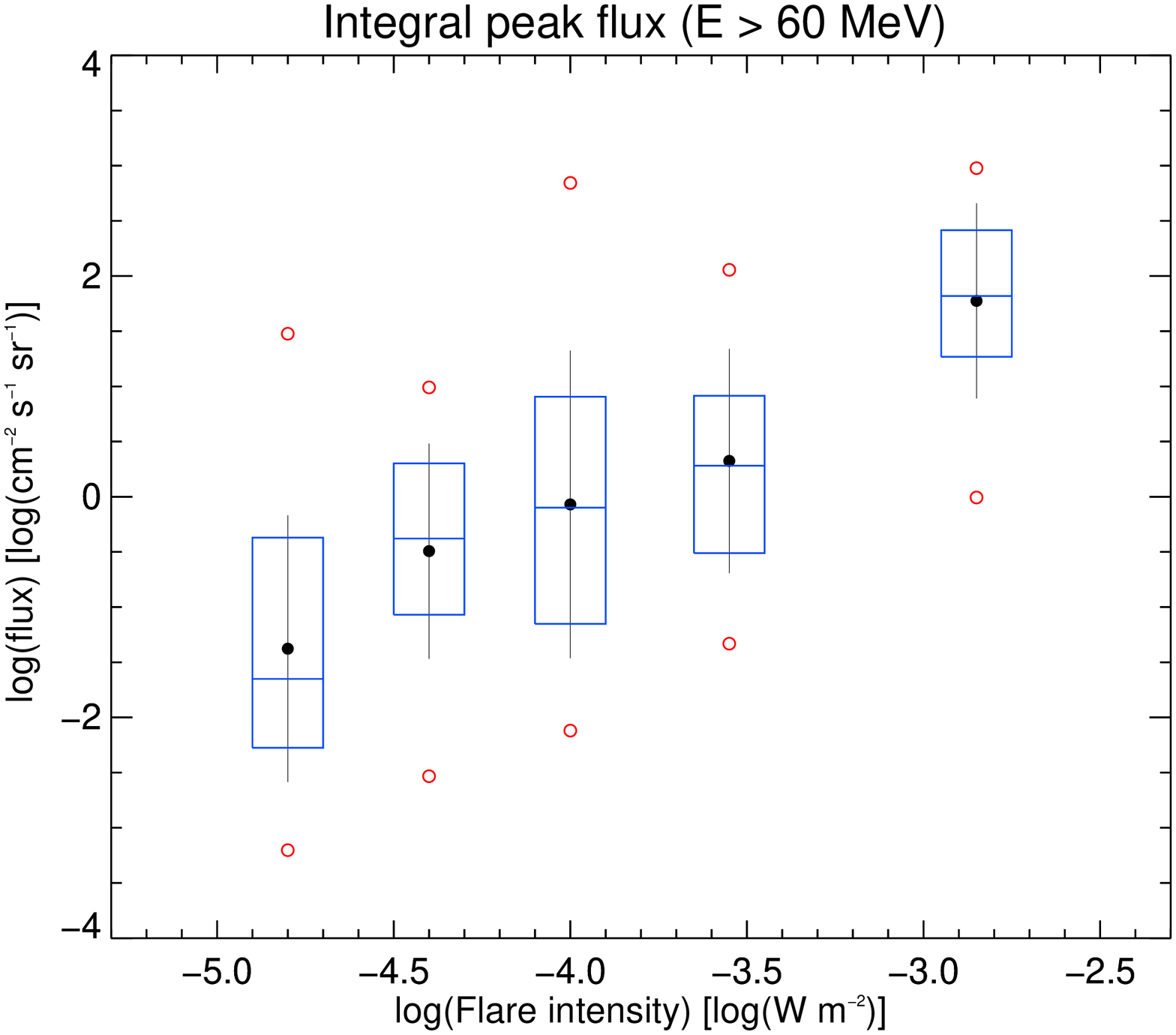}}
\caption{Binned plots of the logarithm of the proton peak flux as a
  function of logarithm of flare intensity for $E>$10~MeV (left) and $E>$60~MeV 
  (right). Black dots show the mean in that bin, while the error bars
  represent the RMS. The horizontal blue line shows the median, and the box
  shows the start of the second quartile and end of the third quartile, so
  that 50\% of the events lie in the box. Red open circles show the outermost
  data points (outliers) in each bin. }\label{fig.flux_vs_flareintens_bins} 
\end{figure}

\begin{figure}[t] 
\mbox{\includegraphics[angle=-90,width=0.49\textwidth]{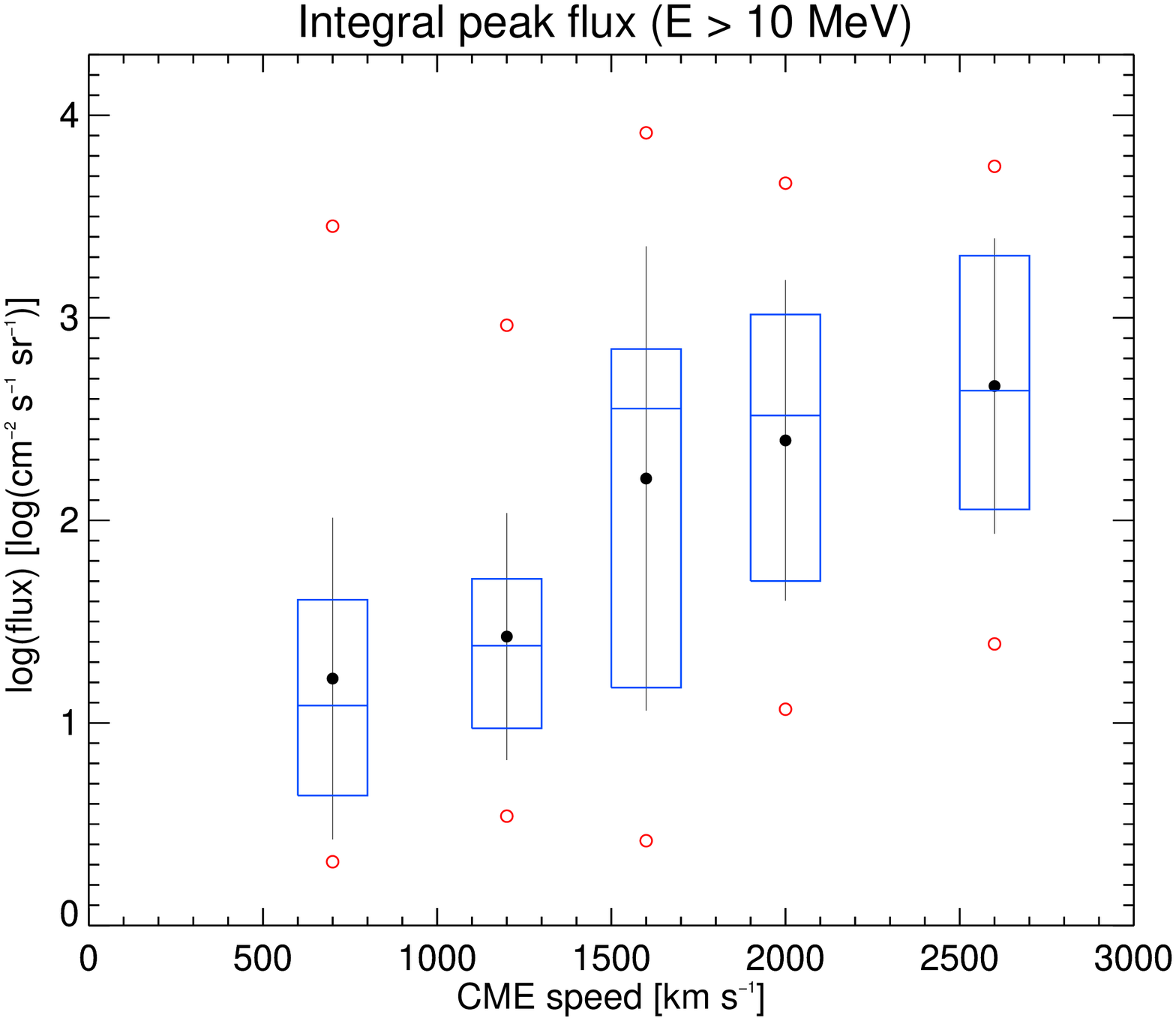}
\includegraphics[angle=-90,width=0.49\textwidth]{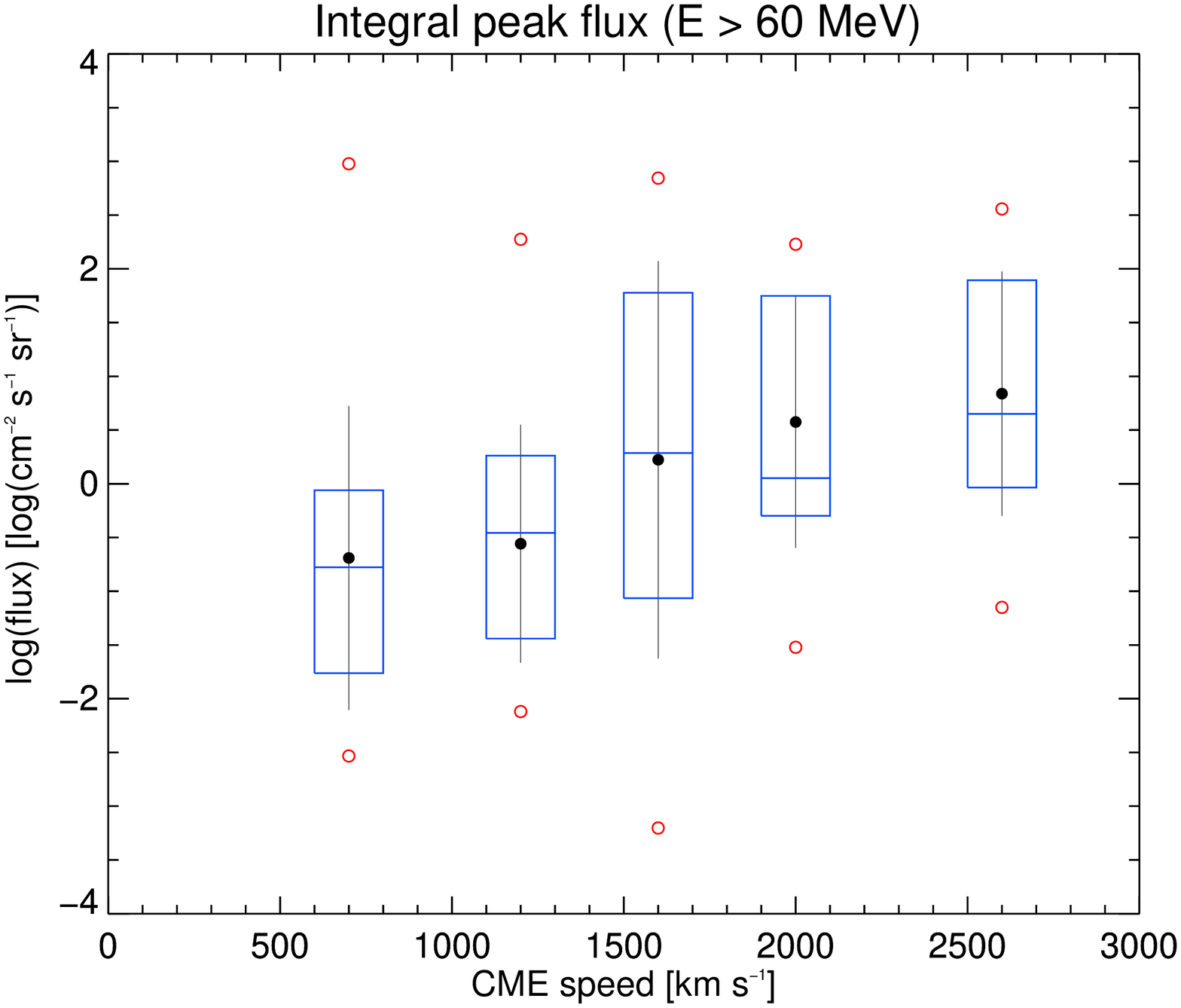}}
\caption{Binned plots of the logarithm of the proton peak flux as a
  function of the CME speed for $E>$10~MeV (left) and $E>$60~MeV
  (right). The convention for the plot is the same as in Figure
  \ref{fig.flux_vs_flareintens_bins}.}\label{fig.flux_vs_cmespeed_bins} 
\end{figure}

As Figures
\ref{fig.flux_vs_flareintens}--\ref{fig.flux_vs_flarelong_cmewidth} and
Table~\ref{table.corrcoeff} indicate, the strongest correlations exist
between the proton peak flux and flare intensity and CME speed. 
To provide a clearer quantification of the dependency of the peak flux on the
solar parameters which could be used
within a forecasting framework, we adopted a binning of the data points of
Figures \ref{fig.flux_vs_flareintens} and \ref{fig.flux_vs_cmespeed}. 
Also for the analysis of the peak fluxes, the binning has been chosen
such that they contain sufficient events and also 
taking into account the shape of the distributions.  
The logarithm of the flare intensity was divided in four equidistant bins
between $-5.0$ and $-3.3$ log(W m$^{-2}$), while all events with larger
values were collected in one bin, resulting in the
following binning: 
[$-5.0$,$-4.6$], [$-4.6$,$-4.2$], [$-4.2$,$-3.8$], [$-3.8$,$-3.3$], 
[$>-3.3$] in units of log(W m$^{-2}$). In terms of flare magnitude, this
corresponds to the bins
[M$1.0$,M$2.5$], [M$2.5$,M$6.3$], [M$6.3$,X$1.6$], [X$1.6$,X$5.0$] and
[$\ge$ X$5.0$].
The CME speed was binned into three equidistant bins between 1000 and
2200~km~s$^{-1}$, and the events associated to slower as well as faster CMEs were
grouped in one bin each, giving the following bins:
$[400,1000]$, $[1000,1400]$, $[1400,1800]$, $[1800,2200]$, $[>2200]$  in
units of km~s$^{-1}$. The lower limit in the first bin is only due to the
absence of SEP events associated to CMEs slower than 400~km~s$^{-1}$.
The mean of the logarithm of the peak fluxes is then
calculated from the data points that fall within each bin.
As an estimate of the spread of the events within a bin, the root mean square 
(RMS) of the values in each bin is calculated. 
The median, $25\%$ and $75\%$ quartiles, and extreme values are also
determined.  
Figures \ref{fig.flux_vs_flareintens_bins} and \ref{fig.flux_vs_cmespeed_bins}
show the binned data for the integrated flux at $E>$10~MeV and $E>$60~MeV
{\it versus} flare intensity and CME speed respectively. As was already
evident from the scatter plots, there is a clear dependence of the peak fluxes
on both flare intensity and CME speed. 
Tables~\ref{table.flarecorr_values} and \ref{table.cmecorr_values} in the
Appendix give values
of the mean and RMS in each bin as a function of flare intensity and
CME speed respectively, for the peak fluxes in the two energy ranges considered.

\subsubsection{Peak Flux as a Function of Multiple Flare and CME Parameters}

We study the dependency of the proton peak fluxes on several combinations of
flare and CME parameters by dividing one of the parameters under study into
bins and determining in each bin the correlation between the integral proton
peak flux (for $E>10$~MeV and $E>60$~MeV) and the other solar parameters. 
Subsequently, the second solar parameter is also binned and the mean, RMS,
quartiles and extremes in each bin are determined.
Only a limited amount of bins is used due to the low statistics.
The M-class flares were binned in two bins equidistant in the logarithm of the
flare intensity, while one bin was used for X-class flares, resulting in the
following flare magnitude bins: 
$\mathrm{M}1 \le I_{\mathrm{f}} < \mathrm{M}3.2$, $\mathrm{M}3.2 \le I_{\mathrm{f}} < \mathrm{X}1$
and $I_{\mathrm{f}} \ge $X$1$.
For the flare longitude three equidistant bins were chosen:
$-90^{\circ} \le L_{\mathrm{f}} < -30^{\circ}$ (east),
$-30^{\circ} \le L_{\mathrm{f}} < 30^{\circ}$ (center) and $30^{\circ}
\le L_{\mathrm{f}} <  90^{\circ}$ (west).
The CME speed  is binned in two equal sized bins for
slower CMEs and one bin for fast CMEs:
$400 \le v_{\mathrm{CME}} < 1200$, $1200 \le v_{\mathrm{CME}} <
2000$ and
$ v_{\mathrm{CME}} \ge 2000$ in units of km~s$^{-1}$, while the width 
is divided into the bins $0^{\circ}\le w_{\mathrm{CME}} < 100^{\circ}$,
$100^{\circ} \le w_{\mathrm{CME}} < 360^{\circ}$ and 
$w_{\mathrm{CME}} = 360^{\circ}$. 
The latter bin corresponds to halo CMEs.
All derived correlation coefficients for the different combinations of
parameters are presented in Table~\ref{table.corrcoeff_2d} and further 
described below.

\begin{table}[tp]
\caption{Correlation coefficients of the peak flux of protons with $E>10$~MeV
(top) and $E>60$ MeV (bottom) with the solar parameters given in the first
column in different bins of flare magnitude $I_{\mathrm{f}}$, flare longitude
location $L_{\mathrm{f}}$ and CME speed $v_{\mathrm{CME}}$.
The number of events used to derive 
each correlation coefficient is given in brackets.} \label{table.corrcoeff_2d} 
\begin{tabular}{cccc}                                
  \hline
  \multicolumn{4}{c}{proton peak flux $E>10$ MeV} \\
  \hline                   
  & M$1.0 \le I_{\mathrm{f}} < $M$3.2$ & M$3.2 \le I_{\mathrm{f}} < $X$1.0$ &  $I_{\mathrm{f}} \ge $X$1.0$ \\
  flare longitude & $0.38\pm0.21$ $(20)$ & $0.30\pm0.14$ $(30)$ & $0.01\pm0.15$ $(38)$ \\
  CME speed       & $0.59\pm0.22$ $(15)$ & $0.40\pm0.12$ $(28)$ & $0.61\pm0.10$ $(35)$ \\
  CME width       & $0.16\pm0.20$ $(15)$ & $0.17\pm0.20$ $(28)$ & $-0.02\pm0.20$ $(35)$ \\
  \hline
  & $-90^{\circ} \le L_{\mathrm{f}} < -30^{\circ}$ & $-30^{\circ} \le L_{\mathrm{f}} <
  30^{\circ}$ & $30^{\circ} \le L_{\mathrm{f}} \le 90^{\circ}$ \\
  flare intensity &  $0.72\pm0.29$ $(7)$& $0.62\pm0.12$ $(38)$& $0.52\pm0.10$ $(43)$ \\
  CME speed       &  $0.27$ $(4)$& $0.72\pm0.07$ $(35)$& $0.45\pm0.12$ $(39)$\\ 
  \hline
  & $ v_{\mathrm{CME}} < 1500$ km~s$^{-1}$  & $v_{\mathrm{CME}} \ge 1500$ km~s$^{-1}$  & \\
  CME width      & $-0.03\pm0.15$ $(42)$& $0.28\pm0.15$ $(36)$&   \\
  \hline
  \hline
  \multicolumn{4}{c}{proton peak flux $E>60$ MeV} \\
  \hline                   
  & M$1.0 \le I_{\mathrm{f}} < $M$3.2$ & M$3.2 \le I_{\mathrm{f}} < $X$1.0$ &  $I_{\mathrm{f}} \ge $X$1.0$ \\
  flare longitude & $0.40\pm0.25$ $(20)$ & $0.44\pm0.13$ $(30)$& $0.12\pm0.14$ $(23)$ \\
  CME speed       & $0.25\pm0.36$ $(15)$ & $0.27\pm0.16$ $(28)$ & $0.38\pm0.17$ $(38)$ \\
  CME width       & $0.06\pm0.22$ $(15)$ & $0.05\pm0.21$ $(28)$ & $0.01\pm0.20$ $(38)$ \\
  \hline
  & $-90^{\circ} \le L_{\mathrm{f}} < -30^{\circ}$ & $-30^{\circ} \le L_{\mathrm{f}} <
  30^{\circ}$ & $30^{\circ} \le L_{\mathrm{f}} \le 90^{\circ}$ \\
  flare intensity &  $0.90\pm0.11^{\dagger}$ $(7)$ & $0.73\pm0.09$ $(38)$ & $0.60\pm0.09$ $(43)$ \\
  CME speed       &  $0.54^{\ddagger}$ $(4)$ & $0.63\pm0.09$ $(35)$ & $0.25\pm0.14$ $(39)$ \\ 
  \hline
  & $ v_{\mathrm{CME}} < 1500$ km~s$^{-1}$  & $v_{\mathrm{CME}} \ge 1500$ km~s$^{-1}$  & \\
  CME width      & $-0.02\pm0.14$ $(42)$ & $0.22\pm0.19$ $(36)$ &   \\
  \hline

\end{tabular}
\end{table}

The correlation coefficient between the peak flux with $E>10$ MeV and 
flare longitude shows a weak correlation for the two M-class flare bins
($0.38\pm0.21$ and $0.30\pm0.14$), while the correlation is non-existent for
X-class flares ($0.01\pm0.15$). 
This difference is quite similar for the integral peak flux above 60 MeV: 
a weak to moderate correlation with longitude ($0.40\pm0.25$ and
$0.44\pm0.13$) for M-class flares, and consistent with zero for the X-class
flares ($0.12\pm0.14$).  
The mean peak fluxes in each of the bins for the combination of flare
magnitude and longitude are shown in
Figure~\ref{fig.flux_vs_fl_long_vs_fl_int} and listed in
Table~\ref{tab.flux_vs_fl_long_vs_fl_int} in the Appendix.
The correlations between the peak flux and flare intensity for the three 
longitude bins are also listed in Table~\ref{table.corrcoeff_2d} and show
no significant dependency on longitude for $E>10$~MeV moving from the east
towards the west side of the Sun ($0.72\pm0.29$, $0.62\pm0.12$ and
$0.52\pm0.10$), but a significant decrease for $E>60$~MeV ($0.90\pm0.11$,
$0.73\pm0.09$ and $0.60\pm0.09$).
It should be noted that the easternmost bin merely contains seven
events. 

\begin{figure}[t] 
\centerline{
  \includegraphics[angle=-90,width=0.33\textwidth]{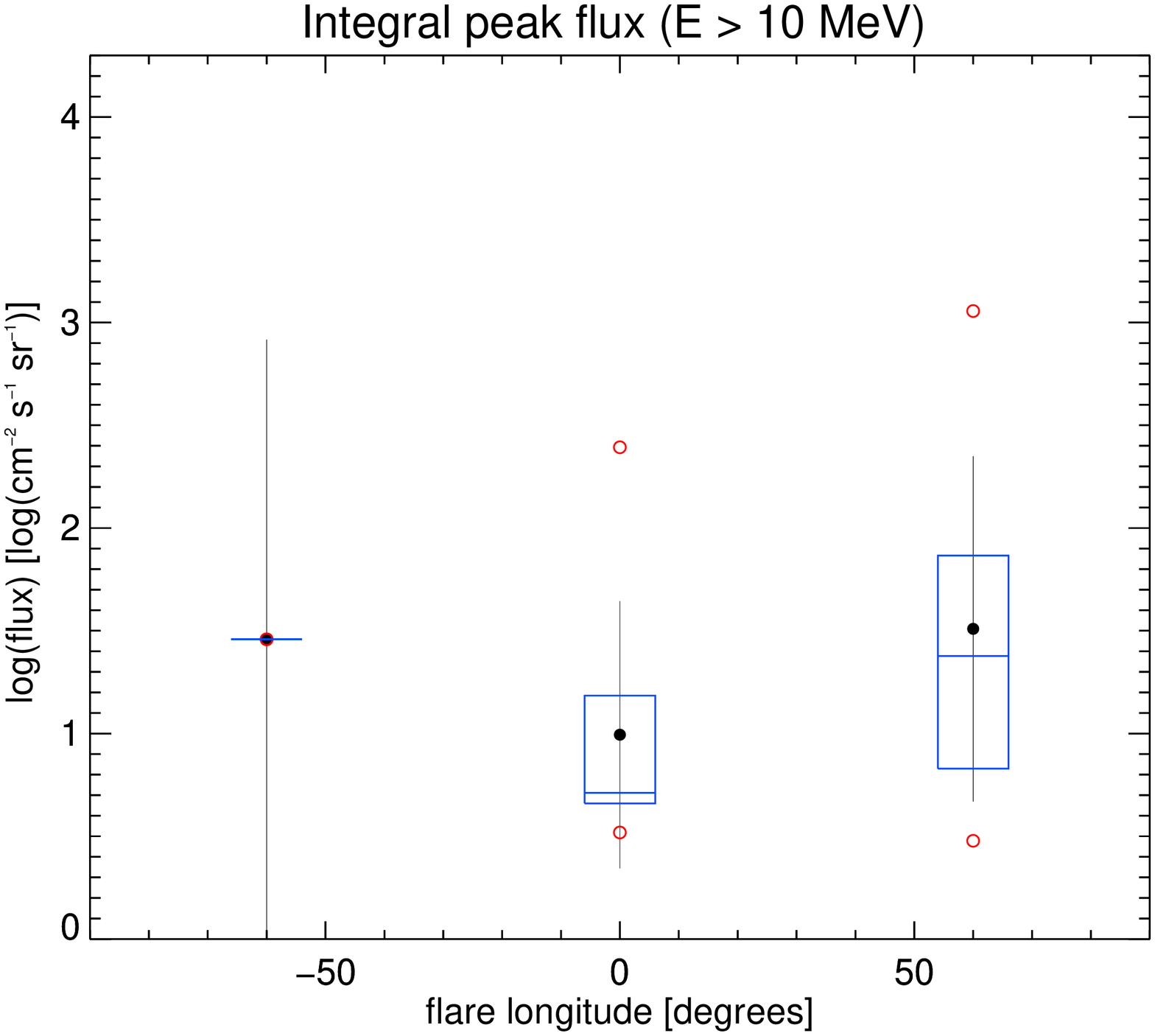}
  \includegraphics[angle=-90,width=0.33\textwidth]{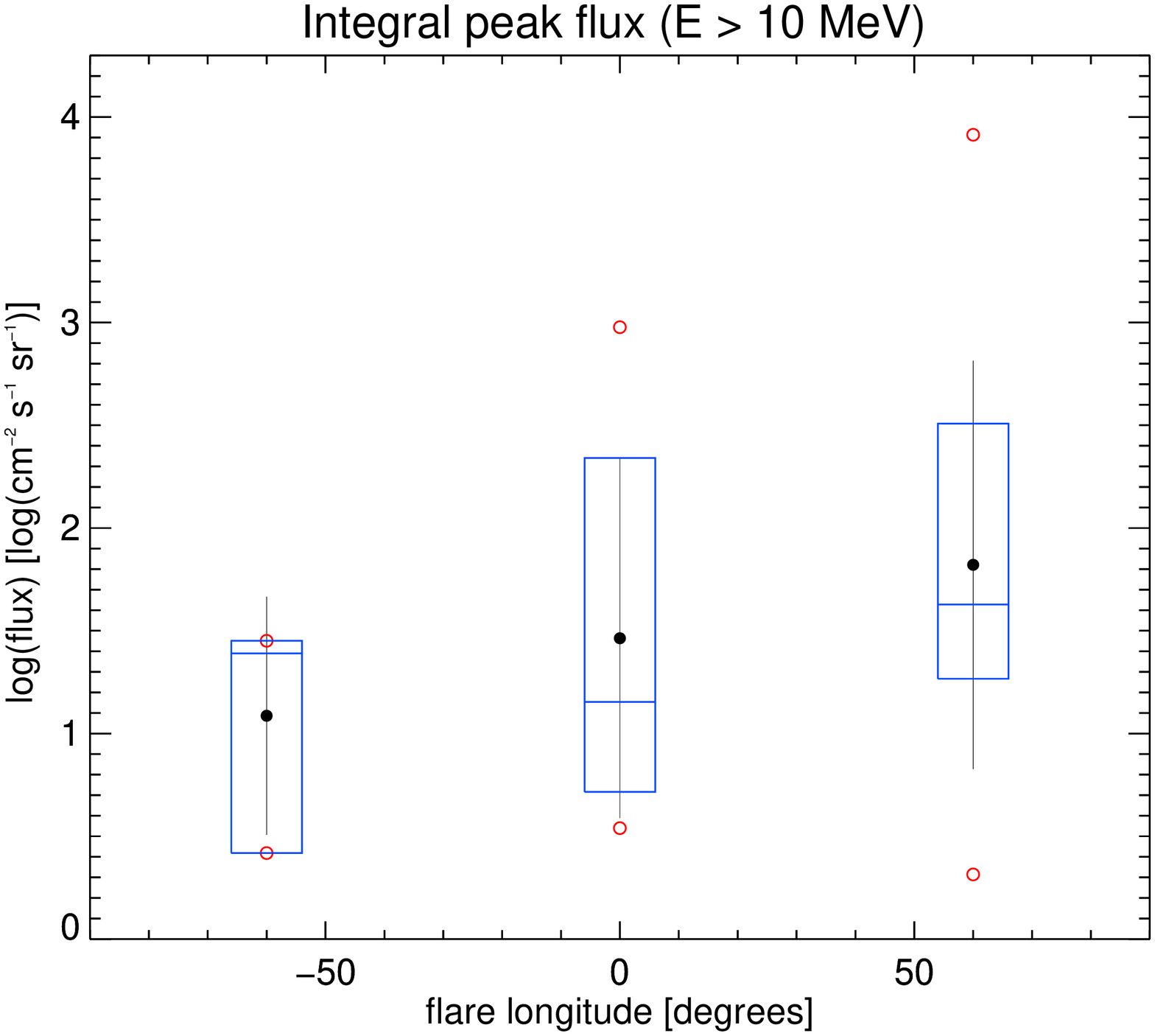}
  \includegraphics[angle=-90,width=0.33\textwidth]{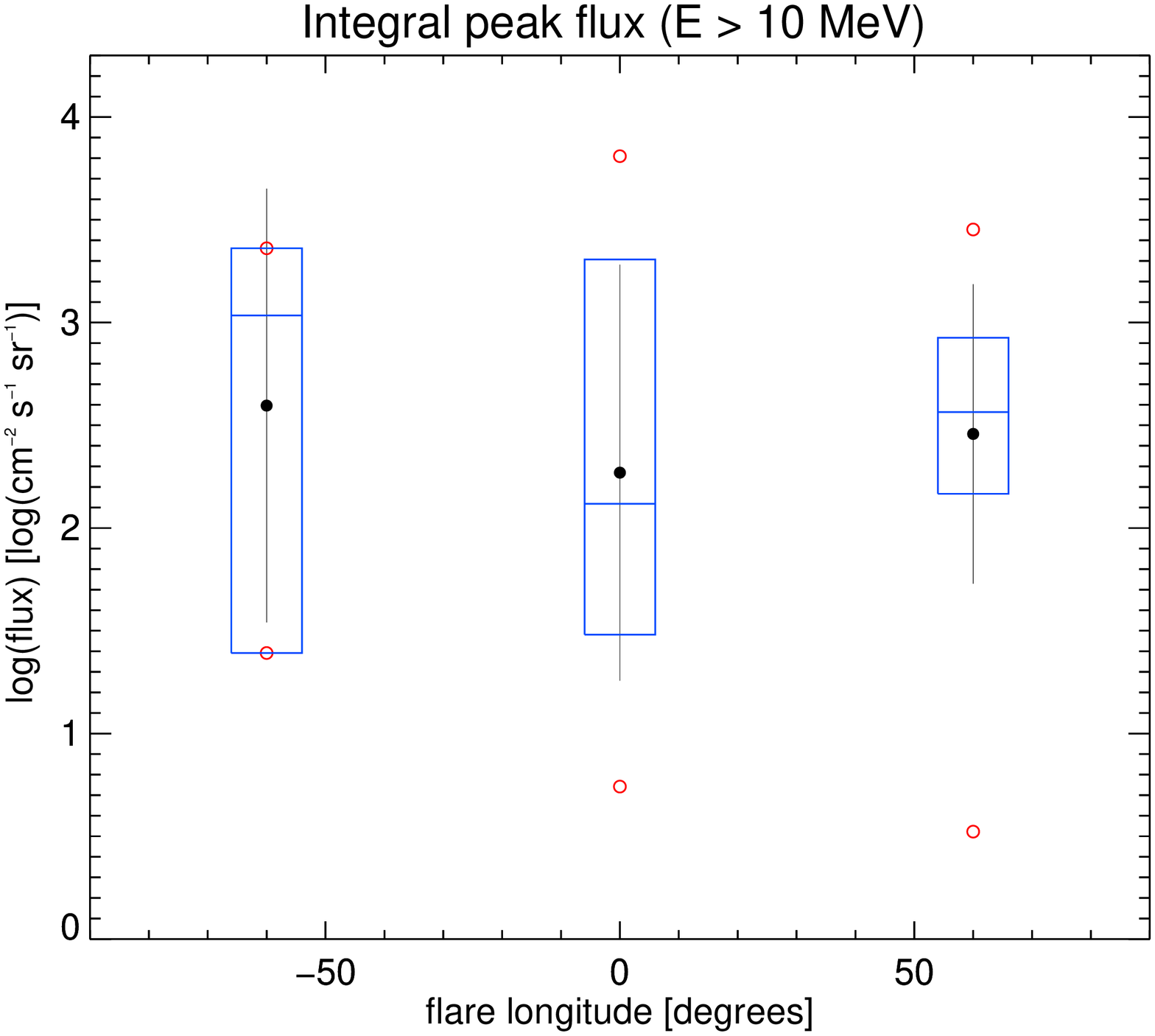}}
\vspace{-0.25\textwidth}   
\centerline{\tiny 
  \hspace{0.04\textwidth}  ${}^{\mathrm{M}1 \le I_{\mathrm{f}} < \mathrm{M}3.2}$  
  \hspace{0.205\textwidth} ${}^{\mathrm{M}3.2 \le I_{\mathrm{f}} < \mathrm{X}1}$
  \hspace{0.205\textwidth}  ${}^{I_{\mathrm{f}} \ge \mathrm{X}1}$
  \hfill}
\vspace{0.23\textwidth}    
\centerline{
  \includegraphics[angle=-90,width=0.33\textwidth]{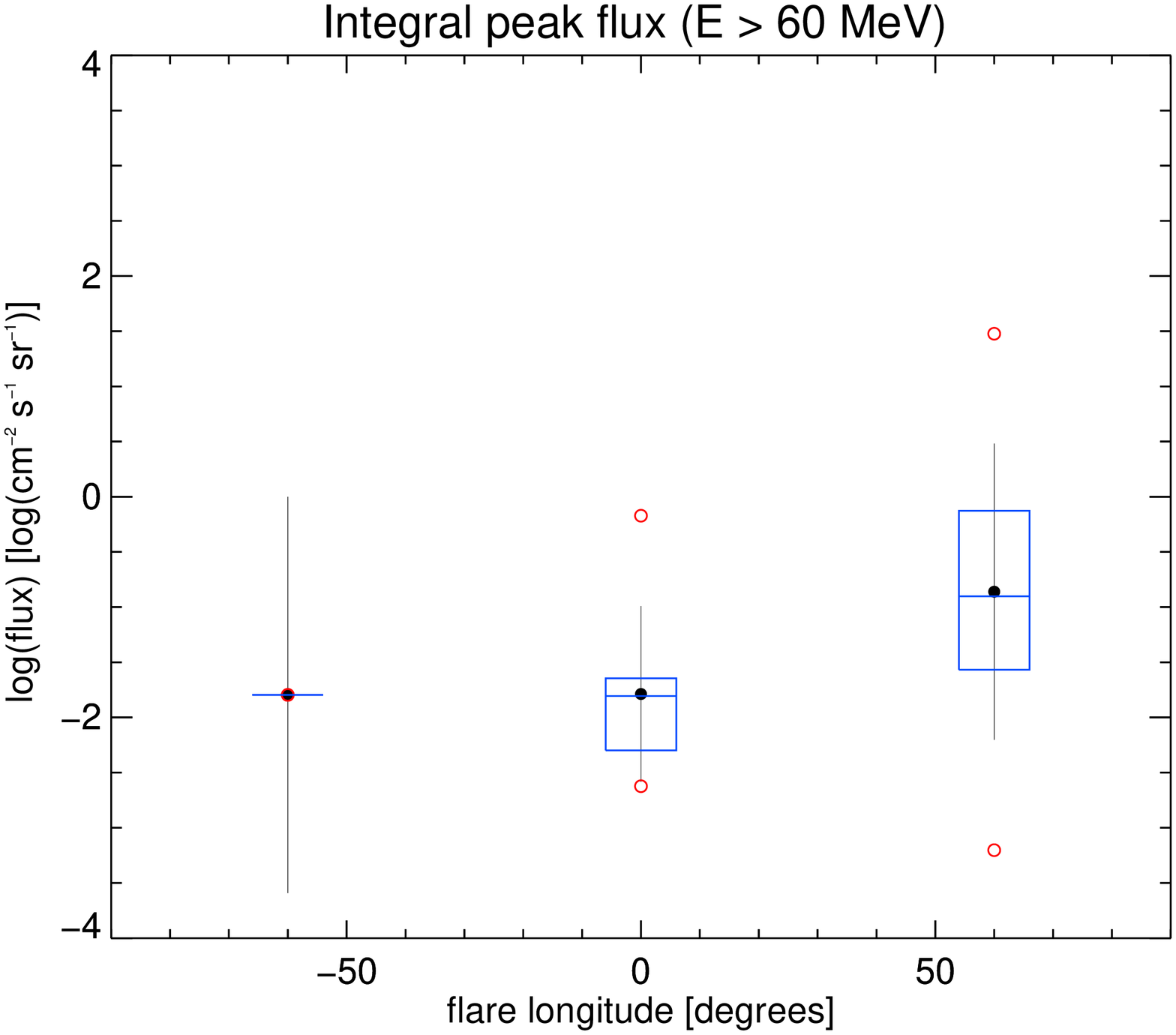}
  \includegraphics[angle=-90,width=0.33\textwidth]{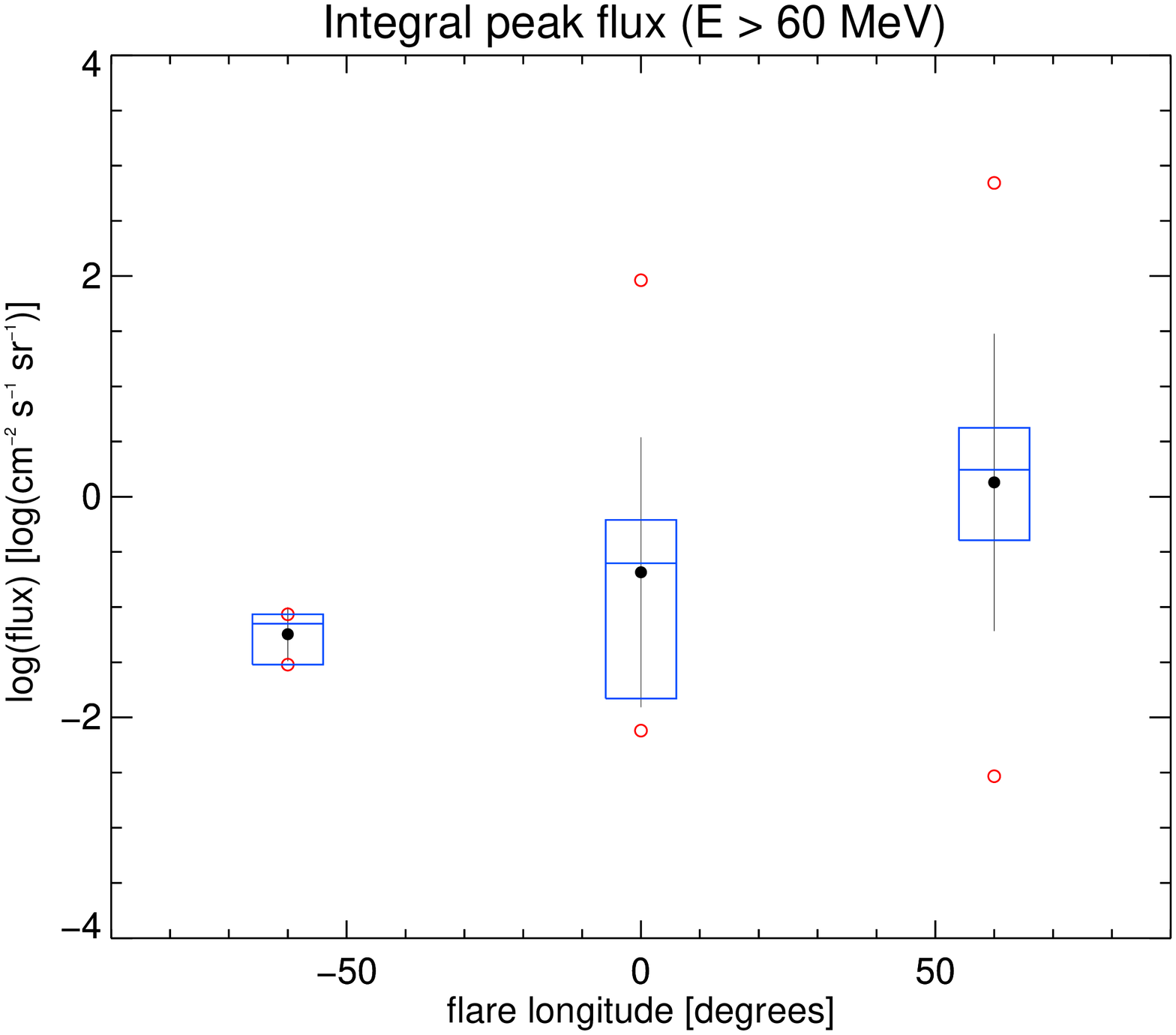}
  \includegraphics[angle=-90,width=0.33\textwidth]{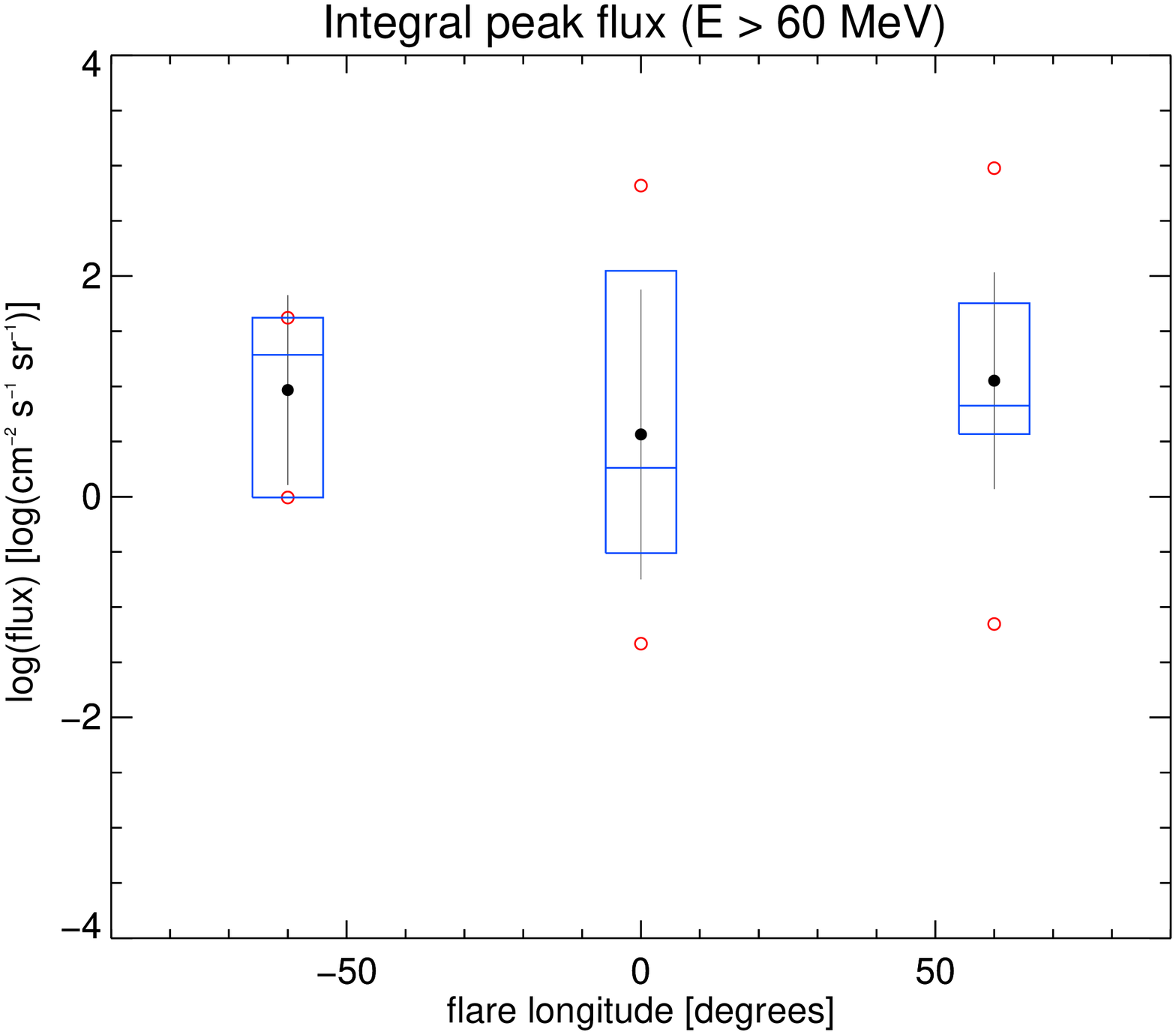}}
\vspace{-0.25\textwidth}   
\centerline{\tiny 
  \hspace{0.04\textwidth}  ${}^{\mathrm{M}1 \le I_{\mathrm{f}} < \mathrm{M}3.2}$  
  \hspace{0.205\textwidth} ${}^{\mathrm{M}3.2 \le I_{\mathrm{f}} < \mathrm{X}1}$
  \hspace{0.205\textwidth}  ${}^{I_{\mathrm{f}} \ge \mathrm{X}1}$
  \hfill}
\vspace{0.23\textwidth}    
\caption{Binned plots of the logarithm of the peak flux for $E>$10~MeV (top)
  and $E>$60~MeV protons (bottom) as a function of the longitude of the flare
  location for three different flare magnitude bins: M$1 \le
  I_{\mathrm{f}} < $M$3.2$ (left),  
  M$3.2 \le I_{\mathrm{f}} < $X$1$ (middle) and $I_{\mathrm{f}} \ge $X1
  (right). The convention the plot is the same as in Figure~\ref{fig.flux_vs_flareintens_bins}.} 
  \label{fig.flux_vs_fl_long_vs_fl_int} 
\end{figure}

The correlations have also been derived for the three flare
magnitude and longitude bins between the proton peak fluxes and the CME speed. 
The correlation with the CME speed for $E>$10~MeV increases from
$0.40\pm0.12$ for the stronger M-class flares to 
$0.61\pm0.10$ for X-class flares. 
The behavior towards smaller flare intensities is unclear due to the
relatively large uncertainty ($0.59\pm0.22$).   
The difference is not significant for the $E>$60~MeV peak fluxes:
changing from $0.25\pm0.36$ and $0.27\pm0.16$ for the two M-class flare bins
to $0.38\pm0.17$ for the X-class flares.
The mean peak fluxes in each of the bins for the combination of flare peak
intensity and CME speed are shown in
Figure~\ref{fig.flux_vs_cme_v_vs_fl_int} and listed in
Table~\ref{tab.flux_vs_cme_v_vs_fl_int} in the Appendix.
Note that there are no SEP events present in the data sample associated with a
flare of magnitude weaker than M$3.2$ and a CME with a speed faster than 2000
km~s$^{-1}$.  
Table~\ref{table.corrcoeff_2d} shows that for both energy channels, the
correlation with the CME speed is the largest for the central location
($0.72\pm0.07$ and $0.63 \pm0.09$) and is significantly 
smaller for longitudes west of $30^{\circ}$ ($0.45\pm0.12$ and $0.25\pm0.14$).
The easternmost bin contains only four events.

\begin{figure}[t]
\centerline{
  \includegraphics[angle=-90,width=0.33\textwidth]{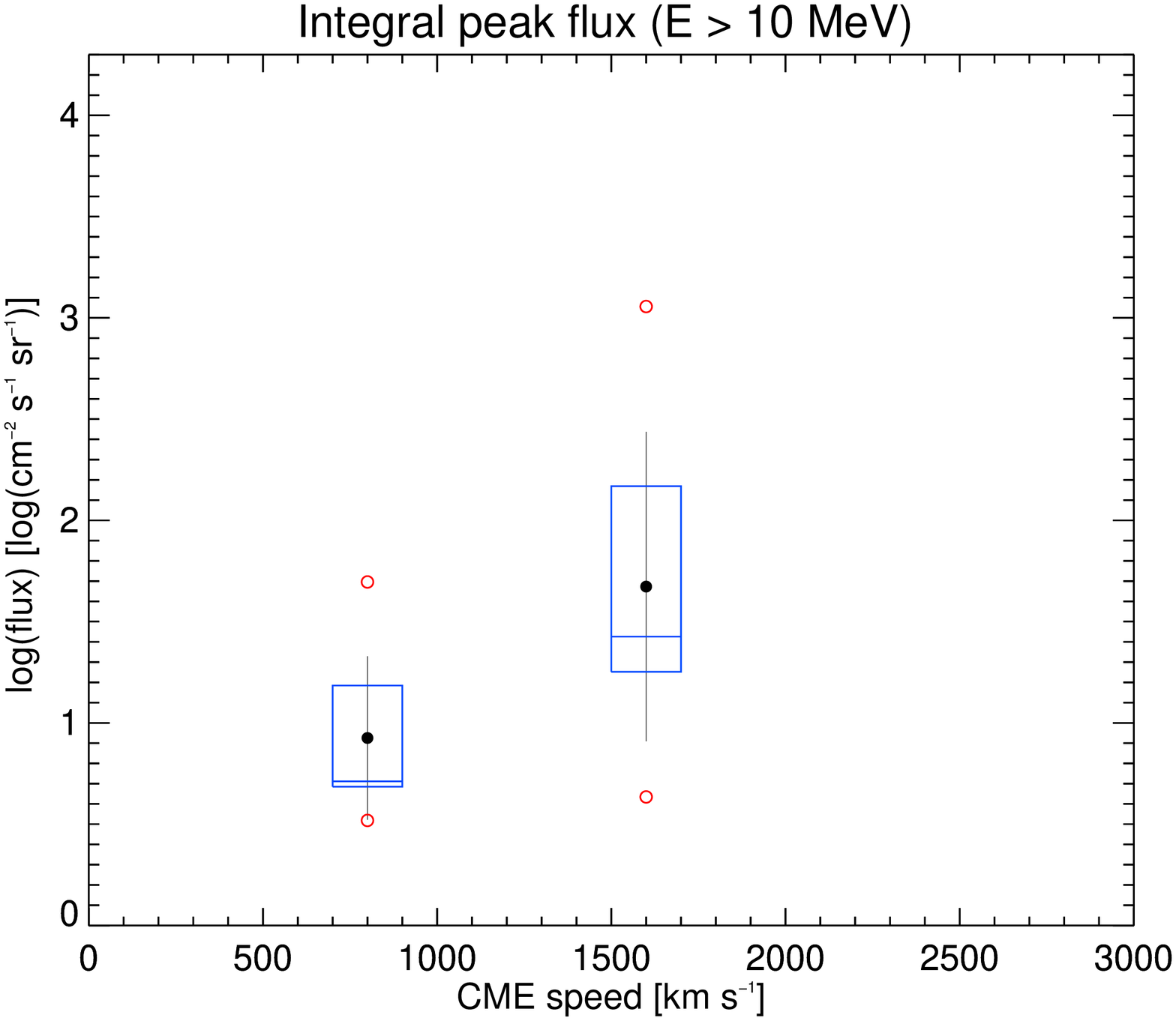}
  \includegraphics[angle=-90,width=0.33\textwidth]{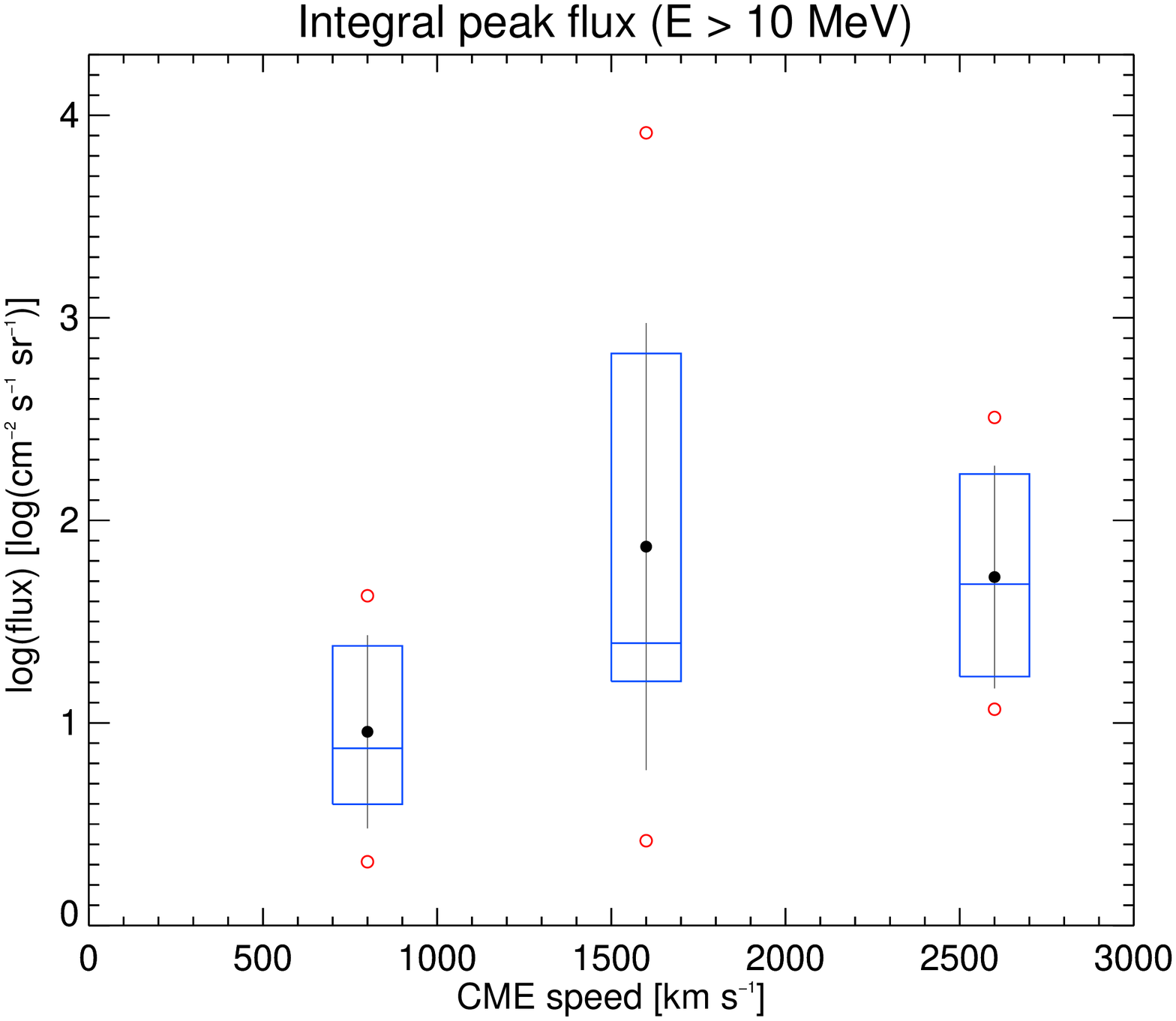}
  \includegraphics[angle=-90,width=0.33\textwidth]{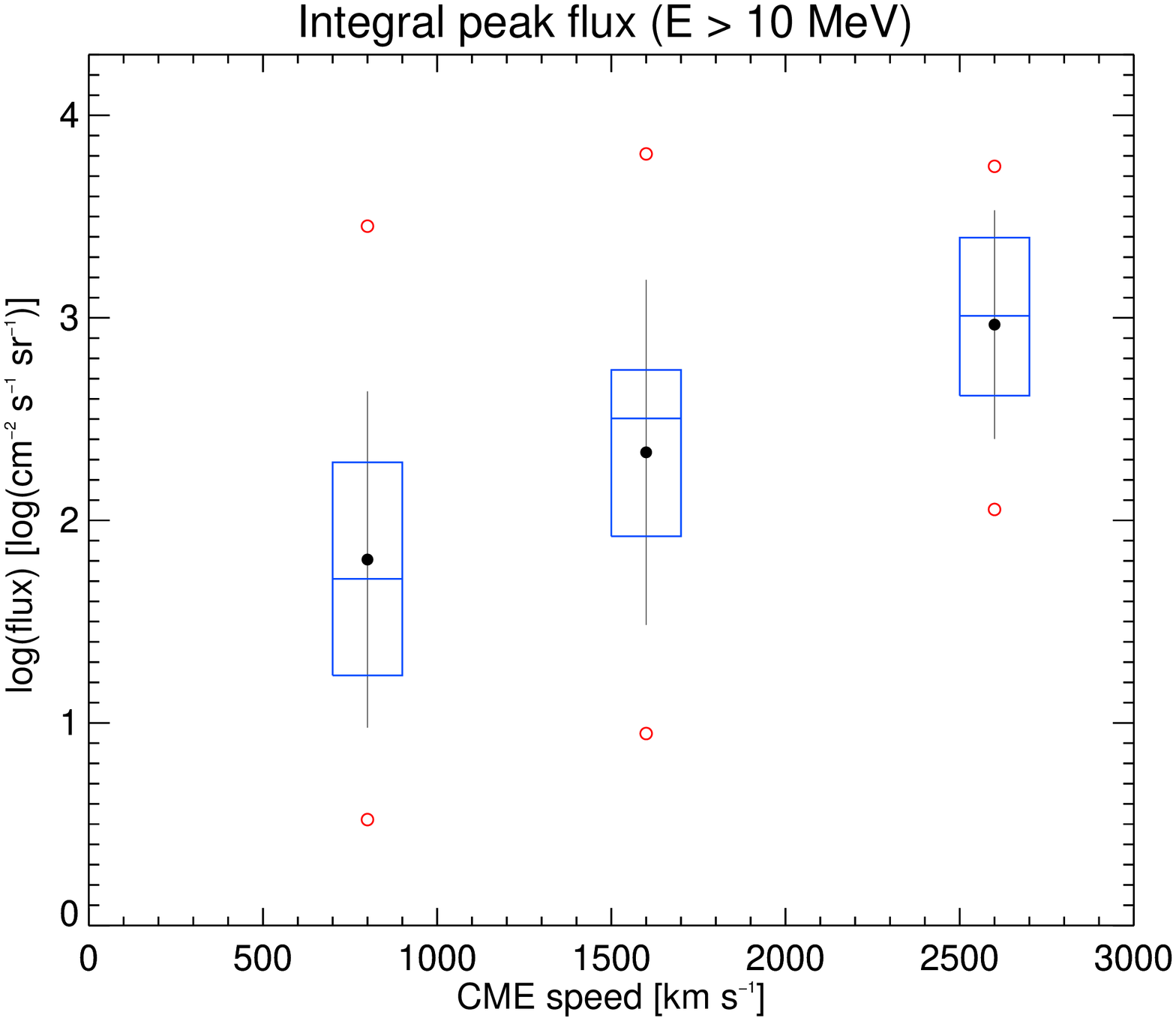}
}
\vspace{-0.25\textwidth}   
\centerline{\tiny 
  \hspace{0.04\textwidth}  ${}^{\mathrm{M}1 \le I_{\mathrm{f}} < \mathrm{M}3.2}$  
  \hspace{0.205\textwidth} ${}^{\mathrm{M}3.2 \le I_{\mathrm{f}} < \mathrm{X}1}$
  \hspace{0.205\textwidth}  ${}^{I_{\mathrm{f}} \ge \mathrm{X}1}$
  \hfill}
\vspace{0.23\textwidth}    
\centerline{
  \includegraphics[angle=-90,width=0.33\textwidth]{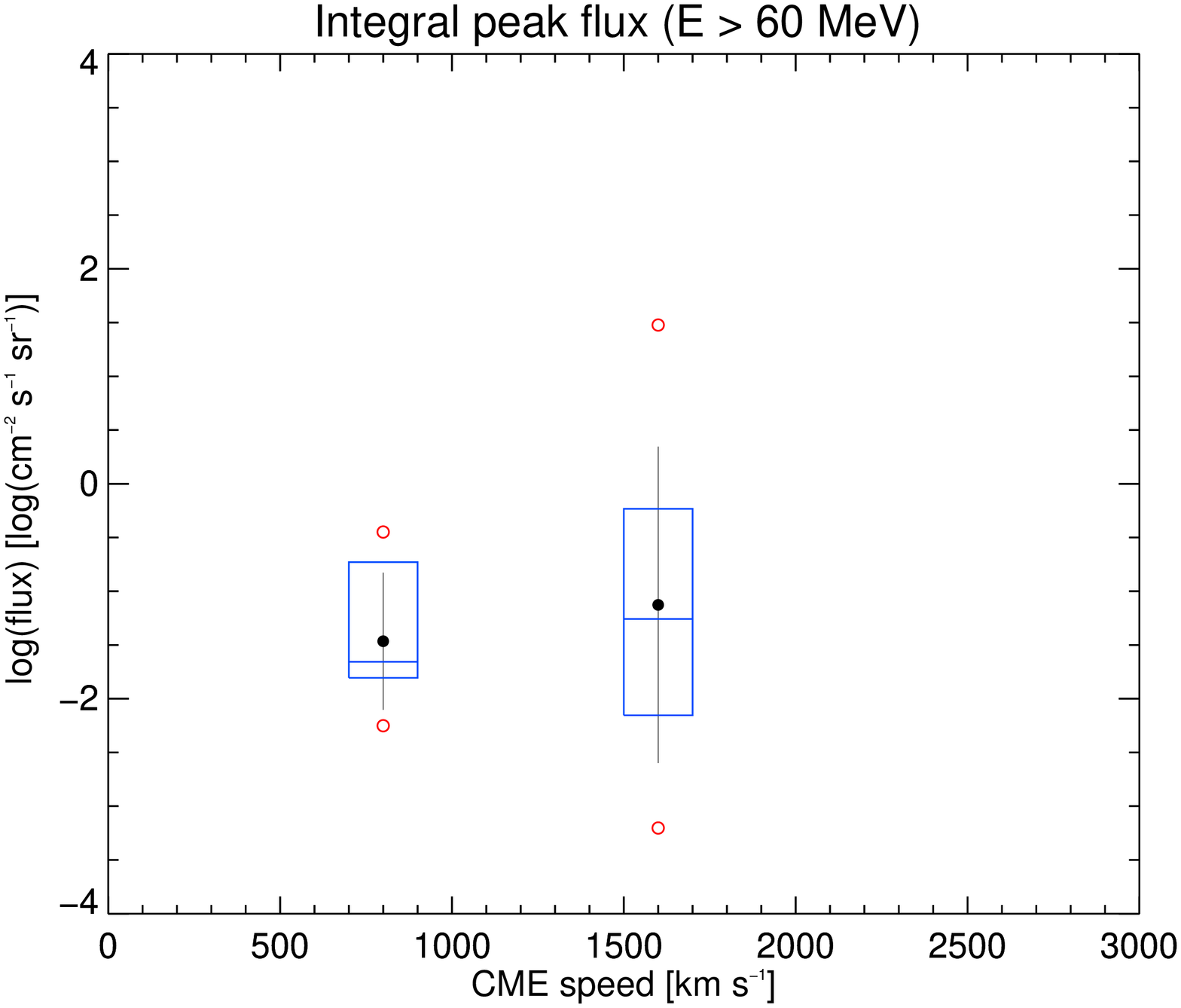}
  \includegraphics[angle=-90,width=0.33\textwidth]{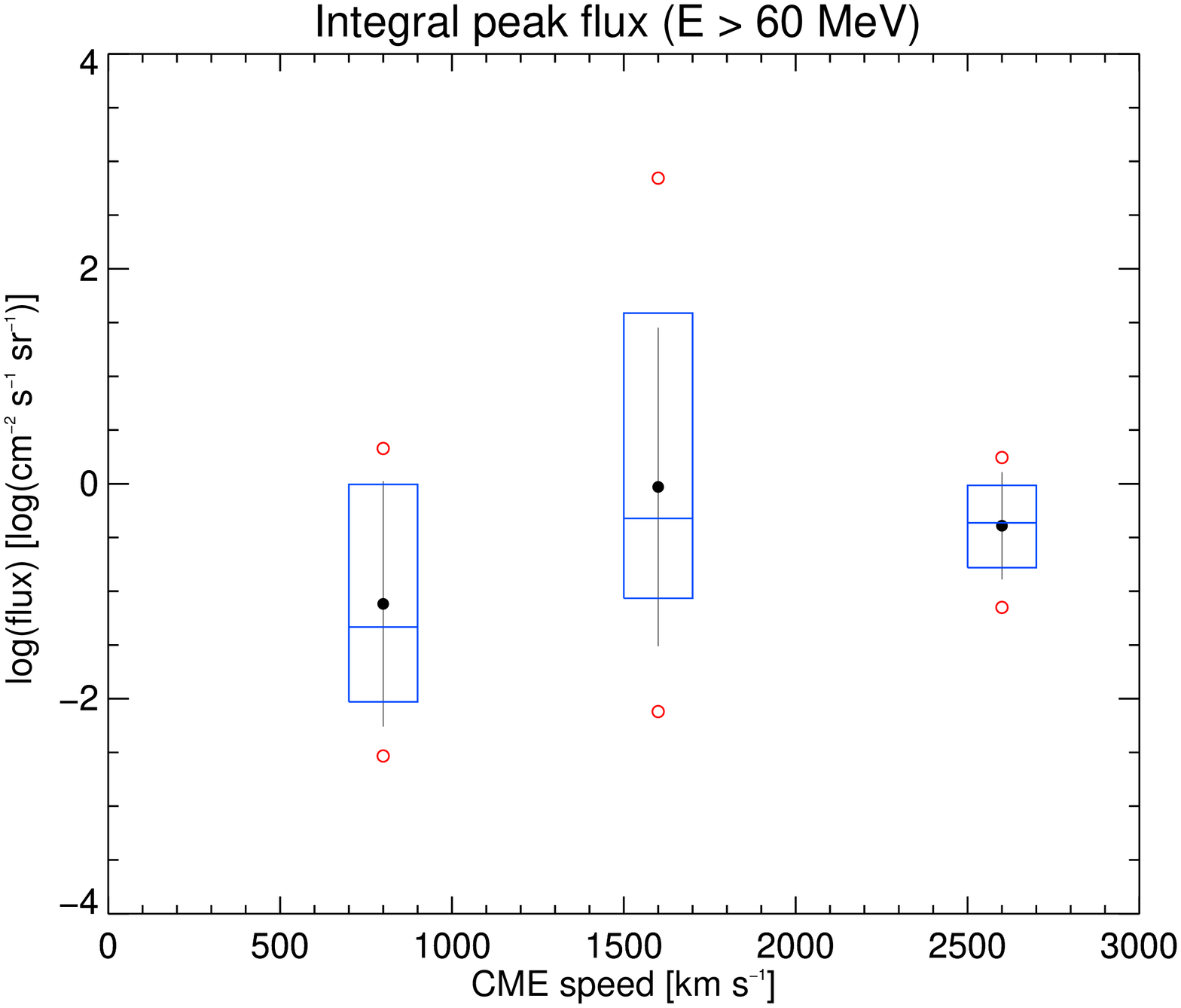}
  \includegraphics[angle=-90,width=0.33\textwidth]{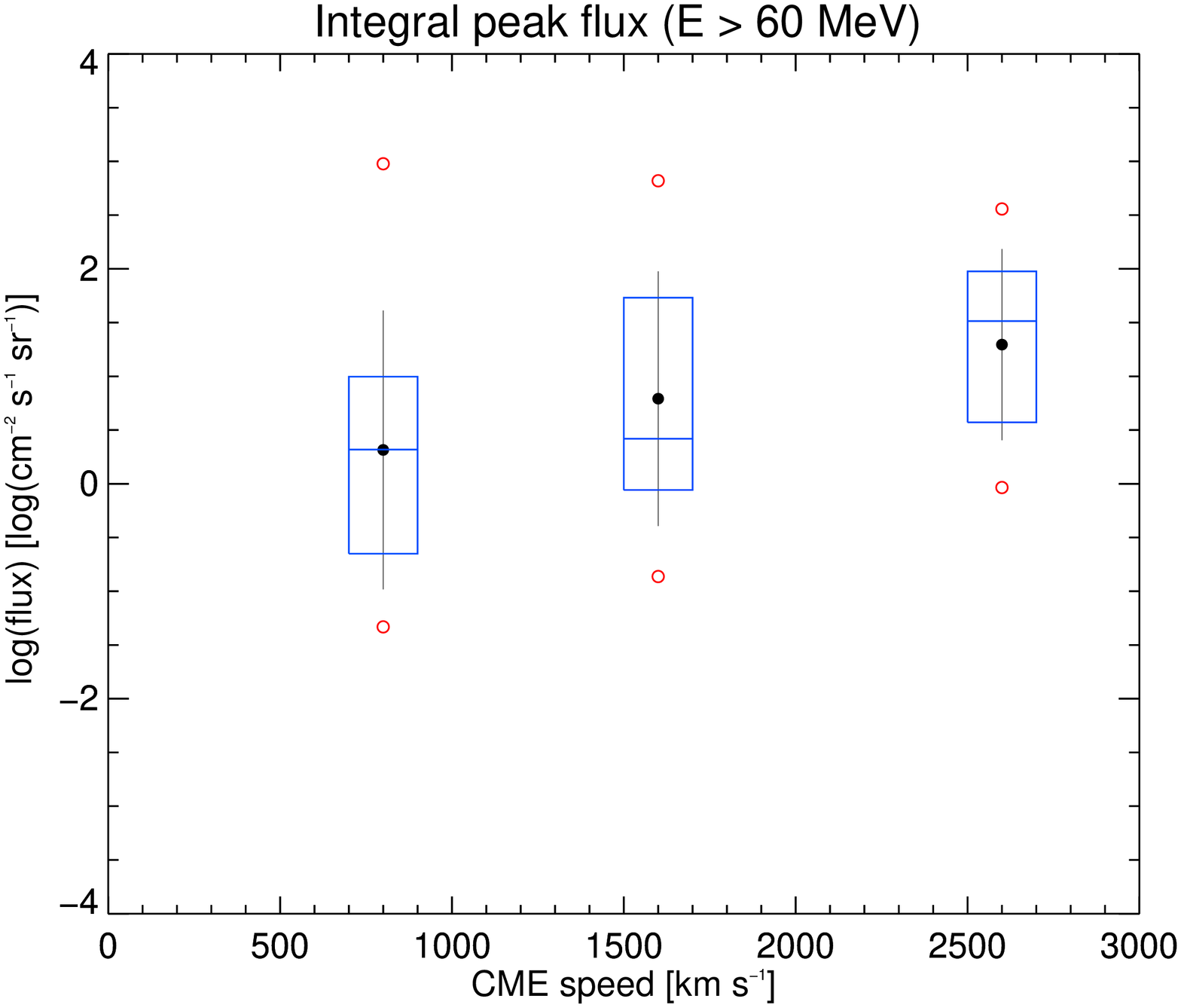}}
\vspace{-0.25\textwidth}   
\centerline{\tiny 
  \hspace{0.04\textwidth}  ${}^{\mathrm{M}1 \le I_{\mathrm{f}} < \mathrm{M}3.2}$  
  \hspace{0.205\textwidth} ${}^{\mathrm{M}3.2 \le I_{\mathrm{f}} < \mathrm{X}1}$
  \hspace{0.205\textwidth}  ${}^{I_{\mathrm{f}} \ge \mathrm{X}1}$
  \hfill}
\vspace{0.23\textwidth}    
\caption{Binned plots of the logarithm of the peak flux for $E>$10~MeV (top)
  and $E>$60~MeV protons (bottom) 
  as a function of the CME speed for three
  different flare magnitude bins: M$1 \le
  I_{\mathrm{f}} < $M$3.2$ (left),  
  M$3.2 \le I_{\mathrm{f}} < $X$1$ (middle) and $I_{\mathrm{f}} \ge $X1
  (right)
  The convention for the plot is the same as in Figure~\ref{fig.flux_vs_flareintens_bins}.}
  \label{fig.flux_vs_cme_v_vs_fl_int} 
\end{figure}

\begin{figure}[tp]
\centerline{
  \includegraphics[angle=-90,width=0.33\textwidth]{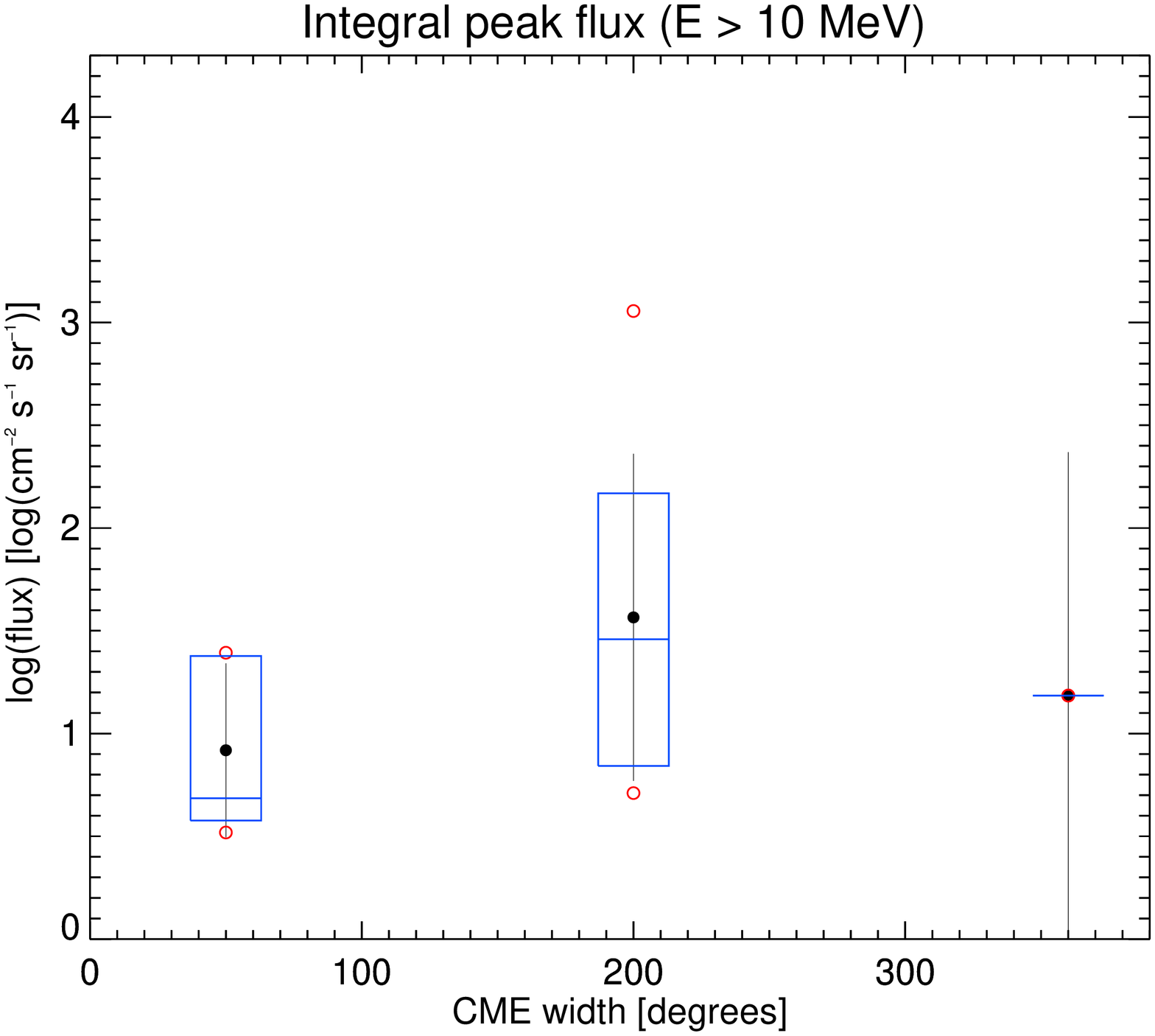}
  \includegraphics[angle=-90,width=0.33\textwidth]{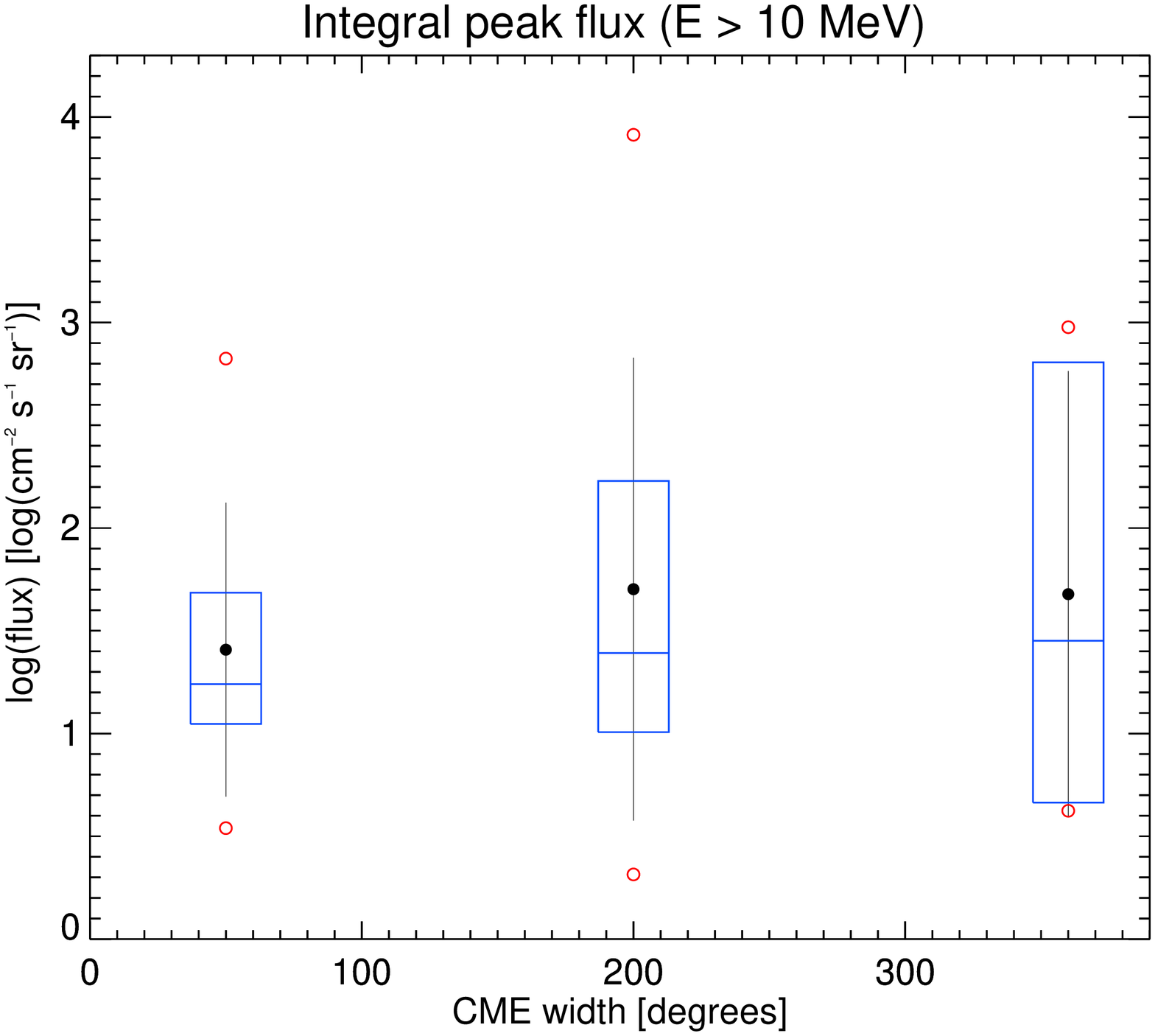}
  \includegraphics[angle=-90,width=0.33\textwidth]{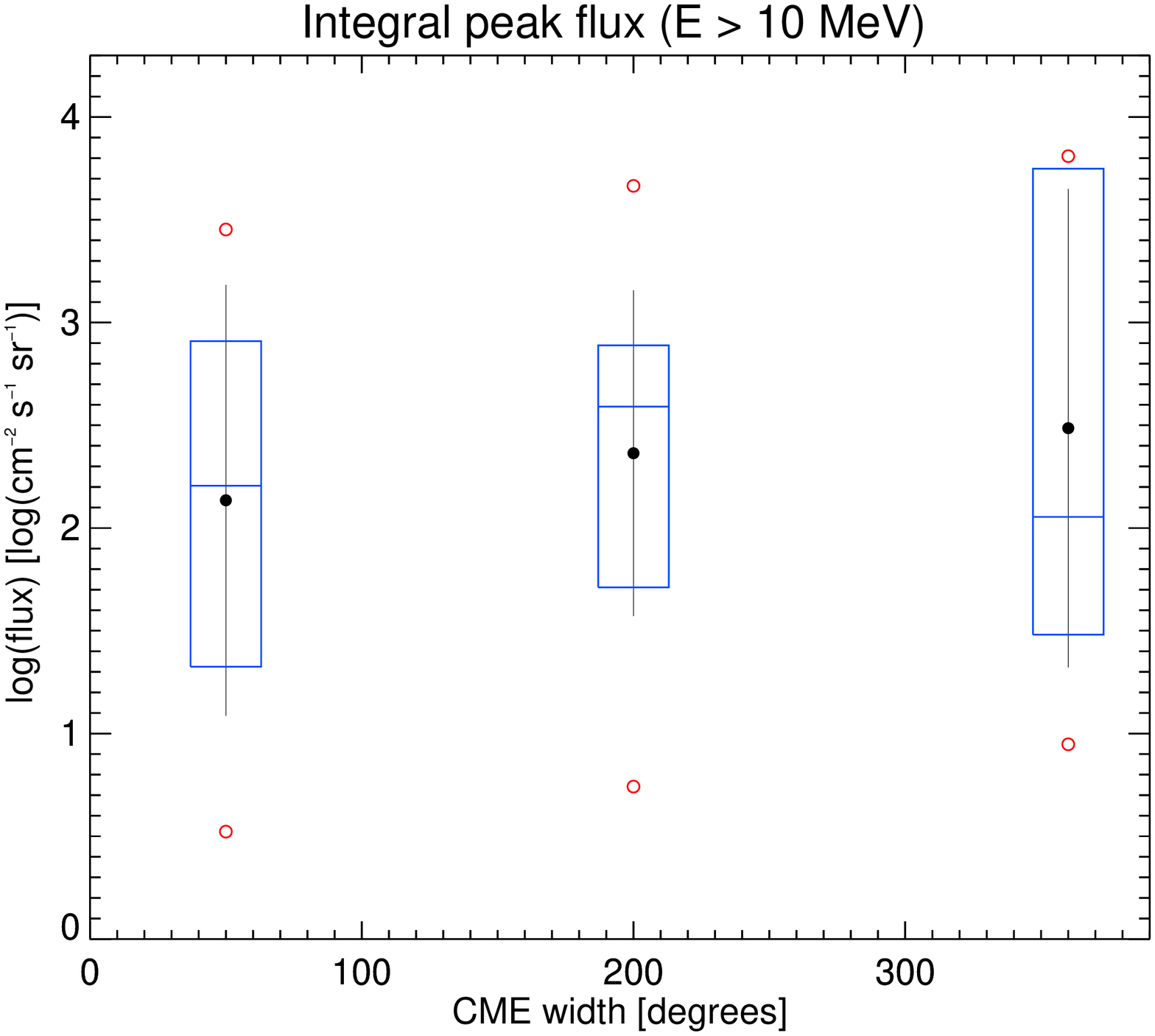}}
\vspace{-0.25\textwidth}   
\centerline{\tiny 
  \hspace{0.04\textwidth}  ${}^{\mathrm{M}1 \le I_{\mathrm{f}} < \mathrm{M}3.2}$  
  \hspace{0.205\textwidth} ${}^{\mathrm{M}3.2 \le I_{\mathrm{f}} < \mathrm{X}1}$
  \hspace{0.205\textwidth}  ${}^{I_{\mathrm{f}} \ge \mathrm{X}1}$
  \hfill}
\vspace{0.23\textwidth}    
\centerline{
  \includegraphics[angle=-90,width=0.33\textwidth]{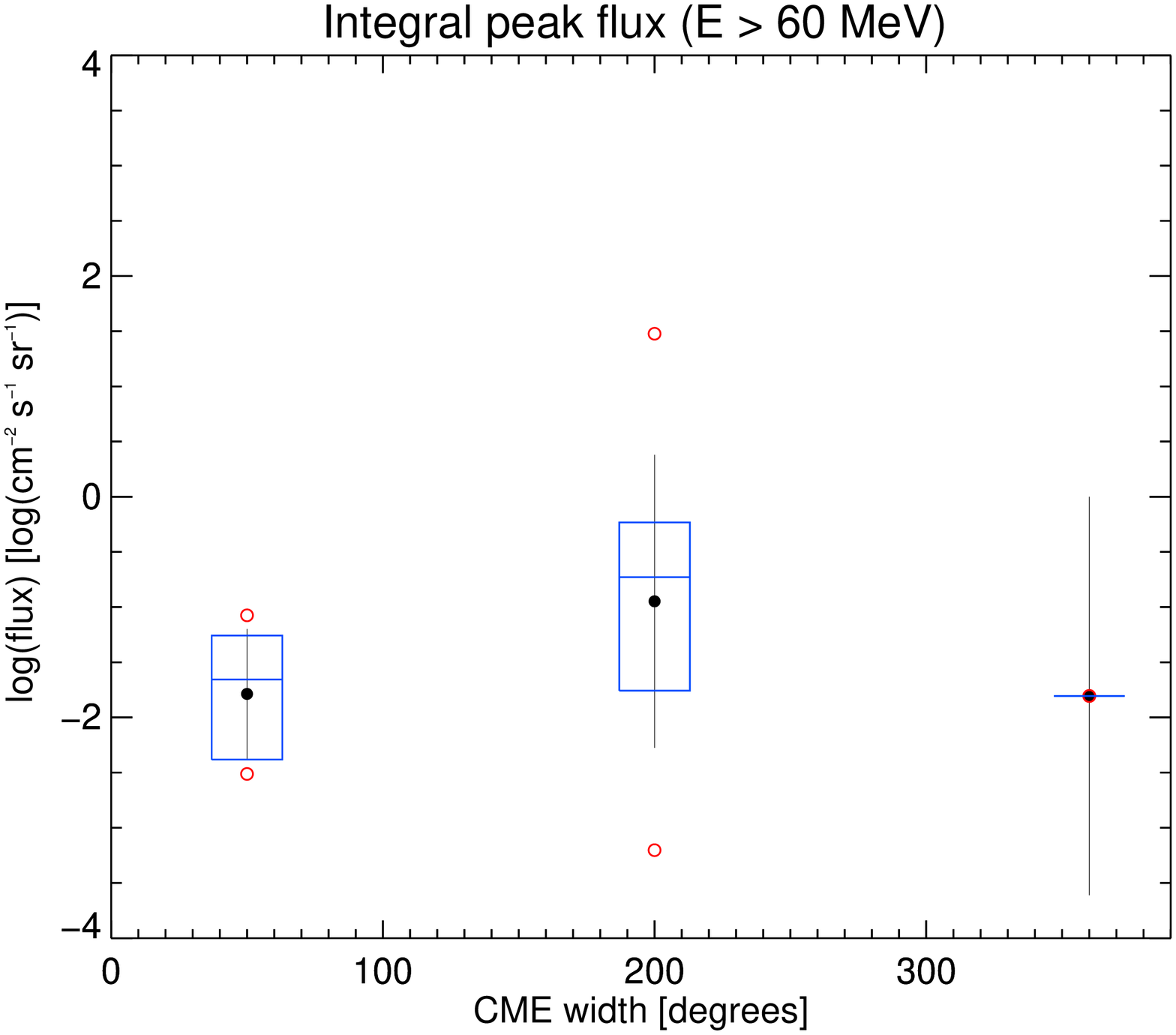}
  \includegraphics[angle=-90,width=0.33\textwidth]{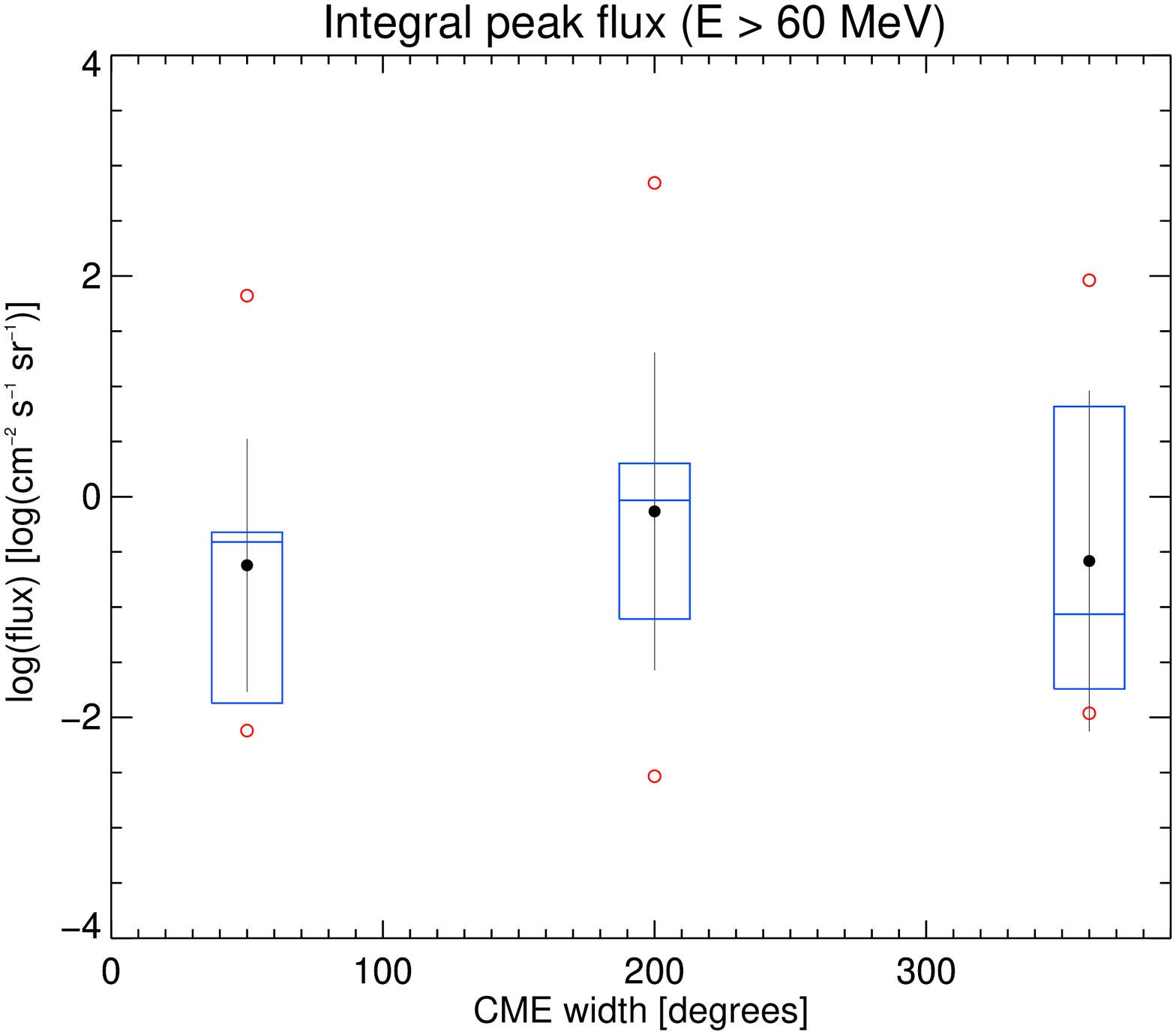}
  \includegraphics[angle=-90,width=0.33\textwidth]{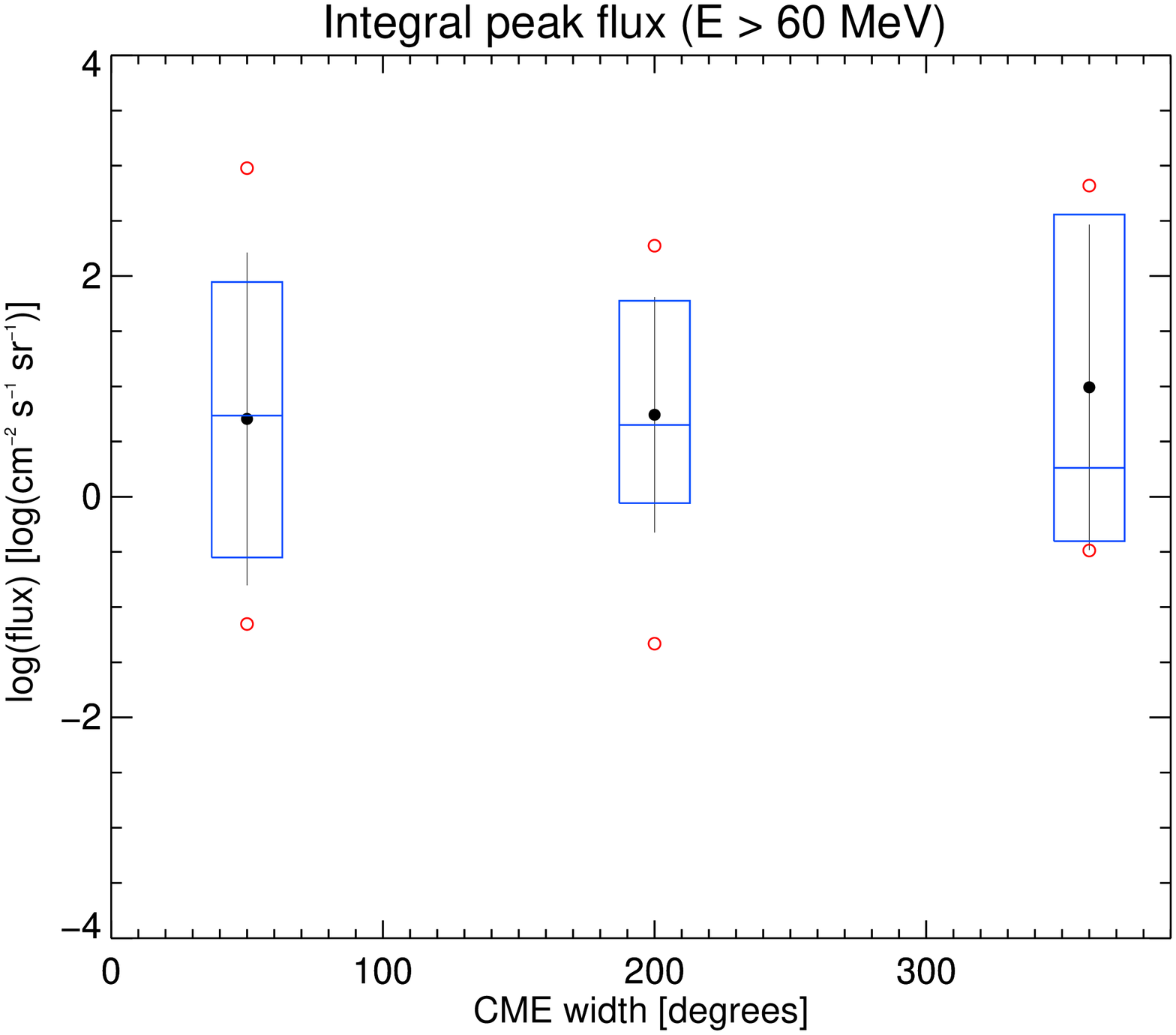}}
\vspace{-0.25\textwidth}   
\centerline{\tiny 
  \hspace{0.04\textwidth}  ${}^{\mathrm{M}1 \le I_{\mathrm{f}} < \mathrm{M}3.2}$  
  \hspace{0.205\textwidth} ${}^{\mathrm{M}3.2 \le I_{\mathrm{f}} < \mathrm{X}1}$
  \hspace{0.205\textwidth}  ${}^{I_{\mathrm{f}} \ge \mathrm{X}1}$
  \hfill}
\vspace{0.23\textwidth}    
\caption{Binned plots of the logarithm of the peak flux for $E>$10~MeV (top)
  and $E>$60~MeV protons (bottom) 
  as a function of the CME width for three
  different flare magnitude bins: $M1 \le I_{\mathrm{f}} < M3.2$ (left), 
  $M3.2 \le I_{\mathrm{f}} < X1$ (middle) and $I_{\mathrm{f}} \ge X1$ (right).
  The convention for the plot is the same as in Figure~\ref{fig.flux_vs_flareintens_bins}.}
  \label{fig.flux_vs_cme_w_vs_fl_int} 
\end{figure}

Finally, the dependence of the peak fluxes on the CME width has been
determined in the three flare magnitude and two CME speed bins.
The correlations between the peak flux for both energy ranges and the CME
width remains within the error close to zero regardless of the flare
magnitude bin.  
Figure~\ref{fig.flux_vs_cme_w_vs_fl_int} shows the mean peak flux in each
bin when flare intensity and CME width are combined and these values
are listed in Table~\ref{tab.flux_vs_cme_w_vs_fl_int} in the Appendix.
The correlation with the CME width has been derived in two bins of CME speed,
namely smaller or greater than 1500~km~s$^{-1}$, and are given in
Table~\ref{table.corrcoeff_2d}.
For both energy ranges, there exists no correlation for slow CMEs
($-0.03\pm0.15$ for $E>$10~MeV and $-0.02\pm0.14$ for $E>$60~MeV), while
a very weak correlation exists for fast CMEs ($0.28\pm0.15$ for $E>$10~MeV and
$0.22\pm0.19$ for $E>$60~MeV). 
The mean peak fluxes in each of the bins for the combination of the CME
speed and width are shown in 
Figure~\ref{fig.flux_vs_cme_w_vs_cme_v} and listed in
Table~\ref{tab.flux_vs_cme_w_vs_cme_v} in the Appendix.

\begin{figure}[tp]
\centering
\centerline{
  \includegraphics[angle=-90,width=0.33\textwidth]{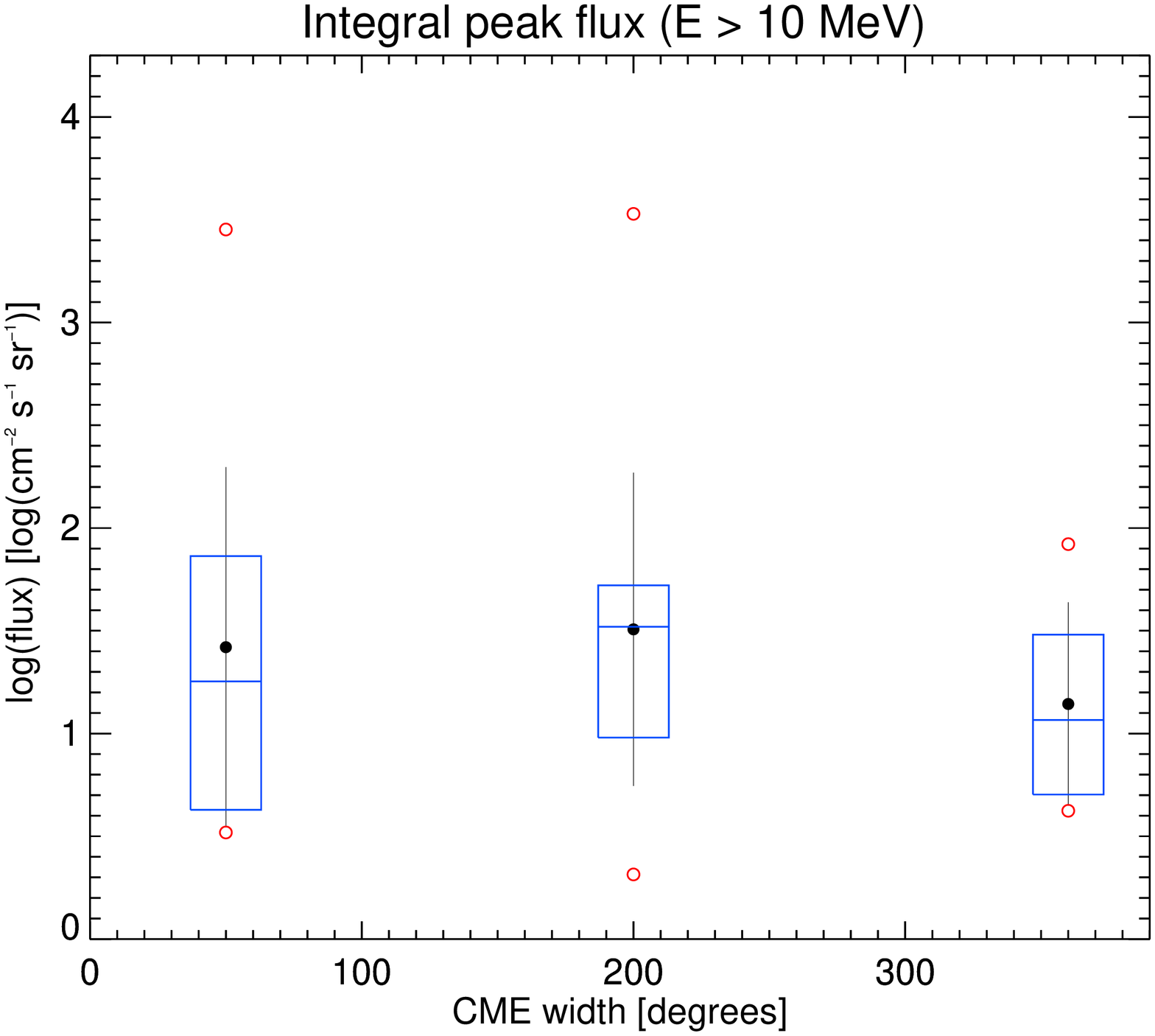}
  \includegraphics[angle=-90,width=0.33\textwidth]{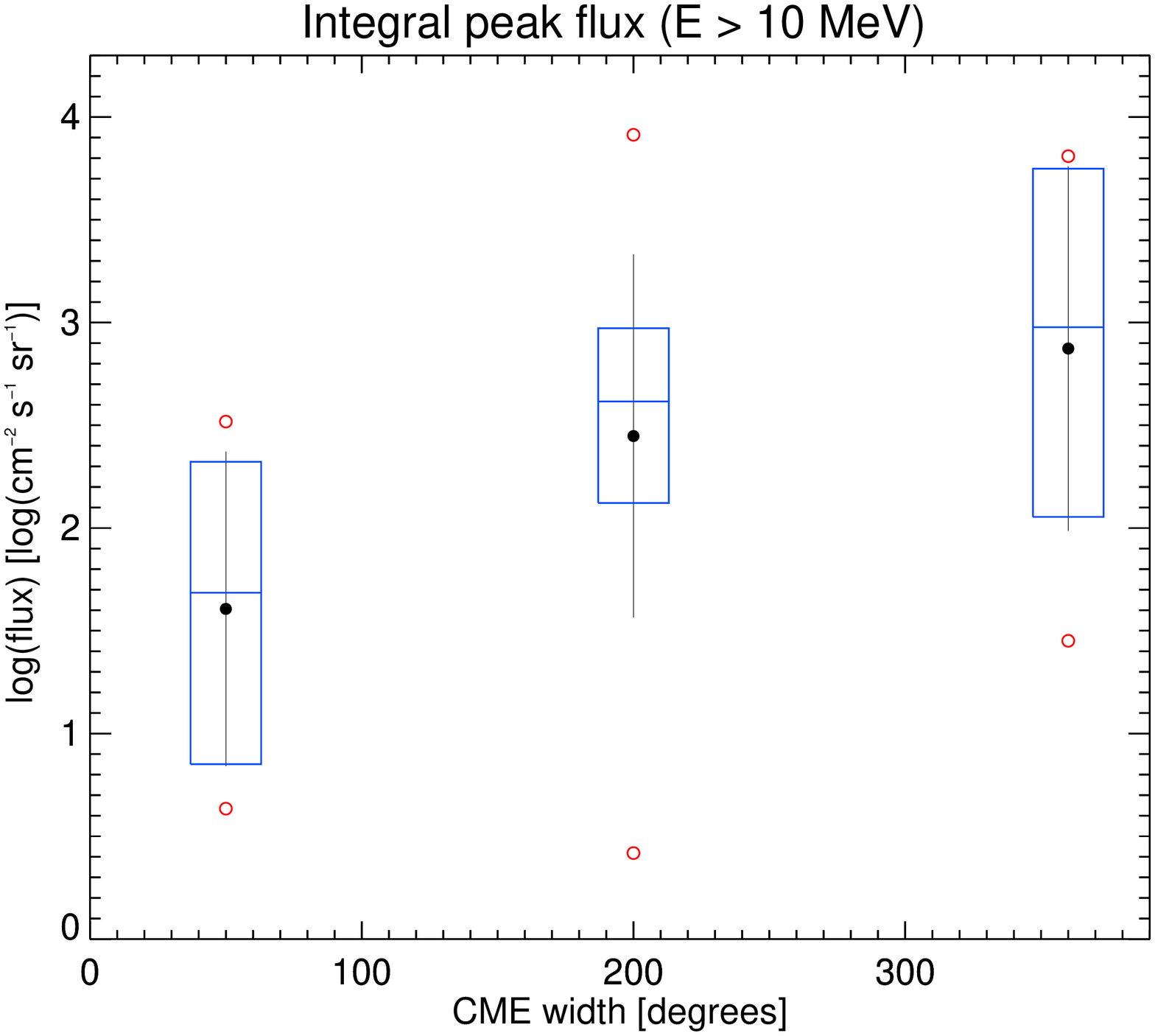}
}
\vspace{-0.258\textwidth}   
\centerline{\tiny 
  \hspace{0.2\textwidth}  ${}^{v_{\textsc{cme}} < 1500 \mathrm{km s^{-1}}}$
  \hspace{0.18\textwidth}   ${}^{v_{\textsc{cme}} \ge 1500 \mathrm{km s^{-1}}}$
  \hfill}
\vspace{0.23\textwidth}    
\centerline{
  \includegraphics[angle=-90,width=0.33\textwidth]{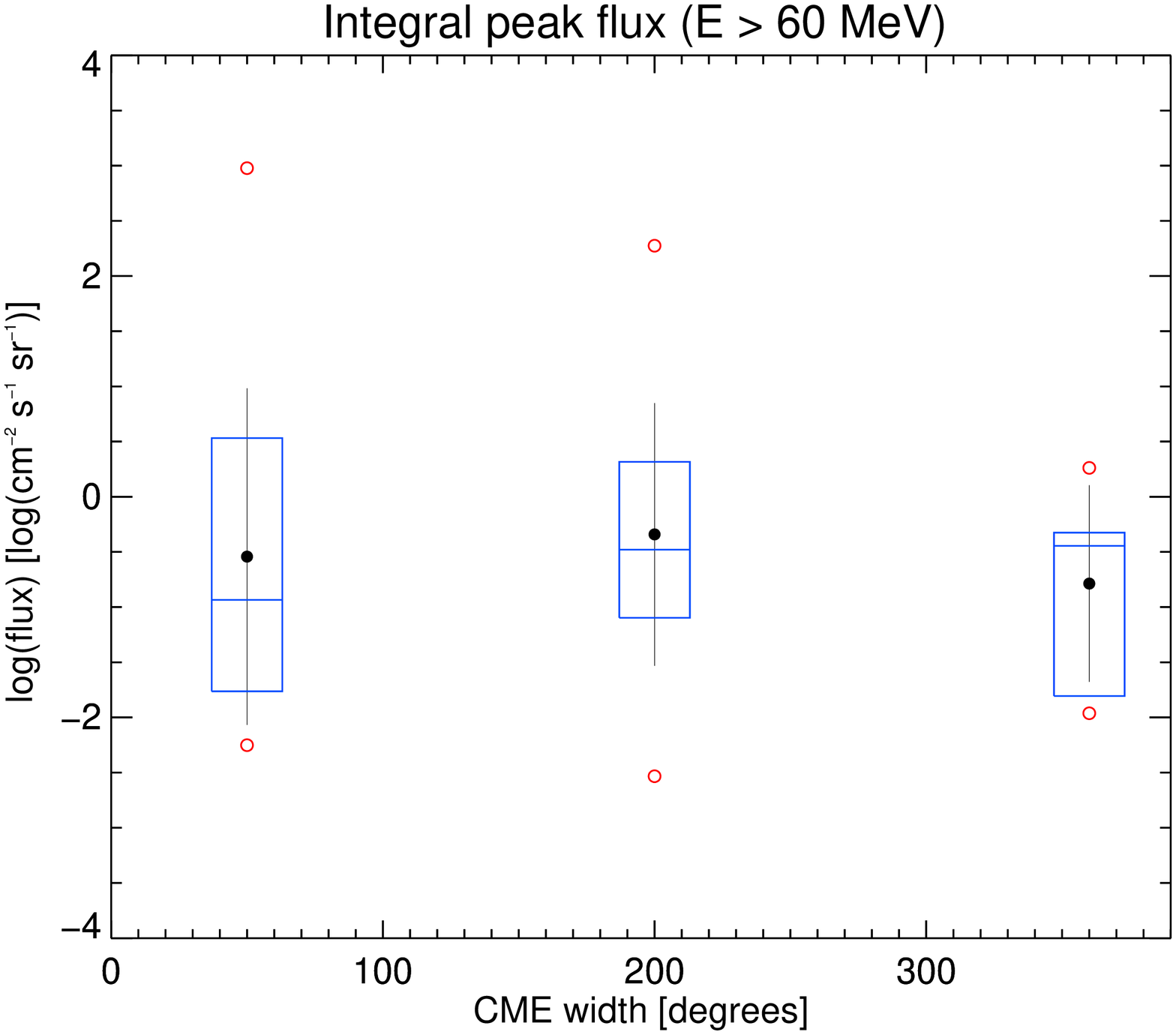}
  \includegraphics[angle=-90,width=0.33\textwidth]{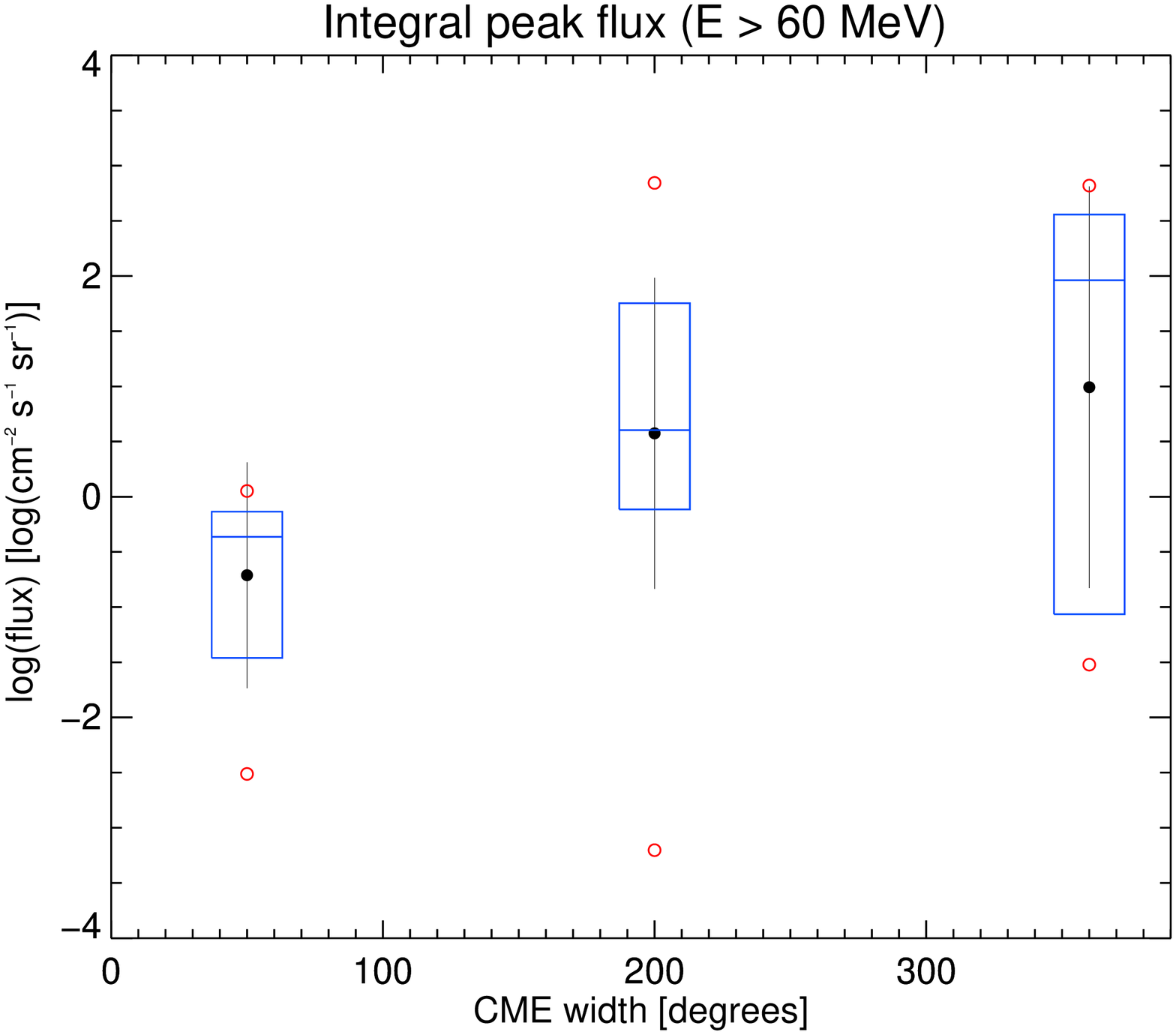}
}
\vspace{-0.258\textwidth}   
\centerline{\tiny 
  \hspace{0.2\textwidth}  ${}^{v_{\textsc{cme}} < 1500 \mathrm{km s^{-1}}}$
  \hspace{0.18\textwidth}   ${}^{v_{\textsc{cme}} \ge 1500 \mathrm{km s^{-1}}}$
  \hfill}
\vspace{0.23\textwidth}    
\caption{Binned plots of the logarithm of the peak flux for $E>$10~MeV (top)
  and $E>$60~MeV protons (bottom) 
  as a function of the CME width for two
  different CME speed bins: $ v_{\mathrm{CME}} < 1500$~km~s$^{-1}$ (left), 
  and $v_{\mathrm{CME}} \ge 1500$~km~s$^{-1}$ (right).
  The convention for the plot is the same as in Figure~\ref{fig.flux_vs_flareintens_bins}.}
\label{fig.flux_vs_cme_w_vs_cme_v} 
\end{figure}

\section{Discussion and Conclusions}\label{sec.discussion}

We have reported on the results of a statistical analysis of the dependence of
SEP occurrence probability and proton peak fluxes at 1~AU 
on both flare (intensity and longitude) and CME (speed and angular
width) characteristics.  
The work presented here goes beyond the usual study of the relationship
between the SEP event and the individual solar parameters by investigating the
effect of combining these parameters. 
Furthermore, the energy dependence of the correlation between the
proton peak flux and the solar parameters was derived and the influence of  
the ESPs on these correlations was explored. 

The results presented here show that the SEP occurrence probabilities
generally increase significantly for events associated with large X-class
flares, western solar longitudes, CMEs with high velocities and halo CMEs. 
It is found that the proton peak fluxes increase with flare intensity
and CME speed as well, but exhibit little correlation with the flare longitude
and CME width.
This is in general agreement with previous works which will be discussed in
detail below.
The observed qualitative behavior of the SEP occurrence probability is similar
irrespective of the data set used for the derivation of these probabilities. 
The results obtained with the CRR2010 and SSE lists show some small
differences in the actual probability values;
this is a consequence of the different energy ranges used, and of the presence
of electron dominated SEP events in the CRR2010 list that are excluded
from the SSE list.   
This demonstrates that the selection criteria for SEP events have an impact on
the derived probability values and that it is important that these criteria
are documented in detail.

We have determined the correlation coefficients between the proton peak flux  
and the solar parameters for each individual
SEPEM energy channel (Figure~\ref{fig.corr_spectrum}).  
The correlation with flare intensity increases with energy, while the
correlation with CME speed shows the opposite behavior. 
Furthermore, the correlation with the CME speed is stronger than the
correlation with the flare intensity below 10~MeV and is weaker 
above 50~MeV. 
While the correlation coefficients derived here do not single out 
a specific acceleration mechanism, they do give an idea of the origin of the  
energetic particles. 
The two correlations coefficients are of similar size in the intermediate 
energy range which indicates a mixed-SEP origin where both flare and CME shock
driven acceleration contribute as proposed by \citet{Kal2003} and supported by
the recent work of \citet{tro2014}.  
Under this assumption, our results suggest that acceleration near the
flare site, or at least the conditions that lead to stronger flares, are
more important for the production of protons at higher energies than CME
related acceleration, while the situation seems opposite for less energetic 
protons.
This conclusion supports earlier findings based on data (see references in 
\citet{tro2014}).
Additionally, a recent model was developed allowing for the simultaneous
existence of acceleration in the corona as the main source for protons
$>50$~MeV together with CME-driven shock acceleration dominating at lower 
proton energies \citep{Koc2014}.
While the energy dependence deserves a more detailed study including more
data, this behavior is an important observation that needs to be taken 
into account when comparing studies based on different proton energy
ranges.
It also provides valuable information for the development of space weather
forecasts in regard to the effects resulting from low and high energy
protons.

We observe a strong rise in both the probability of SEP occurrence 
(Figure~\ref{fig.probflareint} and Table~\ref{tab.probflareint} in the
Appendix) and mean proton peak fluxes (Figure~\ref{fig.flux_vs_flareintens_bins} and
Table~\ref{table.flarecorr_values} in the Appendix) with increasing flare intensities. 
These results are quantitatively and qualitatively
similar to those of \citet{Bel2005} (their Figures~10 and 12).
The results from \citet{Par2010} show the same behavior, but the actual
probabilities differ significantly (their Table~1). 
Their derived correlation
coefficient between the peak flux for $E>$10~MeV and the flare intensity is
0.43, while we obtained $0.55\pm0.07$. 
We believe this difference is due to a difference in the criteria for
SEP event selection and for solar event association. 
While their flare sample is almost three times larger than considered here due to
the larger period covered (1976 to 2006), the number of SEP events are very similar
to ours (166 compared to 160 in the CRR2010 and 90 in the SSE list).
The correlation coefficient presented here is in agreement with the value
derived by \citet{Can2010}.
Our result is much larger than the value determined by \citet{Hwa2010}
(0.37).
They only analyzed SEP events that have both associated CMEs and
flares, while we do not require the 
presence of a CME to derive this correlation. 
Furthermore, their sample includes flares originating beyond the limb which
reduces the correlation 
with the proton peak flux as the observed flare intensity is underestimated.  
\citet{Cli2012} derived a value of 0.52 for events originating from
longitudes between west $20^{\circ}$ to $85^{\circ}$  
and our result is in very good agreement when limited to the same
  region ($0.52\pm0.10$). 
Our value is also in very good agreement with the correlation
  coefficient of $0.59\pm0.07$ obtained by \citet{Mit2013}.
Using proton peak fluxes around 25~MeV from events originating on the western
front-side of the Sun with respect to the 
observing instrument, \citet{Ric2014} found a
value (0.71) which is larger than ours. 
This small difference could be the consequence of the time period they
analyzed (2009 to 2012) and of including C and B flares.

Our analysis shows that the probability for SEP occurrence grows significantly
when the flare occurs more towards the western side of the Sun
(Figure~\ref{fig.probflareintlon} and Table~\ref{tab.probflareintlon} in the
Appendix), while 
the proton peak fluxes show almost no correlation with longitude
(Figure~\ref{fig.flux_vs_flarelong_cmewidth} and Table~\ref{table.corrcoeff}). 
Our results are in agreement with  \citet{Bel2005} which
demonstrated that the SEP occurrence probability is remarkably high for flares
westward of 70$^{\circ}$. 
Our derived probabilities as a function of longitude are comparable with the
longitudinal distribution of SEP-associated flare fractions (Figure 8 of  
their paper) in the M8 to X3 flare strength range.
In an earlier analysis by the same authors \citep{Kur2004}, an order of
magnitude difference was observed in the mean peak flux of protons with $E>$10~MeV
between events originating from longitudes smaller and larger than
$-30^{\circ}$ (their Figure 6). 
Comparing with Figure~\ref{fig.flux_vs_flarelong_cmewidth}, our results do not
confirm this difference 
due to the low statistics of events with longitude $<$$-30^{\circ}$. 
Considering events associated with solar flares located at longitudes $>$$-30^{\circ}$,
they observe an average peak flux of around 1000 cm$^{-2}$s$^{-1}$sr$^{-1}$,
while most of our events are below this value. 
Our observation on the dependence of the probability on
flare longitude shows a good qualitative agreement with the
result from \citet{Par2010} (their Figure~1), but also in this case their 
actual values differ significantly from ours.
Several studies have investigated the dependence of probability of SEP
occurrence on longitude by using the source location of the associated
CME \citep{Gop2008,Par2012} and also concluded that a larger fraction of events
originating at western longitudes result in SEP events.

We determined that the CME speed has a strong influence on the probability,
regardless of flare intensity (Figures~\ref{fig.probflarecmeintnhalo} and
\ref{fig.probcme}, and
Tables~\ref{tab.probflarecmeintnhalo} and \ref{tab.probcme} in the Appendix),
and on the mean 
peak fluxes (Figure~\ref{fig.flux_vs_cmespeed_bins} and
Table~\ref{table.cmecorr_values} in the Appendix).  
Although different selection criteria were used, the large increase of the
probability with the CME speed was also observed by \citet{Gop2008} and
\citet{Par2012}. 
The latter authors also derived a correlation coefficient between the CME speed
and the proton peak flux of 0.57.
Our value ($0.56\pm0.08$ for $E>$10~MeV) agrees well with this result, as
well as with those from \citet{Can2010} (about 0.6) and 
\citet{Mit2013} ($0.63\pm0.05$, although based on the logarithm of the CME
speed).
As was the case for the correlation with flare intensity, 
\citet{Hwa2010} measured a much smaller value ($0.31$) and
\citet{Ric2014} a significantly larger value ($0.66-0.71$) compared to ours.

From Figure~\ref{fig.probcme} (and Table~\ref{tab.probcme} in the Appendix) it
is clear that the 
SEP occurrence probability is significantly higher for halo CMEs compared to
non-halo CMEs, in qualitative agreement with  
the conclusions from \citet{Gop2008} and \citet{Par2012}.
We observed no dependence of the proton peak fluxes on the CME width 
(Figure~\ref{fig.flux_vs_flarelong_cmewidth} and Table~\ref{table.corrcoeff})
and derived a correlation coefficient of $0.16\pm0.12$ (for $E>$10~MeV). 
This is slightly smaller than the value observed by \citet{Mit2013}
($0.29\pm0.12$).  
\citet{Can2010} showed the relation between the proton
peak flux and the CME width (their Figure~10) without quantifying this 
correlation, and no strong dependence was observed which our results also
indicate. 
\citet{Par2010} looked at the correlation for partial and full halo CMEs
separately, finding no evidence for a correlation between the proton peak
flux and the CME width, a conclusion also supported by our results.

Several studies have already investigated the dependence of SEP probabilities
and peak fluxes on multiple solar parameters, but were limited
to a few combinations of two parameters or derived the dependence on one
variable in the specific range of another parameter. 
In this work, we derived the SEP occurrence probability taking fully into
account both flare and CME characteristics.
Figures ~\ref{fig.probflarecmeinthalo}--\ref{fig.probflarecmeintlonnhalo} (and
their respective
Tables~\ref{tab.probflarecmeinthalo}--\ref{tab.probflarecmeintlonnhalo} in the
Appendix) 
clearly indicate a differentiation of SEP occurrence probability when all
flare (location and intensity) and CME (velocity and width) characteristics
are taken into account with respect to probabilities derived from a single
parameter or combinations of specific source characteristics. 
Differences in results from the single parameter and multi-dimensional
analysis are expected, as the SEP occurrence probabilities rely on the
characteristics of events at the source of the SEPs (flare and 
CMEs).  
This conclusion also holds for the dependency of the proton peak fluxes on the
solar parameters as can be seen from
Figures~\ref{fig.flux_vs_fl_long_vs_fl_int}--\ref{fig.flux_vs_cme_w_vs_cme_v}
(and Tables~\ref{tab.flux_vs_fl_long_vs_fl_int}-\ref{tab.flux_vs_cme_w_vs_cme_v} in the
Appendix).  

The correlation coefficients between the proton peak flux and flare longitude
in different flare magnitude ranges for $E>$10~MeV are weak for the two M 
flare bins ($0.29\pm0.25$ and $0.30\pm0.14$) but remain consistent with 
zero for the X flares ($0.01\pm0.15$). 
The difference is even larger for the $E>$60~MeV range, where the correlations
are moderate for M flares ($0.52\pm0.19$ and $0.44\pm0.13$) but again
non-existent for X flares ($0.12\pm0.14$). 
The correlation with the CME speed also shows a dependency on flare
magnitude: it increases from $0.40\pm0.12$ for M flares
$\ge$M3.2 to $0.61\pm0.10$ for X flares for $E>$10~MeV.
This is consistent with the value of 0.69 that \citet{Par2012} derived for
X flares. 
The increase is not significant for $E>$60~MeV (from $0.27\pm0.16$ to
$0.38\pm0.17$). 
The uncertainties on the correlation coefficients for the weaker M flares are
too large to confirm this dependency towards smaller flares.
Within the uncertainties, there is no indication of a dependency on the flare
intensity of the correlations with the CME width.
\citet{Par2010} claimed to observe a decrease of the correlation with the
flare intensity going from eastern to western longitudes (0.65,0.42,0.38), while
\citet{Mit2013} observed no significant change when restricting events to 
western longitudes (their Table~7).
Our results do not show any significant dependency of the correlations on
longitude for $E>$10~MeV ($0.72\pm0.29$, $0.62\pm0.12$, $0.52\pm0.10$), while 
some decrease can be observed for the $E>$60~MeV energy range
($0.90\pm0.11$, $0.73\pm0.09$, $0.60\pm0.09$). 
\citet{Par2012} observed a large difference between the
correlations with the CME speed when the event originates from the central and
western region (0.78 {\it versus} 0.47), while \citet{Mit2013} did not observe a
significant change when analyzing events from more western longitudes (their
Table~7). 
Our derived values also show a significant decrease from $0.72\pm0.07$
($0.63\pm0.09$) to $0.45\pm0.12$ ($0.25\pm0.14$) for
$E>$10~MeV ($E>$60~MeV) going from central to western events. 

For CMEs with a speed slower than 1500~km~s$^{-1}$, the correlation coefficients
for both energy ranges remain consistent with zero, while the
values increase for faster CMEs but are still weak. 

When determining the proton peak flux during a SEP event, we have excluded
any enhancements from the passage of an interplanetary shock wave through the
observation point.
For this purpose, lists of interplanetary shocks observed at the {\it Wind}
and ACE spacecraft were used to identify ESP-like increases in the proton
data.  
We have compared the derived correlation coefficients with the values obtained
when the ESP contribution is not excluded, and found the differences to be
negligible (see Table~\ref{table.corrcoeff_edep}). 
This is not surprising, as only half of the events contain an ESP-like
increase and the secondary enhancements often stay below the initial peak
therefore not affecting the determination of the peak flux.
Only the correlation between the peak flux 
in the lowest SEPEM energy channel ($5.00$--$7.23$ MeV) and the CME speed
shows a small but not very significant decrease when the ESP contribution is
included. 
It can be expected that the effect of the ESP on the correlation coefficients
manifests itself mostly at the lower energies as the shock wave is
more efficient at accelerating particles at lower energies.

In conclusion, our statistical analysis confirms results previously reported
in the literature, and further expands our understanding of SEP occurrence
probability and characteristics, since it involves a multi-parameter study
based on both flare and CME characteristics. 
Furthermore, we observed an energy dependence of the correlation coefficient
between the proton peak flux and both flare intensity and CME speed, 
and only observed a small but not very significant 
decrease in the lowest energy channel when excluding the ESP contribution to
the flux.
Derived results represent valuable input for SEP forecast tools and are
incorporated within the Space Weather Alert System developed as part of the EU
FP7 project COMESEP (\citet{Cro2012}, http://www.comesep.eu).


%
\begin{acks}
This work has received funding from the European Union Seventh Framework
Programme (FP7/2007--2013) under grant agreement n. 263252 [COMESEP]. 
We also acknowledge the ESA SEPEM reference proton dataset.
\end{acks}

%
\appendix   

\begin{landscape}
\begin{center}
  \begin{longtable}[h]{ccrrrrrr}
    \caption{The SSE list and associated solar parameters as described in
      Section~\ref{subsec.eventlists} containing the following information:
      the SEP onset date and time, the integral peak flux 
      for E$>$10~MeV and E$>$60~MeV, the flare magnitude and location, and
      the CME speed and width. } \label{tab.sselist} \\

    \hline
    \multicolumn{2}{c}{SEP Onset} & 
    \multicolumn{2}{c}{$\quad$ Proton Peak Flux} & 
    \multicolumn{2}{c}{$\quad$ Flare Parameters} & 
    \multicolumn{2}{c}{$\quad$ CME Parameters} \\
    \multicolumn{1}{c}{Date} & 
    \multicolumn{1}{c}{Time} &
    \multicolumn{1}{r}{$\quad$ E$>$10~MeV} & 
    \multicolumn{1}{r}{E$>$60~MeV} & 
    \multicolumn{1}{r}{$\quad$ Magnitude} &
    \multicolumn{1}{r}{$\,$ Location} &
    \multicolumn{1}{r}{$\quad$ Speed} & 
    \multicolumn{1}{r}{Width} \\
    \multicolumn{1}{c}{YYYY-MM-DD} & 
    \multicolumn{1}{c}{HH:SS} &
    \multicolumn{1}{r}{[cm$^{-2}$s$^{-1}$sr$^{-1}$]} & 
    \multicolumn{1}{r}{[cm$^{-2}$s$^{-1}$sr$^{-1}$]} & 
    \multicolumn{1}{r}{} &
    \multicolumn{1}{r}{} &
    \multicolumn{1}{r}{[km~s$^{-1}$]} & 
    \multicolumn{1}{r}{[degrees]} \\

    \hline
    \endfirsthead
    
    \multicolumn{8}{c}%
    {{\bfseries \tablename\ \thetable{}} -- The SSE list and associated
      solar parameters (continued)} \\
    \multicolumn{8}{c}{}\\
    \hline
    \multicolumn{2}{c}{SEP Onset} & 
    \multicolumn{2}{c}{$\quad$ Proton Peak Flux} & 
    \multicolumn{2}{c}{$\quad$ Flare Parameters} & 
    \multicolumn{2}{c}{$\quad$ CME Parameters} \\
    \multicolumn{1}{c}{Date} & 
    \multicolumn{1}{c}{Time} &
    \multicolumn{1}{r}{$\quad$ E$>$10~MeV} & 
    \multicolumn{1}{r}{E$>$60~MeV} & 
    \multicolumn{1}{r}{$\quad$ Magnitude} &
    \multicolumn{1}{r}{$\,$ Location} &
    \multicolumn{1}{r}{$\quad$ Speed} & 
    \multicolumn{1}{r}{Width} \\
    \multicolumn{1}{c}{YYYY-MM-DD} & 
    \multicolumn{1}{c}{HH:SS} &
    \multicolumn{1}{r}{[cm$^{-2}$s$^{-1}$sr$^{-1}$]} & 
    \multicolumn{1}{r}{[cm$^{-2}$s$^{-1}$sr$^{-1}$]} & 
    \multicolumn{1}{r}{} &
    \multicolumn{1}{r}{} &
    \multicolumn{1}{r}{[km~s$^{-1}$]} & 
    \multicolumn{1}{r}{[degrees]} \\
    \hline
    \endhead

    \hline \multicolumn{8}{r}{{Continued on next page}} \\ \hline
    \endfoot

    \hline
    \endlastfoot

    1997-11-04 & 06:50 &      54 &     3.8 & X2.1 & 33W,14N & 785 & 110 \\ 
    1997-11-06 & 12:00 &     290 &      54 & X9.4 & 63W,18N & 1556 & 115 \\ 
    1998-04-20 & 12:55 &    1100 &      30 & M1.4 & 90W, -- & 1863 & 150 \\ 
    1998-04-29 & 23:15 &     3.5 &  0.0076 & M6.8 & 20E,18S & 1374 & 90 \\ 
    1998-05-02 & 13:55 &     130 &      12 & X1.1 & 15W,15N & 938 & 130 \\ 
    1998-05-06 & 08:25 &     160 &     8.2 & X2.7 & 65W,11N & 1099 & 90 \\ 
    1998-08-24 & 23:00 &     150 &     5.2 & X1.0 & 09E,35S & -- & -- \\ 
    1998-09-30 & 14:25 &     910 &     9.8 & M2.8 & 85W,19N & -- & -- \\ 
    1998-11-06 & 03:45 &      11 &  0.0080 & M8.4 & 18W,26N & 1118 & 60 \\ 
    1999-05-04 & 14:45 &     2.6 &   0.086 & M4.4 & 32E,15S & 1584 & 110 \\ 
    1999-06-04 & 08:00 &      48 &    0.43 & M3.9 & 69W,17N & 2221 & 80 \\ 
    2000-06-07 & 04:00 &      51 &   0.047 & X2.3 & 14E,20S & 1119 & 180 \\ 
    2000-06-10 & 18:00 &      42 &     2.1 & M5.2 & 40W,22N & 1108 & 120 \\ 
    2000-06-25 & 16:40 &     4.3 &  0.0031 & M1.9 & 55W,16N & 1617 & 70 \\ 
    2000-07-13 & 00:30 &     2.1 &  0.0029 & M5.7 & 64W,16N & 820 & 101 \\ 
    2000-07-14 & 11:00 &    6500 &     660 & X5.7 & 07W,22N & 1674 & 360 \\ 
    2000-07-22 & 12:00 &      18 &    0.40 & M3.7 & 56W,14N & 1230 & 80 \\ 
    2000-09-12 & 14:50 &     250 &    0.67 & M1.1 & 09W,17N & 1550 & 100 \\ 
    2000-11-08 & 23:45 &    8200 &     700 & M7.4 & 75W,10N & 1738 & 120 \\ 
    2000-11-24 & 07:00 &     8.9 &    0.40 & X2.0 & 03W,22N & 1289 & 360 \\ 
    2000-11-24 & 16:00 &      84 &     1.8 & X2.3 & 07W,22N & 1245 & 360 \\ 
    2000-11-25 & 18:00 &      25 &   0.071 & M8.2 & 50E,07S & 2519 & 120 \\ 
    2001-01-22 & 01:15 &     3.0 &   0.033 & M1.1 & 32W,05N & -- & -- \\ 
    2001-01-28 & 19:30 &      50 &    0.36 & M1.5 & 59W,04N & 916 & 120 \\ 
    2001-03-29 & 13:50 &      30 &    0.33 & X1.7 & 12W,16N & 942 & 360 \\ 
    2001-04-02 & 11:20 &     3.3 &   0.070 & X1.1 & 62W,16N & 992 & 50 \\ 
    2001-04-02 & 23:00 &     770 &      18 & X20 & 78W,17N & 2505 & 100 \\ 
    2001-04-09 & 16:00 &     5.0 &    0.47 & M7.9 & 04W,21N & 1192 & 360 \\ 
    2001-04-10 & 07:00 &     110 &    0.92 & X2.3 & 09W,23N & 2411 & 360 \\ 
    2001-04-12 & 12:00 &      42 &     2.1 & X2.0 & 42W,20N & 1184 & 120 \\ 
    2001-04-14 & 18:00 &     1.6 &  0.0010 & M1.0 & 71W,17N & 830 & 50 \\ 
    2001-04-15 & 14:00 &     920 &     190 & X14 & 84W,20N & 1199 & 110 \\ 
    2001-04-27 & 04:55 &     4.2 &   0.011 & M7.8 & 31W,17N & 1006 & 360 \\ 
    2001-09-24 & 12:00 &    2000 &      78 & X2.6 & 23E,16S & 2402 & 120 \\ 
    2001-10-09 & 14:00 &     5.1 &   0.019 & M1.4 & 10E,25S & 973 & 120 \\ 
    2001-10-19 & 04:55 &     5.5 &    0.17 & X1.6 & 18W,16N & 558 & 180 \\ 
    2001-10-19 & 16:00 &     9.7 &    0.29 & X1.6 & 29W,15N & 901 & 160 \\ 
    2001-10-22 & 04:00 &    0.14 & 0.00051 & M1.0 & 57W,17N & 772 & 20 \\ 
    2001-10-22 & 15:00 &      23 &    0.80 & M6.7 & 18E,21S & 1336 & 140 \\ 
    2001-11-04 & 16:55 &     950 &      92 & X1.0 & 18W,06N & 1810 & 360 \\ 
    2001-11-17 & 19:55 &      29 &   0.016 & M2.8 & 42E,13S & 1379 & 160 \\ 
    2001-11-22 & 21:00 &      19 &     2.0 & M3.8 & 67W,25N & 1443 & 120 \\ 
    2001-11-23 & 01:00 &    3400 &      39 & M9.9 & 34W,15N & 1437 & 270 \\ 
    2001-11-28 & 20:00 &     3.7 &   0.013 & M6.9 & 16E,04S & 500 & 90 \\ 
    2001-12-26 & 05:55 &     670 &      66 & M7.1 & 54W,08N & 1446 & 90 \\ 
    2002-03-16 & 08:05 &      15 &   0.016 & M2.2 & 03W,08N & 957 & 360 \\ 
    2002-03-22 & 12:00 &      14 & 0.00063 & M1.6 & 90W, -- & 1750 & 130 \\ 
    2002-04-17 & 11:30 &      23 &   0.084 & M2.6 & 34W,14N & 1240 & 70 \\ 
    2002-04-21 & 00:00 &    2300 &      60 & X1.5 & 84W,14N & 2393 & 120 \\ 
    2002-07-16 & 13:20 &      38 &    0.31 & X3.0 & 01E,14S & 1151 & 100 \\ 
    2002-08-14 & 03:35 &      25 &   0.036 & M2.3 & 54W,09N & 1309 & 60 \\ 
    2002-08-16 & 22:10 &     5.4 &   0.016 & M5.2 & 20E,14S & 1585 & 160 \\ 
    2002-08-18 & 20:00 &     5.1 &   0.023 & M2.2 & 19W,12N & 682 & 100 \\ 
    2002-08-19 & 10:00 &     3.3 &  0.0056 & M2.0 & 25W,12N & 549 & 80 \\ 
    2002-08-22 & 03:10 &      33 &     2.1 & M5.4 & 62W,07N & 998 & 80 \\ 
    2002-08-24 & 00:00 &     340 &      34 & X3.1 & 81W,02N & 1913 & 150 \\ 
    2002-11-09 & 17:10 &     330 &    0.48 & M4.6 & 29W,12N & 1838 & 90 \\ 
    2003-05-28 & 07:55 &      12 &    0.14 & X3.6 & 20W,06N & 1366 & 220 \\ 
    2003-05-31 & 03:00 &      25 &     1.1 & M9.3 & 65W,07N & 1835 & 150 \\ 
    2003-06-18 & 20:00 &      28 &   0.030 & M6.8 & 55E,07S & 1813 & 360 \\ 
    2003-10-23 & 00:45 &     4.6 &  0.0024 & M1.2 & 17E,03S & -- & -- \\ 
    2003-10-26 & 18:05 &     510 &     3.6 & X1.2 & 38W,02N & 1537 & 130 \\ 
    2003-10-28 & 12:00 &    5600 &     360 & X17 & 08E,16S & 2459 & 360 \\ 
    2003-10-29 & 22:00 &    2700 &     170 & X10 & 02W,15N & 2029 & 360 \\ 
    2003-11-02 & 18:00 &    1400 &      80 & X8.3 & 56W,14N & 2598 & 130 \\ 
    2003-11-04 & 21:00 &     390 &     4.5 & X28 & 83W,19N & 2657 & 130 \\ 
    2003-11-20 & 07:15 &      17 &    0.16 & M9.6 & 08W,03N & 669 & 90 \\ 
    2004-07-25 & 16:00 &      88 &    0.51 & M2.2 & 33W,04N & 1333 & 130 \\ 
    2004-09-19 & 18:05 &      61 &     1.1 & M1.9 & 58W,03N & -- & -- \\ 
    2004-11-07 & 02:50 &     6.1 &  0.0050 & M1.9 & 02W,07N & -- & -- \\ 
    2004-11-07 & 17:00 &     550 &     1.9 & X2.0 & 17W,09N & 1759 & 150 \\ 
    2004-11-09 & 19:00 &      89 &    0.53 & M8.9 & 51W,07N & 2000 & 130 \\ 
    2004-11-10 & 03:00 &     450 &     5.3 & X2.5 & 49W,09N & 2000 & 120 \\ 
    2005-01-15 & 08:25 &      12 &    0.39 & M8.6 & 06E,11S & 2049 & 90 \\ 
    2005-01-15 & 23:00 &     440 &     3.1 & X2.6 & 05W,15N & 2861 & 130 \\ 
    2005-01-17 & 10:00 &    4600 &     110 & X3.8 & 25W,15N & 2094 & 110 \\ 
    2005-01-20 & 06:00 &    2800 &     950 & X7.1 & 61W,14N & 882 & 80 \\ 
    2005-05-06 & 12:00 &     4.8 &   0.022 & M1.3 & 76W,04N & 1144 & 30 \\ 
    2005-05-13 & 21:00 &     430 &   0.086 & M8.0 & 11E,12S & 1689 & 360 \\ 
    2005-06-16 & 20:50 &      47 &     4.2 & M4.0 & 85W,09N & -- & -- \\ 
    2005-07-13 & 18:10 &      16 &    0.25 & M5.0 & 78W,11N & 1423 & 70 \\ 
    2005-07-14 & 11:00 &     130 &     1.1 & X1.2 & 89W,08N & 2115 & 80 \\ 
    2005-08-22 & 03:35 &     9.4 &    0.19 & M2.6 & 48W,09N & 1194 & 160 \\ 
    2005-08-22 & 18:00 &     320 &     1.8 & M5.6 & 60W,12N & 2378 & 100 \\ 
    2005-09-07 & 23:25 &    1100 &      19 & X17 & 89E,06S & -- & -- \\ 
    2005-09-13 & 22:00 &     200 &    0.88 & X1.5 & 10E,09S & 1866 & 130 \\ 
    2006-12-05 & 17:35 &      25 &    0.98 & X9.0 & 68E,07S & -- & -- \\ 
    2006-12-06 & 22:00 &    2300 &      42 & X6.5 & 64E,06S & -- & -- \\ 
    2006-12-13 & 02:00 &     740 &     110 & X3.4 & 23W,06N & 1774 & 180 \\ 
    2006-12-14 & 22:00 &     230 &     5.4 & X1.5 & 46W,07N & 1042 & 70 \\ 

  \end{longtable}
\end{center}
\end{landscape}

\begin{table}[tp]
  \caption{Probabilities of SEP occurrence and their respective errors as a function
    of flare magnitude derived from the CRR2010 list (roman fonts)
      and the SSE list (italic fonts).} 
  \label{tab.probflareint}
  \begin{tabular}{ccccc}
    \hline
    \multicolumn{5}{c}{Flare magnitude} \\
    M1-M3.9 & M4-M6.9 & M7-M9.9 & X1-X4.9 & $\ge$X5 \\
    \hline
    0.062$\pm$0.008 & 0.20$\pm$0.03 & 0.30$\pm$0.07 & 0.44$\pm$0.05 & 0.67$\pm$0.11 \\
    {\it 0.024$\pm$0.005} & {\it 0.10$\pm$0.03} & {\it 0.24$\pm$0.06} & {\it
      0.28$\pm$0.05} & {\it 0.67$\pm$0.11} \\
    \hline
  \end{tabular}
\end{table}

\begin{table}[tp]
  \caption{Probabilities of SEP occurrence and their respective
    errors as a function of flare magnitude
    and flare longitude derived from the CRR2010 list (roman fonts) and
      the SSE list (italic fonts).} 
  \label{tab.probflareintlon}
  \begin{tabular}{cccccc}
    \hline
    Flare longitude & \multicolumn{5}{c}{Flare magnitude} \\
    & M1-M3.9 & M4-M6.9 & M7-M9.9 & X1-X4.9 & $\ge$X5 \\
    \cline{2-6}
    [-90$^{\circ}$,-71$^{\circ}$] & 0.03$\pm$0.02 & 0. & 0. & 0.3$\pm$0.19 &
    0.43$\pm$0.27 \\

    & {\it 0.} & {\it 0.} & {\it 0.} & {\it 0.} & {\it 0.33$\pm$0.27} \\

    [-70$^{\circ}$,-31$^{\circ}$] & 0.014$\pm$0.008 & 0.17$\pm$0.07 &
    0.18$\pm$0.12 & 0.09$\pm$0.06 & 0.40$\pm$0.22 \\
    
    & {\it 0.005$\pm$0.005} & {\it 0.07$\pm$0.05} & {\it 0.09$\pm$0.09} &
    {\it 0.} & {\it 0.40$\pm$0.22} \\

    [-30$^{\circ}$,29$^{\circ}$] & 0.06$\pm$0.01 & 0.22$\pm$0.06 &
    0.30$\pm$0.10 & 0.60$\pm$0.08 & 0.75$\pm$0.22 \\

    & {\it 0.020$\pm$0.008} & {\it 0.10$\pm$0.04} & {\it 0.22$\pm$0.09} & {\it
      0.51$\pm$0.08} & {\it 0.75$\pm$0.22} \\

    [30$^{\circ}$,69$^{\circ}$] & 0.10$\pm$0.02 & 0.21$\pm$0.07 &
    0.50$\pm$0.16 & 0.48$\pm$0.10 & 1. \\

    & {\it 0.05$\pm$0.01} & {\it 0.12$\pm$0.06} & {\it 0.50$\pm$0.16} & {\it
      0.30$\pm$0.10} & {\it 1.} \\

    [70$^{\circ}$,90$^{\circ}$] & 0.10$\pm$0.03 & 0.42$\pm$0.14 &
    0.20$\pm$0.18 & 0.58$\pm$0.14 & 1. \\

    & {\it 0.032$\pm$0.016} & {\it 0.25$\pm$0.13} & {\it 0.20$\pm$0.18} & {\it
      0.25$\pm$0.13} & {\it 1.}  \\

    \hline
  \end{tabular}
\end{table}

\begin{table}[p]
  \caption{Probabilities of SEP occurrence and their respective
    errors as a function of flare magnitude for halo CMEs.}
  \label{tab.probflarecmeinthalo}
  \begin{tabular}{ccc}
    \hline
    \multicolumn{3}{c}{Flare magnitude} \\
    M1-M3.9 & M4-M9.9 & $\ge$X1\\
    \hline
    0.48$\pm$0.07 & 0.64$\pm$0.08 & 0.62$\pm$0.07 \\
    \hline
  \end{tabular}
\end{table}

\begin{table}[p]
  \caption{Probabilities of SEP occurrence and their respective
    errors as a function of flare magnitude and CME velocity for non-halo
    CMEs.}\label{tab.probflarecmeintnhalo}
  \begin{tabular}{cccc}
    \hline
    & \multicolumn{3}{c}{CME velocity} \\
    & 0-499 km~s$^{-1}$ & 500-999 km~s$^{-1}$ & $\ge$1000 km~s$^{-1}$ \\
    \cline{2-4}
    M1-M3.9 & 0.03$\pm$0.02 & 0.10$\pm$0.03 & 0.39$\pm$0.10 \\

    M4-M9.9 & 0.11$\pm$0.11 & 0.22$\pm$0.09 & 0.25$\pm$0.15 \\

    $\ge$X1  & 0.09$\pm$0.09 & 0.25$\pm$0.15 & 0.60$\pm$0.16 \\
    \hline
  \end{tabular}
\end{table}

\begin{table}[p]
  \caption{Probabilities of SEP occurrence and their respective
    errors as a function of flare magnitude and longitude for halo
    CMEs.}\label{tab.probflarecmeintlonhalo}
  \begin{tabular}{cccc}
    \hline
    & \multicolumn{3}{c}{Flare magnitude} \\
    & M1-M3.9 & M4-M9.9 & $\ge$X1\\
    \cline{2-4}
    [-90$^{\circ}$,-1$^{\circ}$] & 0.26$\pm$0.09 & 0.47$\pm$0.12 & 0.25$\pm$0.11 \\

    [0$^{\circ}$,90$^{\circ}$] & 0.67$\pm$0.09 & 0.82$\pm$0.09 & 0.79$\pm$0.07 \\
    \hline
  \end{tabular}
\end{table}

\begin{table}[p]
  \caption{Probabilities of SEP occurrence  and their respective
    errors as a function of flare magnitude, flare longitude and CME velocity
    for non-halo CMEs. }
  \label{tab.probflarecmeintlonnhalo}
  \begin{tabular}{cccc}
    \hline
    & \multicolumn{3}{c}{CME velocity} \\
    & 0-499 km~s$^{-1}$ & 500-999 km~s$^{-1}$ & $\ge$1000 km~s$^{-1}$ \\
    \cline{2-4}
    M1-M3.9 and [-90$^{\circ}$,-1$^{\circ}$] & 0 & 0.02$\pm$0.02 & 0.11$\pm$0.11 \\

    M4-M9.9 and [-90$^{\circ}$,-1$^{\circ}$] & 0 & 0.13$\pm$0.12 & 0.25$\pm$0.15 \\

    $\ge$X1 and [-90$^{\circ}$,-1$^{\circ}$] & 0 & 0.20$\pm$0.18 & 0.25$\pm$0.22 \\

    M1-M3.9 and [0$^{\circ}$,90$^{\circ}$] & 0.05$\pm$0.30 & 0.20$\pm$0.06 &
    0.57$\pm$0.13 \\

    M4-M9.9 and [0$^{\circ}$,90$^{\circ}$] & 0.20$\pm$0.18 & 0.29$\pm$0.12 & 0 \\

    $\ge$X1 and [0$^{\circ}$,90$^{\circ}$] & 0.20$\pm$0.18 & 0.33$\pm$0.27 &
    0.83$\pm$0.15 \\
    \hline
  \end{tabular}
\end{table}

\begin{table}[p]
  \caption{Probabilities of SEP occurrence and their respective
    errors as a function of CME velocity for all CMEs, non-halo and halo 
    CMEs.}\label{tab.probcme}
  \begin{tabular}{cccc}
    \hline
    & \multicolumn{3}{c}{CME velocity} \\
    & 0-499 km~s$^{-1}$ & 500-999 km~s$^{-1}$ & $\ge$1000 km~s$^{-1}$ \\
    \cline{2-4}
    all CMEs & 0.06$\pm$0.02 & 0.22$\pm$0.03 & 0.56$\pm$0.04 \\

    non-halo CMEs & 0.04$\pm$0.02 & 0.13$\pm$0.03 & 0.41$\pm$0.08 \\

     halo CMEs    & 0.30$\pm$0.14 & 0.51$\pm$0.08 & 0.64$\pm$0.05 \\
    \hline
  \end{tabular}
\end{table}

\begin{table}[t]
\caption{Mean and RMS of the logarithm of the proton peak flux for $E>$10~MeV and
  $E>$60~MeV in the five flare magnitude  bins. Peak fluxes are
  expressed as log(cm$^{-2}$s$^{-1}$sr$^{-1}$).} 
\label{table.flarecorr_values}  
\begin{tabular}{ccc}                                
  \hline                   
Flare intensity bin & Peak flux   & Peak flux  \\ 
          &  $E>$10~MeV  & $E>$60~MeV \\ 
\hline
M$1.0 \le I_{\mathrm{f}} < $M$2.5$ & $ 1.24\pm 0.75$ & $-1.38\pm 1.21$ \\ 
M$2.5 \le I_{\mathrm{f}} < $M$6.3$ & $ 1.47\pm 0.73$ & $-0.49\pm 0.97$ \\ 
M$6.3 \le I_{\mathrm{f}} < $X$1.6$ & $ 1.88\pm 1.00$ & $-0.07\pm 1.39$ \\ 
X$1.6 \le I_{\mathrm{f}} < $X$5.0$ & $ 2.02\pm 0.83$ & $\phantom{+}0.32\pm 1.02$ \\ 
 $ I_{\mathrm{f}} \ge$ X$5.0$       & $ 3.02\pm 0.66$ & $\phantom{+}1.77\pm 0.88$ \\ 
  \hline
\end{tabular}

\end{table}

\begin{table}[t]
\caption{Mean and RMS of the logarithm of the proton peak flux for $E>$10~MeV and
  $E>$60~MeV in the five CME speed bins. Peak fluxes
  are expressed as log(cm$^{-2}$s$^{-1}$sr$^{-1}$). } \label{table.cmecorr_values} 
\begin{tabular}{ccc}                                 
  \hline                   
CME speed bin & Peak flux   & Peak flux  \\ 
 &  $E>$10~MeV  & $E>$60~MeV \\ 
\hline
$400 \le v_{\mathrm{CME}} < 1000$ km~s$^{-1}$  & $1.21 \pm 0.79$ & $-0.69 \pm 1.41$ \\ 
$1000 \le v_{\mathrm{CME}} < 1400$ km~s$^{-1}$ & $1.43 \pm 0.61$ & $-0.56 \pm 1.11$ \\ 
$1400 \le v_{\mathrm{CME}} < 1800$ km~s$^{-1}$ & $2.21 \pm 1.15$ & $\phantom{+}0.22 \pm
1.85$ \\ 
$1800 \le v_{\mathrm{CME}} < 2200$ km~s$^{-1}$ & $2.39 \pm 0.79$ & $\phantom{+}0.57 \pm
1.17$ \\ 
$ v_{\mathrm{CME}} \ge 2200$ km~s$^{-1}$ & $2.66 \pm 0.73$ & $\phantom{+}0.84 \pm
1.14$ \\ 
  \hline
\end{tabular}
\end{table} 

\begin{table}[t]
\caption{Mean and RMS of the logarithm of the proton peak flux
  in the nine flare magnitude and location bins. Peak fluxes are
  expressed as
  log(cm$^{-2}$s$^{-1}$sr$^{-1}$) for $E>$10~MeV and $E>$60~MeV.}
  \label{tab.flux_vs_fl_long_vs_fl_int} 
\begin{tabular}{cccc}
  \hline
  & M$1.0 \le I_{\mathrm{f}} < $M$3.2$ & M$3.2 \le I_{\mathrm{f}} < $X$1.0$ &  $I_{\mathrm{f}} \ge $X$1.0$ \\
  \hline
  & \multicolumn{3}{c}{proton peak flux $E>$10~MeV} \\
  $-90^{\circ} \le L_{\mathrm{f}} < -30^{\circ}$ & $\phantom{+}1.46 \pm 1.46$ &
  $\phantom{+}1.09 \pm 0.58$ & $\phantom{+}2.60 \pm 1.06$ \\
  $-30^{\circ} \le L_{\mathrm{f}} < \phantom{+}30^{\circ}$  & $\phantom{+}0.99 \pm 0.65$
  & $\phantom{+}1.46 \pm 0.88$ & $\phantom{+}2.27 \pm 1.01$ \\
  $\phantom{+}30^{\circ} \le L_{\mathrm{f}} < \phantom{+}90^{\circ}$ &  $\phantom{+}1.51
  \pm 0.84$ & $\phantom{+}1.82 \pm 0.99$  & $\phantom{+}2.46 \pm 0.73$  \\
  \hline
  & \multicolumn{3}{c}{proton peak flux $E>$60~MeV} \\
  $-90^{\circ} \le L_{\mathrm{f}} < -30^{\circ}$ & $-1.80 \pm 1.80$ &
  $-1.25 \pm 0.24$ &$\phantom{+}0.97 \pm 0.86$   \\
  $-30^{\circ} \le L_{\mathrm{f}} < \phantom{+}30^{\circ}$  & $-1.79 \pm 0.80$ 
  & $-0.68\pm 1.22$ & $\phantom{+}0.56 \pm 1.31$  \\
  $\phantom{+}30^{\circ} \le L_{\mathrm{f}} < \phantom{+}90^{\circ}$ & $-0.86
  \pm 1.34$ & $\phantom{+}0.13 \pm 1.35$ & $\phantom{+}1.05 \pm 0.98$  \\
  \hline

\end{tabular}
\end{table}

\begin{table}[t]
\caption{Mean and RMS of the logarithm of the proton peak flux
  in the nine flare magnitude and CME speed bins. Peak fluxes  are
  expressed as 
  log(cm$^{-2}$s$^{-1}$sr$^{-1}$) for $E>$10~MeV and $E>$60~MeV.}
  \label{tab.flux_vs_cme_v_vs_fl_int} 
\begin{tabular}{cccc}
  \hline
  & M$1.0 \le I_{\mathrm{f}} < $M$3.2$ & M$3.2 \le I_{\mathrm{f}} < $X$1.0$ &  $I_{\mathrm{f}} \ge $X$1.0$ \\
  \hline
  &\multicolumn{3}{c}{proton peak flux $E>$10~MeV} \\
%
  $400 \le v_{\mathrm{CME}} < 1200~\mathrm{km s^{-1}}$& $\phantom{+}0.93 \pm 0.40$ &
  $\phantom{+}0.96 \pm 0.48$ & $\phantom{+}1.81\pm 0.83$   \\
  $1200 \le v_{\mathrm{CME}} < 2000~\mathrm{km s^{-1}}$& $\phantom{+}1.67 \pm 0.76$
  & $\phantom{+}1.87 \pm 1.10$ & $\phantom{+}2.34 \pm 0.85$  \\
  $v_{\mathrm{CME}} \ge 2000~\mathrm{km s^{-1}}$ & $-$ & 
  $\phantom{+}1.72 \pm 0.55$ & $\phantom{+}2.97 \pm 0.56$  \\
  \hline
  &\multicolumn{3}{c}{proton peak flux $E>$60~MeV} \\
%
  $400 \le v_{\mathrm{CME}} < 1200~\mathrm{km s^{-1}}$ & $-1.47 \pm 0.64$ &
  $-1.12 \pm 1.14$ &$\phantom{+}0.31 \pm 1.30$   \\
  $1200 \le v_{\mathrm{CME}} < 2000~\mathrm{km s^{-1}}$ & $-1.13 \pm 1.47$ & $-0.03
  \pm 1.48$ & $\phantom{+}0.79 \pm 1.18$  \\
  $v_{\mathrm{CME}} \ge 2000~\mathrm{km s^{-1}}$ & $-$ & 
  $-0.39 \pm 0.50$ & $\phantom{+}1.30 \pm 0.89$  \\ 
  \hline

\end{tabular}
\end{table}

\begin{table}[tp]
\caption{Mean and RMS of the logarithm of the proton peak flux
  in the nine flare magnitude and CME width bins. Peak fluxes are
  expressed as 
  log(cm$^{-2}$s$^{-1}$sr$^{-1}$) for $E>$10~MeV and $E>$60~MeV.}
\label{tab.flux_vs_cme_w_vs_fl_int} 
\begin{tabular}{cccc}
  \hline
  & M$1.0 \le I_{\mathrm{f}} < $M$3.2$ & M$3.2 \le I_{\mathrm{f}} < $X$1.0$ &  $I_{\mathrm{f}} \ge $X$1.0$ \\
  \hline
  & \multicolumn{3}{c}{proton peak flux $E>$10~MeV} \\
  $0^{\circ}\le w_{\mathrm{CME}} < 100^{\circ}$ & $\phantom{+}0.92 \pm 0.42$ &
  $\phantom{+}1.41 \pm 0.72$ &$\phantom{+}2.13 \pm 1.05$   \\
  $100^{\circ} \le w_{\mathrm{CME}} < 360^{\circ}$  & $\phantom{+}1.57 \pm
  0.80$ & $\phantom{+}1.70 \pm 1.13$ & $\phantom{+}2.36 \pm 0.79$  \\
  $w_{\mathrm{CME}} = 360^{\circ}$  & $\phantom{+}1.18 \pm 1.18$ &
  $\phantom{+}1.68 \pm 1.09$ & $\phantom{+}2.49 \pm 1.16$  \\
  \hline
  & \multicolumn{3}{c}{proton peak flux $E>$60~MeV} \\
  $0^{\circ}\le w_{\mathrm{CME}} < 100^{\circ}$ & $-1.79 \pm 0.59$ &
  $-0.62 \pm 1.15$ &$\phantom{+}0.71 \pm 1.51$   \\
  $100^{\circ} \le w_{\mathrm{CME}} < 360^{\circ}$  & $-0.95 \pm 1.33$ &
  $-0.13 \pm 1.44$ & $\phantom{+}0.74 \pm 1.07$  \\
  $w_{\mathrm{CME}} = 360^{\circ}$  & $-1.81 \pm 1.81$ & $-0.58 \pm 1.55$ &
  $\phantom{+}0.99 \pm 1.47$  \\
  \hline
\end{tabular}
\end{table}

\begin{table}[tp]
\caption{Mean and RMS of the logarithm of the proton peak flux
  in the six CME speed and width bins. Peak fluxes are expressed as 
  log(cm$^{-2}$s$^{-1}$sr$^{-1}$) for $E>$10~MeV and $E>$60~MeV.}
\label{tab.flux_vs_cme_w_vs_cme_v} 
\begin{tabular}{ccc}
  \hline
  & $ v_{\mathrm{CME}} < 1500$~km~s$^{-1}$  &  $v_{\mathrm{CME}} \ge 1500$~km~s$^{-1}$ \\
  \hline
  & \multicolumn{2}{c}{proton peak flux $E>$10~MeV} \\
  $0^{\circ}\le w_{\mathrm{CME}} < 100^{\circ}$ & $\phantom{+}1.42 \pm 0.88$ &
  $\phantom{+}1.61 \pm 0.75$ \\
  $100^{\circ} \le w_{\mathrm{CME}} < 360^{\circ}$  & $\phantom{+}1.51 \pm
  0.76$ & $\phantom{+}2.45 \pm 0.88$ \\
  $w_{\mathrm{CME}} = 360^{\circ}$  & $\phantom{+}1.14 \pm 0.49$ &
  $\phantom{+}2.87 \pm 0.89$ \\
  \hline
  & \multicolumn{2}{c}{proton peak flux $E>$60~MeV} \\
  $0^{\circ}\le w_{\mathrm{CME}} < 100^{\circ}$ & $-0.54 \pm 1.52$ &
  $-0.71 \pm 1.02$ \\
  $100^{\circ} \le w_{\mathrm{CME}} < 360^{\circ}$  & $-0.34 \pm
  1.19$ & $\phantom{+}0.57 \pm 1.41$ \\
  $w_{\mathrm{CME}} = 360^{\circ}$  & $-0.79 \pm 0.89$ &
  $\phantom{+}0.99 \pm 1.82$ \\
  \hline
\end{tabular}
\end{table}

\clearpage

%
%

\bibliographystyle{spr-mp-sola}
\bibliography{sep_analysis_references}  
%
%
%
%

\end{article} 
\end{document}